\pdfoutput=1
\newcommand*{\ATLASLATEXPATH}{}
\documentclass[cernpreprint, atlasdraft=true, texlive=2016, UKenglish]{\ATLASLATEXPATH atlasdoc}
 
\usepackage{todonotes}
\DeclareBoolOption[true]{LANGEDIT}
\LEcontact{Chip Brock brockr@msu.edu}

\usepackage{pifont}
\usepackage{xcolor}
\usepackage{pdflscape}
\usepackage{\ATLASLATEXPATH atlaspackage}
\usepackage{\ATLASLATEXPATH atlasbiblatex}
 
\usepackage{\ATLASLATEXPATH atlasphysics}
\usepackage{multirow}
 
\graphicspath{{logos/}{figures/}}
 
\usepackage{ANA-EXOT-2019-13-PAPER-defs}

\AtlasTitle{Search for pairs of scalar leptoquarks decaying into quarks and electrons or muons in $\sqrt{s}=13$~TeV  $pp$ collisions with the ATLAS detector}
 
\AtlasAbstract{
A search for new-physics resonances decaying into a lepton and a jet performed by the ATLAS experiment is presented.
Scalar leptoquarks pair-produced in \(pp\) collisions at $\sqrt{s}=13$~TeV at the Large Hadron Collider
are considered using an integrated luminosity of 139~fb$^{-1}$, corresponding to the full Run 2 dataset.
They are searched for in events with two electrons or two muons and
two or more jets, including jets identified as arising from the fragmentation of $c$- or $b$-quarks.
The observed yield in each channel is consistent with the Standard Model background expectation.
Leptoquarks with masses below 1.8~TeV and 1.7~TeV are excluded in the electron and muon channels, respectively,
assuming a branching ratio into a charged lepton and a quark of 100\%, with minimal dependence on the quark flavour.
Upper limits on the aforementioned branching ratio are also given as a function of the leptoquark mass. }
 
\author{The ATLAS Collaboration}
 
\AtlasRefCode{EXOT-2019-13}

\AtlasJournal{JHEP}
\AtlasJournalRef{JHEP 10 (2020) 112}
\AtlasDOI{https://doi.org/10.1007/JHEP10(2020)112}
\journal{JHEP}
\PreprintIdNumber{CERN-EP-2020-084}
 
\AtlasCoverSupportingNote{Supporting documentation}{https://cds.cern.ch/record/2689896}
\AtlasCoverSupportingNote{Note showing selection change in $q\ell$ channels}{https://https://cds.cern.ch/record/2718124}

\hypersetup{pdftitle={ATLAS document},pdfauthor={The ATLAS Collaboration}}
 
\begin{document}
 
\maketitle

\section{Introduction}
\label{sec:intro}
 
Leptoquarks (LQs) are hypothetical colour-triplet particles that carry both baryon and lepton quantum numbers ($B\ne0, \ L\ne0$).
As such, LQs couple simultaneously to both quarks and leptons, enabling direct transitions
between the two. The spin of a LQ state is either 0 (scalar LQ) or 1 (vector LQ), and only the former is considered in this paper.
Because of their SU(3) and SU(2) charge (colour and weak isospin, respectively), LQs can mediate flavour-changing neutral currents,
and enable the violation of lepton flavour universality, which has been suggested as an explanation of recent
measurements of $B$-meson decays~\cite{Hiller:2014yaa,Gripaios:2014tna,Freytsis:2015qca,Bauer:2015knc,DiLuzio:2017chi,Buttazzo:2017ixm,Aaij:2014ora}.
New-physics models involving LQs might also resolve several interesting physical phenomena observed in nature.
For instance, LQs can be used to explain the origins of the neutrino masses~\cite{Pas:2015hca,Dorsner:2017wwn,Popov:2016fzr,Babu:2010vp},
as well as the origins of CP violation, thereby explaining the observed matter/antimatter asymmetry in the
universe~\cite{Barger:2000gv,PhysRevD.68.116004}. 
In addition, LQs could provide a satisfying connection between the apparent symmetry of lepton and quark generations,
as well as unification of the electromagnetic and weak forces at high energy~\cite{PhysRevLett.32.438, FRITZSCH1975193}.
 
\begin{figure}[htbp]
\centering
\includegraphics[scale = 0.55]{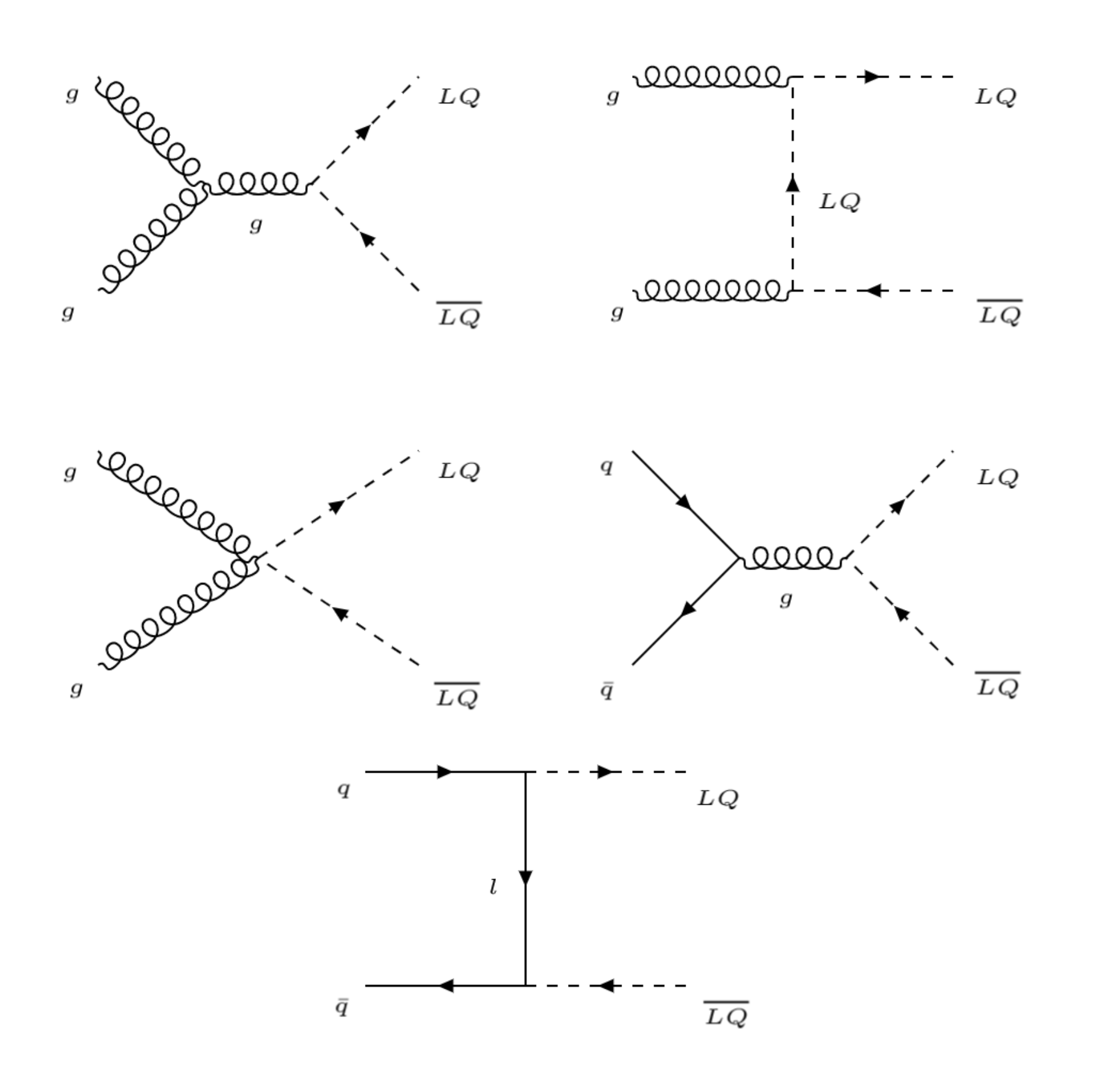}
\caption{The primary mechanisms by which LQs can be pair produced at the LHC, gluon--gluon and quark--antiquark initiated, are shown.}
\label{fig:ProductionMechanisms}
\end{figure}

At the LHC, the pair production of LQs is possible via gluon--gluon fusion and quark--antiquark annihilation,
as shown in Figure~\ref{fig:ProductionMechanisms}, including strong and lepton $t$-channel exchange
production~\cite{Dorsner:2018ynv}. Only the lowest order is shown for the primary mechanisms by which LQs
can be pair produced at the LHC, gluon--gluon and quark--antiquark initiated.
The production cross-section largely depends only on the mass of the LQ, $m_{\mathrm{LQ}}$.
The cross-section is taken to be equivalent to that calculated~\cite{Beenakker:2016lwe,Beenakker:1997ut,Beenakker:2010nq,Beenakker:2016gmf}
for the direct pair production of top squarks ($\tilde{t}$),
the supersymmetric partners of the top quark, as both are massive, coloured, scalar particles with the same production modes.\footnote{Recent calculations~\cite{Borschensky:2020hot} show that diagrams involving $t$-channel lepton exchange might lead to corrections to the total cross-section at the percent level. Those are not taken into account for the interpretation of the results, but effects are expected to be within the uncertainties of the calculated cross-sections~\cite{Beenakker:2016lwe,Beenakker:1997ut,Beenakker:2010nq,Beenakker:2016gmf}. }
Single production in association with a lepton is also possible, but the cross-section is model-dependent and it is not considered in this paper.
 
LQs are assumed to couple to the quark--lepton pair via a single Yukawa interaction, with decays involving either
charged leptons or neutrinos. 
The couplings are determined by two parameters, the model parameter $\beta$ and the coupling parameter $\lambda$.
The coupling to the charged lepton is given by $\sqrt{\beta \lambda}$ and the coupling to the neutrino by
$\sqrt{1-\beta\lambda}$. Only the case of decays via electrons and muons is addressed in this paper.
A traditional approach to LQ decay (such as in the Buchm\"{u}ller--R\"{u}ckl--Wyler model~\cite{BUCHMULLER1987442}), is to assume that LQs interact only with leptons and quarks of the same generation. This paper relaxes that restriction and considers cross-generational LQ decays.
While the results are interpreted assuming one decay mode at a time (100\% branching ratio, $\mathcal{B}=1$), LQs
with cross-generational decays might provide a possible solution to the anomalies in $B$-meson decays as observed
by LHCb~\cite{Aaij:2019kcg} if mixed decays into charged leptons (e.g.\ $\mathrm{LQ} \to b \mu$ and $\to s \mu$) are allowed.
The couplings to leptons and quarks are small such that LQs have narrow decay
widths ($<10\%$ of $m_{\mathrm{LQ}}$) and on-shell production dominates.
 
This paper presents a dedicated search for the pair production of LQs using the complete Run 2
dataset of $139$~fb$^{-1}$ of proton--proton (\(pp\)) collision data with $\sqrt{s}= 13$~\TeV.
Events are selected by requiring an oppositely charged electron or muon ($\ell = e, \mu$) pair and at least two jets
that may be identified as originating from the fragmentation of $c$- or $b$-quarks (referred
to as $c$-jets and $b$-jets, respectively) using dedicated tagging
algorithms. The LQ decay channels that are searched for are therefore
$eq$, $\mu q$, $ec$, $\mu c$, $eb$, and $\mu b$, where $q$ is a $u$-, $d$- or
$s$-quark.
The results are presented as a function of $m_{\mathrm{LQ}}$.
This paper reports the first dedicated ATLAS search for cross-generational
LQ decays using $c$- and $b$-jet identification.
 
The most recent searches for scalar leptoquark pairs from ATLAS and
CMS were performed using $36.1$~fb$^{-1}$ of integrated luminosity at
a 13~\TeV\ centre-of-mass energy. A search by ATLAS for first- and second-generation
LQs~\cite{EXOT-2016-08} did not use $b$-tagging in the
signal regions and so excluded LQs decaying with 100\% branching ratio
($\mathcal{B}$) into $eQ$ or $\mu Q$, where $Q=u$, $d$, $s$, $c$ or $b$,
below a mass of 1400~\GeV. CMS has also searched for first-generation~\cite{CMS-EXO-17-009}
and second-generation~\cite{CMS-EXO-17-003} LQ pairs, excluding masses below 1435~\GeV\
and 1530~\GeV\ respectively for $\mathcal{B}=1$.  ATLAS has searched
for third-generation up- and down-like LQ pairs, decaying into $t\nu/b\tau$ or
$b\nu/t\tau$~\cite{Aaboud:2019bye} with limits on LQ
masses up to 1100~\GeV.
CMS has excluded third-generation LQs decaying into $\tau t$~\cite{CMS-B2G-16-028}
for $m_{\mathrm{LQ}}<900$~\GeV\ and $\tau b$~\cite{CMS-B2G-16-016} for $m_{\mathrm{LQ}}<1020$~\GeV, and
cross-generational LQ decays into $\mu t$~\cite{CMS-B2G-16-027} for $m_{\mathrm{LQ}}<1420$~\GeV.
Searches for new physics in $\ell$+$b$-jets events have also  been performed by ATLAS using
$36.1$~fb$^{-1}$ of Run 2 data, targeting $B-L$ $R$-parity-violating
supersymmetric models, and top
squarks in particular~\cite{Aaboud:2017opj}. As the production cross-section and decay
modes of top squarks are equivalent to those of LQs, the exclusion limits on
$m^{}_{\tilde{t}}$ can be directly translated into $m_{\mathrm{LQ}}$
constraints. That search excludes top squarks with masses between 600 and 
1500~\GeV\ depending on the branching ratio into charged leptons and
$b$-quarks.

\section{The ATLAS detector}
\label{sec:detector}
 
The ATLAS detector~\cite{PERF-2007-01, } is a multipurpose particle physics detector with a forward--backward symmetric cylindrical
geometry and nearly 4$\pi$ coverage in solid angle.\footnote{ATLAS uses a right-handed coordinate system with its origin at the nominal
interaction point in the centre of the detector. The positive $x$-axis is defined by the direction from the interaction point to the centre
of the LHC ring, with the positive $y$-axis pointing upwards, while the beam direction defines the $z$-axis. Cylindrical coordinates
$(r,\phi)$ are used in the transverse plane, $\phi$ being the azimuthal angle around the $z$-axis. The component of momentum
in the transverse plane is denoted by \pt. The pseudorapidity $\eta$ is
defined in terms of the polar angle $\theta$ by $\eta=-\ln\tan(\theta/2)$. Rapidity is defined as $y$ = $0.5 \ln[(E + p_z)/(E-p_z)]$
where $E$ denotes the energy, and $p_z$ is the component of the momentum along the beam direction.
The separation of two objects in $\eta$--$\phi$ space is given by $\Delta R=\sqrt{(\Delta \eta)^2 + (\Delta \phi)^2}$.}
The inner tracking detector consists of silicon pixel and  microstrip detectors
covering the pseudorapidity region $|\eta| < 2.5$, surrounded by a transition radiation tracker
which enhances electron identification in the region $|\eta| < 2.0$.
Between Run 1 and Run 2, a new inner pixel layer, the insertable B-layer~\cite{ATLAS-TDR-2010-19, PIX-2018-001}, was added at a mean sensor radius of
3.3~cm.
The inner detector (ID) is surrounded by a thin superconducting solenoid providing an axial 2~T magnetic field, and by a fine-granularity
lead/liquid-argon (LAr) electromagnetic calorimeter covering $|\eta| < 3.2$. A steel/scintillator-tile calorimeter provides hadronic coverage
in the central pseudorapidity range ($|\eta| < 1.7$).
The endcap and forward regions ($1.5 < |\eta| < 4.9$) of the hadronic calorimeter are made of LAr active layers with either copper or tungsten
as the absorber material.
An extensive muon spectrometer (MS) with an air-core toroidal magnet system surrounds the calorimeters.
Three layers of high-precision tracking chambers provide coverage in the range $|\eta| < 2.7$, while
dedicated fast chambers allow triggering in the region $|\eta| < 2.4$.
The ATLAS trigger system consists of a hardware-based level-1 trigger followed by a software-based high-level trigger~\cite{TRIG-2016-01}.
 
\section{Data and Monte Carlo samples}
\label{sec:datamc}
The data analysed in this study correspond to 139~fb$^{-1}$ of \(pp\) collision
data collected by the ATLAS detector between 2015 and 2018 with a centre-of-mass energy of 13~\TeV\ and a
25\,ns proton bunch crossing interval. The uncertainty in the combined 2015--2018 integrated
luminosity is 1.7\%~\cite{ATLAS-CONF-2019-021},
obtained using the LUCID-2 detector~\cite{LUCID2} for the primary luminosity measurements.
All detector subsystems were required to be operational during data taking and to fulfil data quality requirements.
The presence of additional interactions in the same bunch crossing, referred to as pile-up, is characterised
by the average number of such interactions, \averagemu, which was 33.7 for the combined dataset.
 
Candidate events were recorded by either single-muon or single-electron triggers~\cite{TRIG-2016-01} with various
transverse momentum \pt (muons) or transverse energy \et (electrons) thresholds.
The lowest \pt (\et) threshold without trigger prescaling was 24 (26)~\GeV\
and included a requirement on the energy in a cone around the lepton, referred to as `isolation', that
was not applied for triggers with higher thresholds. A trigger matching requirement~\cite{TRIG-2016-01} was applied,
where the lepton must lie in the vicinity of the corresponding trigger-level object.
 
Dedicated Monte Carlo (MC) simulated samples are used to model SM
processes and to estimate the expected signal yields.  All samples were
produced using the ATLAS simulation infrastructure~\cite{SOFT-2010-01}
and $\GEANT4$~\cite{Agostinelli:2002hh}. A subset of samples use a faster
simulation based on a parameterisation of the calorimeter response and
$\GEANT4$ for the other detector systems~\cite{SOFT-2010-01}. 
The simulated events are reconstructed with the same algorithms as
used for data, and contain a realistic modelling of pile-up
interactions. The pile-up profiles in the simulation match those of each dataset
between 2015 and 2018, and are obtained by overlaying minimum-bias events,
simulated using the soft QCD processes of
\PYTHIA 8~\cite{Sjostrand:2014zea} using the NNPDF2.3LO set of
PDFs~\cite{Ball:2012cx} and a set of tuned
parameters called the A3 tune~\cite{ATL-PHYS-PUB-2016-017}.
 
Signal event samples with LQs pair produced via the strong interaction\footnote{It should be noted that $t$-channel lepton exchange
production is not included in these samples.} were generated at next-to-leading order (NLO) with
\madgraph~\cite{Alwall:2014hca}~v2.6.0 and interfaced to \PYTHIA~8.230 for the modelling of parton showers (PS),
hadronisation, and the underlying event with the A14 tune~\cite{ATL-PHYS-PUB-2014-021}.
The matrix element (ME) calculation was performed at tree level and includes the emission of up to two additional partons.
The ME--PS matching was done using the CKKW-L~\cite{Lonnblad:2012ix} prescription, with a matching scale set
to one quarter of the LQ mass.
The NNPDF2.3 LO~\cite{Ball:2012cx} parton distribution function (PDF) set was used.
Samples with LQ mass set between 400 GeV and 2000 GeV were generated at mass intervals of 50 GeV within the range 800-1600 GeV, 100 GeV otherwise.
Signal cross-sections are considered equivalent to those of pair-produced top squarks.
They are calculated to approximate next-to-next-to-leading order (NNLO) in the strong coupling constant,
adding the resummation of soft gluon emission at next-to-next-to-leading-logarithm
(approximate NNLO+NNLL) accuracy~\cite{Beenakker:2016lwe,Beenakker:1997ut,Beenakker:2010nq,Beenakker:2016gmf}.
The nominal cross-section and its uncertainty are derived using the PDF4LHC15\_mc PDF set, following the
recommendations of Ref.~\cite{Butterworth:2015oua}. For LQ masses between 400~\GeV\ and 2.0~\TeV,
the cross-sections range from 2.1\,pb to 0.02\,fb.
 
Background samples were simulated using different MC event generators depending on the process. These include top-quark pair ($t\bar{t}$)
and single ($Wt$, $s$- and $t$-channel) production, associated production of $W$ or $Z$ bosons or Drell-Yan with jets ($W$+jets, $Z$/Drell-Yan+jets) and diboson production.
All background processes are normalised to the best available theoretical calculation of their respective cross-sections.
The event generators, the accuracy of theoretical cross-sections, the underlying-event parameter tunes, and the PDF sets
used in simulating the SM background processes most relevant for this analysis are summarised in \Tab{\ref{tab-MCsamples}}.
For all samples, except those generated using \SHERPA, 
the \evtgen~v1.2.0~\cite{Lange:2001uf} program was used to simulate the properties of the $b$- and $c$-hadron decays.
 
\begin{table}[t!]
\centering
\scalebox{0.69}{
\begin{tabular}{lllllll}
\hline
\multicolumn{2}{c}{\multirow{2}{*}{Process}} & \multirow{2}{*}{Generator} & \multirow{2}{*}{PDF set}  & PS and & \multirow{2}{*}{UE tune}   & Cross-section\\
& & & & fragmentation/hadronisation &  & order \\
\hline
\multicolumn{2}{c}{Top pair ($t\bar{t}$)} & \POWHEGBOXV{v2}~\cite{Alioli:2010xd} &  NNPDF 3.0~\cite{Ball:2014uwa}& \PYTHIA 8 & A14 & NNLO+NNLL~\cite{Czakon:2011xx} \\
\multirow{2}{*}{Single-top$\bigg\{$} & $t$-channel  & \POWHEGBOXV{v1} & NNPDF 3.0 & \PYTHIA 8 & A14 &NNLO+NNLL~\cite{Kidonakis:2011wy} \\
& $s$- and $Wt$-channel & \POWHEGBOXV{v2} & NNPDF 3.0 & \PYTHIA 8 & A14 &NNLO+NNLL~\cite{Kidonakis:2010ux,Kidonakis:2010tc} \\
\multicolumn{2}{c}{$W$+jets, $Z$/Drell--Yan+jets} & \SHERPA~2.2.1~\cite{Gleisberg:2008ta,Gleisberg:2008fv,Cascioli:2011va,Schumann:2007mg,Hoeche:2012yf} & NNPDF 3.0 & \SHERPA & Default & NNLO~\cite{Catani:2009sm}\\
\multicolumn{2}{c}{Diboson} & \SHERPA~2.2.1 -- 2.2.2 & NNPDF 3.0 & \SHERPA & Default & NLO~\cite{Gleisberg:2008ta}\\
\hline
\end{tabular}
}
\caption{List of generators used for the different background
processes. Information is given about the underlying-event (UE)
tunes, the PDF sets and the perturbative QCD highest-order accuracy (NLO, NNLO, and NNLL)
used for the normalisation of the different samples.}
\label{tab-MCsamples}
\end{table}

\section{Event reconstruction and object definitions}
\label{sec:analysis}
 
An event is selected if it passes at least one of the single-lepton trigger requirements described
in the previous section.
The event quality is checked to remove events with noise bursts or coherent noise in the calorimeters.
At least one \(pp\) interaction vertex is required to be reconstructed in an event.
The primary vertex is chosen to be the vertex with the highest summed $\pt^2$ of tracks with transverse
momentum $\pt>0.5\,$~\GeV\ which are associated with that vertex.
 
Electron candidates are reconstructed by matching inner-detector tracks
to clusters of energy deposited in the EM calorimeter. Electrons must
have $\pt^e>20$~\GeV\ and $|\eta_e|<2.47$. The associated track must have
$|d_0|/\sigma_{d_0}<3$ and $|z_0|\sin \theta<0.5$~mm, where $d_0$
($z_0$) is the transverse (longitudinal) impact parameter relative to
the primary vertex and $\sigma_{d_0}$ is the associated error in $d_0$. Candidates are identified with a likelihood method
and must satisfy the `medium' identification
criteria according to Ref.~\cite{PERF-2018-01}. The likelihood relies on the shape of
the EM shower measured in the calorimeter, the quality of the track
reconstruction, and the quality of the match between the track and the
cluster. To suppress candidates originating from photon
conversions, hadron decays, or jets misidentified as electrons,
candidates are required to satisfy the gradient isolation criteria based on
tracking and calorimeter measurements~\cite{PERF-2018-01}.
 
Muon candidates are reconstructed in the range $|\eta_\mu|<2.5$ by
combining tracks in the ID with tracks in the MS.  For
$2.5<|\eta_\mu|<2.7$, muons may be reconstructed solely from the MS track
and a loose requirement on the compatibility of originating from the
interaction point is applied.
An additional category of muons, called calorimeter-tagged muons,
are used in the region $|\eta_\mu|<0.1$, where the MS
is only partially instrumented. For these muons the ID
track must be compatible with energy deposits in the calorimeter
consistent with a minimum-ionising particle.
 
All muon candidates must have $\pt^\mu>20$~\GeV, $|d_0|/\sigma_{d_0}<3$, and
$|z_0|\sin \theta<0.5$~mm. Muons from
hadron decays are suppressed by imposing a track-based isolation requirement~\cite{PERF-2015-10}.
In order to improve the momentum resolution, further quality
requirements are placed on the muons. The `medium' quality requirements
described in Ref.~\cite{PERF-2015-10} are used for candidates with
$\pt^\mu<800$~\GeV. The main requirements are a minimum of three hits in the
muon detector in a minimum of two layers (except for $|\eta_\mu|<0.1$, where there is a minimum
of one hit) and for the difference between the momentum measurements in the
ID and MS to have a $q/p$ significance of less than~$7.0$. The significance is
defined as the absolute value of the difference between the ratio
of the charge and momentum of the muons measured in the ID and MS divided
by the sum in quadrature of the corresponding uncertainties.
As muons with $\pt^\mu>800$~\GeV\ have poorer momentum resolution, the
more stringent `high-$\pt$' quality requirements are imposed: muons with
$|\eta_\mu|>2.5$ without an inner-detector track are rejected; candidates must
have hits in each of the three layers of the muon detector; and regions
where the alignment is suboptimal are removed. The `high $\pt$' quality
requirements remove 20\% of muons but improve the $\pt^\mu$ resolution
by approximately 30\%~\cite{PERF-2015-10}
above 1.5~\TeV\  and
suppress backgrounds.
 
Jets in the range $|\eta_j|<4.5$ and $\pt^j>20$~\GeV\ are reconstructed
from energy deposits in the calorimeter~\cite{PERF-2014-07}, using the
anti-$k_t$ algorithm~\cite{Cacciari:2008gp,Fastjet} with a radius
parameter of $0.4$. To suppress jets arising from pile-up, a jet-vertex-tagging
technique using a multivariate likelihood~\cite{PERF-2014-03} is applied to jets with $\pt^j \leq 60$~\GeV,
requiring that at least 60\% of the total \pt of tracks in
the jet be associated with the event's primary vertex.
 
To resolve the reconstruction ambiguities among electrons, muons, and jets,
an overlap removal procedure is applied.
First, any electron with the same ID track as a muon is rejected, unless it is a calorimeter-tagged muon,
in which case the muon is removed. If the electron shares the same ID track with another electron, the one with
lower \pt~is discarded.
Next, candidate jets with fewer than three associated tracks
are discarded if they lie within a cone of $\Delta R = 0.2$ around a muon candidate, irrespective of the track requirement for the electron candidates.
Subsequently, electrons within a cone of size $\Delta R = {\mathrm{min}} (0.4, 0.04 + 10\,\GeV/\pt)$ around a
jet are removed. Last, muons within a cone, defined in the same way as for electrons,
around any remaining jet are removed.

Jets in the range $|\eta_j|<2.5$ are 
categorised as $b$-tagged or $c$-tagged jets, exploiting a multivariate algorithm
that uses calorimeter and tracking information~\cite{Aad:2019aic}.
Jets are first tested using the $b$-tagging algorithm, which has an
efficiency of about 70\% for true $b$-jets with a rejection factor of about 8 for charm jets and
about 300 for light-flavour jets~\cite{PERF-2016-05}. In the $c\ell$
channels, jets that are not $b$-tagged are tested with
the $c$-tagging algorithm, which has an efficiency of about 27\% for
true $c$-jets and approximate rejection factors of 12 for $b$-jets and 59
for light-flavour jets. The $c$-tagging algorithm is not used in the other channels.
 
When the selection requires two $b$-tagged jets, the substantial rejection rate of
the tagging algorithm results in a significant statistical uncertainty
for simulated Drell--Yan (DY) events containing only light-flavour jets or $c$-jets.
Hence, instead of applying the $b$-tagging requirement, all events with $c$-jets or light-flavour
jets are weighted by the probability that these jets pass it. 
This procedure, documented in Ref.~\cite{HIGG-2016-29}, significantly
increases the number of simulated events present after the full event
selection, reducing the statistical uncertainty of the Drell--Yan background
by up to a few orders of magnitude.
 
The event's missing transverse momentum (its modulus referred to \etmiss) is computed as the negative vectorial
sum of the transverse momenta of leptons and jets. 
The \etmiss calculation also includes a track-based soft term~\cite{PERF-2016-07}
accounting for the contribution from particles from the primary vertex that
are not already included in the \etmiss calculation.
 
\section{Event selection}
\label{strategy}
 
The event selection prioritises events consistent with scalar leptoquark production in high signal-to-background
kinematic regions and has been optimised to reject signatures consistent with reducible backgrounds or poorly
modelled event reconstruction.
 
Events are required to have exactly two electrons or two muons,
oppositely charged and with transverse momenta greater than $27$~\GeV,
ensuring the full efficiency of the trigger. At least two jets with
$\pt^j>45$~\GeV\ and $|\eta_j|<2.5$ are required.  Selections on the
dilepton pair invariant mass, $m_{\ell \ell}>130$~\GeV, and transverse
momentum, $\pt^{\ell \ell}>75$~\GeV, are made to suppress background
from the DY production and on-shell $Z$ boson production.
If there are more than two jets in the event, firstly those jets with $b$- or $c$-tags are
chosen as the candidate jets arising from the decays of the
leptoquarks. 
For events with one tagged jet, the highest-$\pt$ untagged jet is chosen as the second 
candidate. For events with zero tagged jets, the two highest-$\pt$ jets are chosen.
Events with more than two tagged jets are likely to be background and are rejected
for the $c\ell$ and $b\ell$ channels.
Background from $t\bar{t}$ production is suppressed by requiring
$\etmiss/\sqrt{H_\mathrm{T}} < 3.5$~\GeV$^{1/2}$, where $H_\mathrm{T}$
is the scalar sum of the transverse momenta of all lepton candidates
and selected jets in the event. This selection is preferable to
a simple $\etmiss$ selection as it is looser at higher \pt where
the resolutions for the leptons and jets are worse.
 
Leptoquark candidates are identified from the two possible lepton--jet
combinations by selecting the pairs closest in lepton--jet
invariant mass, $m_{\ell j}$. SM background contributions are suppressed by requiring that
\[
m^{\mathrm{asym}}= \frac{m_{\ell j}^{\mathrm{max}} - m_{\ell j}^{\mathrm{min}}}{m_{\ell j}^{\mathrm{max}} +
m_{\ell j}^{\mathrm{min}}} < 0.4,
\]
where $m_{\ell j}^{\mathrm{max}}$ and $m_{\ell j}^{\mathrm{min}}$ are the reconstructed masses
of the two LQ candidates, ordered such that $m_{\ell j}^{\mathrm{max}}>m_{\ell j}^{\mathrm{min}}$.
The selected region is further divided into a signal region (SR), requiring
$m^{\mathrm{asym}}<0.2$, and a sideband region (SB) where $0.2<m^{\mathrm{asym}}<0.4$.
The SB is used in the maximum-likelihood fit, as described in Section~\ref{sec:result}, to
help constrain  the normalisation of the main backgrounds in a region with a low signal expectation.
The results are presented as a function of the average of the two reconstructed leptoquark masses,
$m_{{\ell j}}^{\mathrm{Av}}=(m_{\ell j}^{\mathrm{max}} + m_{\ell j}^{\mathrm{min}})/2$.
The reconstructed mass resolution is found to not exceed 7\% of the LQ mass in all channels.
 
A summary of the event selections for signal and SB regions is
given in Table~\ref{tab:Selection} before any specification of the flavour tags.
The main selections for the Top control regions (CRs), used to aid in the estimation of
the $t\bar{t}$~background and described in detail in Section~\ref{sec:background}, are also reported.
 
\begin{table}[t!]
\begin{center}
\scalebox{0.9}{
\begin{tabular}{c|c|c}
\hline\hline
\multicolumn{3}{c}{Preselection}     \\
\hline
\multicolumn{3}{c}{2 oppositely charged leptons ($e$, $\mu$)}\\
\multicolumn{3}{c}{2 or more jets}\\
\multicolumn{3}{c}{$\pt^e>27~\GeV$, $|\eta_e|<2.47$; $\pt^\mu>27~\GeV$, $|\eta_\mu|<2.7$ }\\
\multicolumn{3}{c}{$\pt^j>45~\GeV$, $|\eta_j|<2.5$}\\
\multicolumn{3}{c}{$\pt^{\ell \ell} > 75~\GeV$}\\
\multicolumn{3}{c}{$\etmiss/\sqrt{H_\mathrm{T}}  < 3.5~\GeV$$^{1/2}$}\\
\multicolumn{3}{c}{$m_{\ell \ell}>130~\GeV$}\\
\hline\hline
SB & SR  & Top CR      \\
\hline
\multicolumn{2}{c|}{$ee$ or $\mu\mu$}  & $e\mu$ \\
\hline
$ 0.2 < m^{\mathrm{asym}} < 0.4$& \multicolumn{2}{|c} {$m^{\mathrm{asym}}  <0.2$}\\
\hline\hline
\end{tabular}
}
\end{center}
\caption{Summary of the preselection and region-specific selections applied before flavour tagging.}
\label{tab:Selection}
\end{table}
 
The SR and SB are further categorised to isolate kinematic regions that
separate events consistent with light, charm and bottom quark production.
These regions are defined by the number of $b$- and $c$-tagging jets in the events.
An inclusive selection (referred to as pretag) is used for the $q\ell$ channels.
Although the $\mathrm{LQ} \to q\ell$ interpretations do not
benefit from jet tagging when assuming that $\mathcal{B} = 1$ as considered in this paper,
a selection of this kind might also provide sensitivity to cross-generational LQ decays if
mixed decays into charged leptons are possible.
The $b$-tagging selections are used in the $b\ell$ channels and target $\mathrm{LQ} \to b\ell, \mathcal{B} = 1$ models,
while both the $b$-tagging and $c$-tagging selections are used in the $c\ell$ channels,
targeting $\mathrm{LQ} \to c\ell, \mathcal{B} = 1$ models.
In the $b\ell$ channels the events are split into those with zero, one, or two
$b$-tags (referred to as 0-tag, 1-tag, and 2-tag, respectively). 
In the $c\ell$ channels the events are split into those with zero tags
(untagged), at least one $c$-tag ($c$-tag), and at least one $b$-tag
($b$-tag). Events with one $c$-tagged jet and one $b$-tagged jet are
placed in the $c$-tag category.
 
The signal and SB regions are not mutually exclusive between the search channels.
The acceptance times detector efficiency for LQ events after all selections is highest in
the electron channel $qe$ for LQ masses around 1.3~\TeV\ (62\%) and between
45\% and 55\% for the mass range 400--2000~\GeV\ in all channels.
For muon-based selections, this is reduced to a maximum of 53\% for LQ masses (around 900~\GeV)
in $q\mu$ channels and to about 30\% for high LQ masses overall,
due to the low efficiency of the high-\pt muon selection and the poorer efficiency
of the $\etmiss/\sqrt{H_\mathrm{T}}$ selection for muons than for electrons.

\section{Background determination}
\label{sec:background}
The backgrounds in the analysis are estimated from simulated samples
described in Section~\ref{sec:datamc}, with the aid of control and
sideband regions for checks and estimates of systematic uncertainties.
The dominant background in the pretag ($q\ell$), untagged ($c\ell$) and 0-tag ($b\ell$) SRs
arises from DY production in association with two or more jets, followed by
$t\bar{t}$ background. In the 1-tag, 2-tag, $c$-tag, and $b$-tag
categories, the $t\bar{t}$ background dominates whilst DY background
is subdominant.  The DY background is
further split into three categories, referred to as DY+light-jets,
DY+$c$-jets, and DY+$b$-jets, based on the flavour of the heaviest quark
as determined from simulation in either of the jets selected to reconstruct the
LQ candidates.
 
To compensate for the limited number of events at high values of the average mass of
the LQ candidates, a fit is made to the smoothly falling distributions for $t\bar{t}$
samples, and extrapolated to high $m_{{\ell j}}^\mathrm{Av}$ with the following function
\[
f^{t\bar{t}} (m_{{\ell j}}^\mathrm{Av}) = a (m_{{\ell j}}^\mathrm{Av})^b,
\]
where $a$ and $b$ are the fitted parameters. In all cases, checks are performed to guarantee
that the function reproduces the event yields at lower $m_{\ell j}^\mathrm{Av}$ values and
that its cumulative distribution (starting from the highest $m_{\ell j}^\mathrm{Av}$ values)
is consistent with the small integrated event yields available in the MC samples.
Other SM processes, dibosons ($WW, WZ, ZZ$) and single-top production
(mostly $Wt$), contribute less than 10\% in all SRs and are estimated directly from MC samples.
Rare processes such as $t\bar{t}V$, with $V=W,Z$ or Higgs bosons, and tribosons are negligible.
Contributions from events where one or both electrons or muons are misidentified jets or non-promptly produced
(referred to as `fake' background) are checked using a same-sign lepton control region
mirroring the SR selections. They are found to be dominated by single-electron fake contributions
and well described by the $W (\rightarrow \ell \nu)+\text{jets}$ simulated samples, which are
used for this estimate. A systematic error is assigned which covers
any disagreement between the data and MC simulation in the same-sign region as described in Section~\ref{sec:sys}.

The shape of the DY background is taken from simulation, while the systematic
uncertainty on the shape, as described in Section~\ref{sec:sys}, is determined by comparing data with the predictions
in two control regions dominated by the DY process.
The first of these is an extended SB region, which has $m^{\mathrm{asym}}>0.4$ and is used to validate
the off-shell DY prediction;  the second is an on-shell
$Z$ control region defined by inverting the selection on $m_{\ell\ell}$ ($<130$~\GeV)
and removing the $m^{\mathrm{asym}}$ requirement.
 
An additional set of control regions is used to constrain the normalisation of
the $t\bar{t}$ production background (Top CRs) in the pretag and tagged categories.
The regions are identical to the default SR selections, corresponding to pretag for the $q\ell$ channels,
$c$-tag and $b$-tag for the $c\ell$ channels, and 1-tag and 2-tag
for the $b\ell$ channels, except that an electron--muon pair is taken in place of
the same-flavour lepton pair.
The 0-tag/untagged categories do not utilise this region as $t\bar{t}$ production is not dominant.
 
\begin{figure}[t!]
\centering
\includegraphics[width=0.47\textwidth]{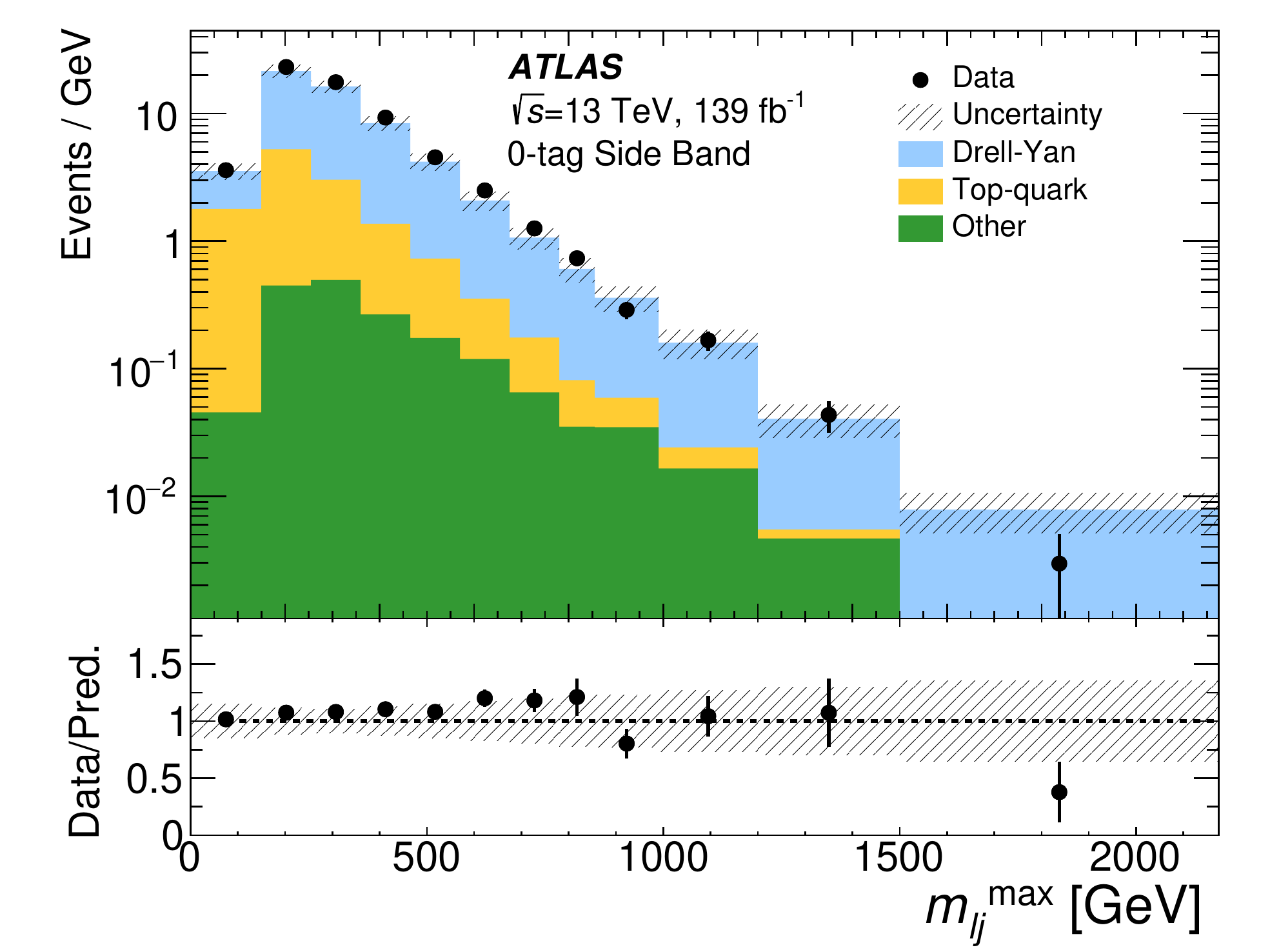}
\includegraphics[width=0.47\textwidth]{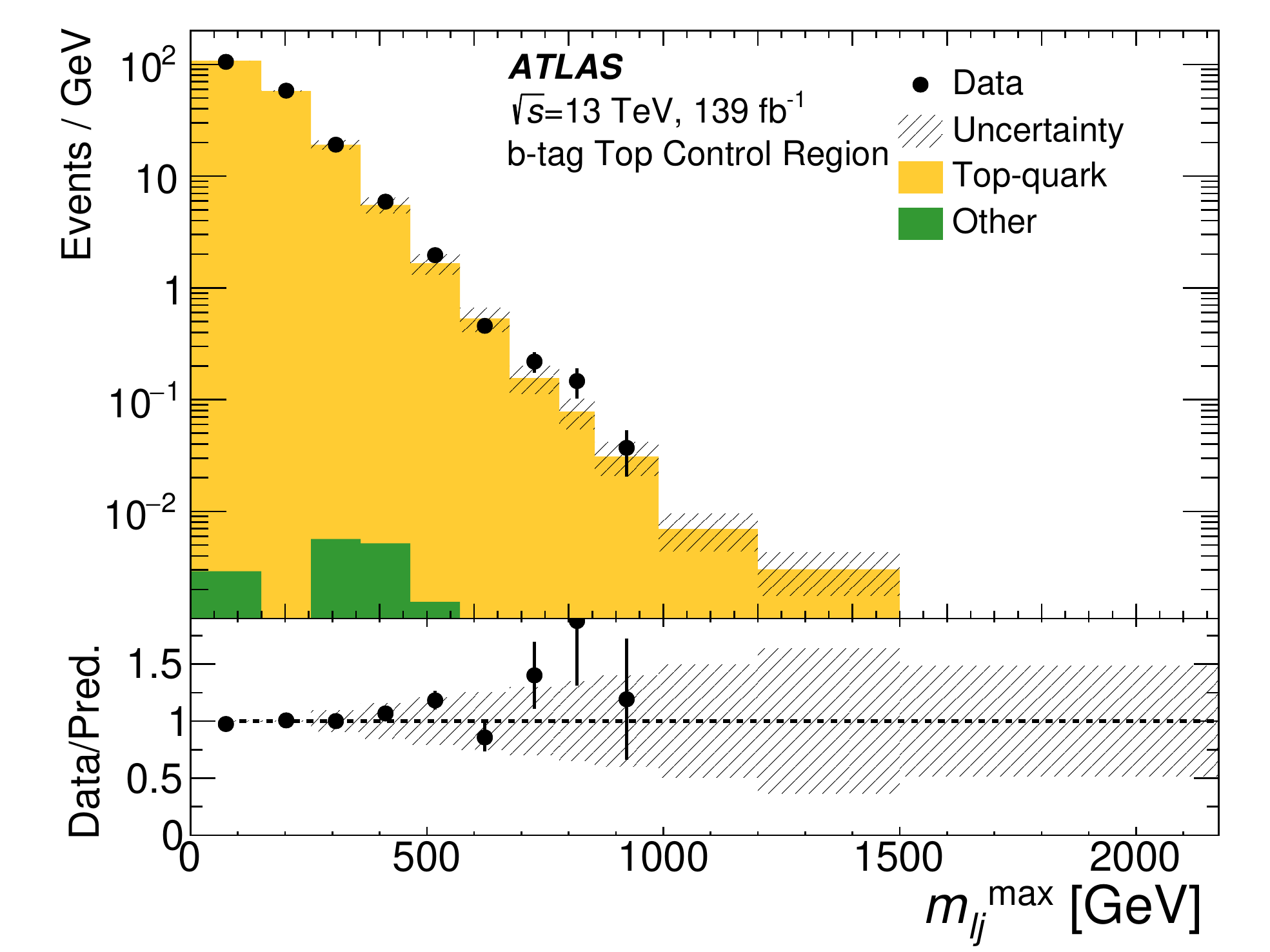}
 
\caption{Distributions of $m_{\ell j}^{\mathrm{max}}$ in  the  combined $be$ and $b\mu$ 0-tag SB for the $b\ell$ channels (left) and the $b$-tag Top CR for the $c\ell$ channels (right).
The total modelling uncertainty combined with the MC statistical error is shown as the hatched band as explained in Section~\ref{sec:sys}. The category `Other' represents the sum of all SM background contributions except those from top-quark processes ($t\bar{t}$ and single-top) and, for the 0-tag SB distribution,
Drell--Yan processes. }
\label{fig:syst:model}
\end{figure}

Figure~\ref{fig:syst:model} shows distributions of $m_{\ell j}^{\mathrm{max}}$ in the 0-tag SB region used
to validate the DY predictions, and the $b$-tag Top CR for the $c\ell$ channels used for $t\bar{t}$ background contributions.
Distributions are depicted before the profile-likelihood fit described below.
Differences between data and MC predictions are used to estimate modelling uncertainties
for these SM background processes, as explained in Section~\ref{sec:sys}.
The $m_{\ell j}^{\mathrm{max}}$ variable is used instead of $m_{{\ell j}}^\mathrm{Av}$ to retain more statistics
in the tail of the distributions.
 
\section{Systematic uncertainties}
\label{sec:sys}
 
Several sources of experimental and theoretical systematic uncertainty in the signal and background
estimates are considered.
 
For the LQ processes, experimental uncertainties in the signal yields are dominated by the uncertainty
arising from lepton identification and jet energy scale and resolution ($q\ell$ channels)
and from the $b$- and $c$-tagging efficiencies and mis-tagging rates ($c\ell$ and $b\ell$ channels).
The uncertainties in the jet energy scale and resolution are
based on their respective measurements in data~\cite{PERF-2016-04, PERF-2011-04} and account for up to 2\% of
the signal yields.
Uncertainties in electron identification efficiency, trigger efficiency, isolation efficiency,
energy scale, and resolution amount to less than 6\%~\cite{PERF-2015-10,PERF-2018-01}, while
the muon uncertainties are less than 10\%~\cite{PERF-2018-01,PERF-2015-10}.
The $b$- and $c$-tagging uncertainties are estimated by varying the $\eta$-, \pt- and flavour-dependent
scale factors applied to each jet in the simulation within a range that reflects the systematic uncertainty
in the measured tagging efficiency and mis-tag rates in data~\cite{Aad:2019aic}.
Uncertainties in $b$- and $c$-tagging are found to be less than 16\% for the $c\ell$ channels and
19\% for the $b\ell$ channels.
The uncertainty due to the pile-up modelling~\cite{STDM-2015-05} is typically less than 1\%.
Overall, the experimental uncertainties in the signals are between 1\% and 20\% of the yields, including the 1.7\% uncertainty
in the combined 2015–2018 integrated luminosity.

Theoretical uncertainties in the yields predicted using the approximate NNLO+NNLL cross-section are calculated for each
LQ mass~\cite{Beenakker:2016lwe,Beenakker:1997ut,Beenakker:2010nq,Beenakker:2016gmf}.
They are dominated by the uncertainties in the renormalisation and factorisation scales followed
by the uncertainty in the PDFs, and range between 7\% and 22\% for LQ masses between 400~\GeV\ and 2000~\GeV.
Additional uncertainties in the acceptance and efficiency in simulated signal samples are also taken into account.
They are dominated by uncertainties due to the modelling of initial- and final-state radiation
and renormalisation and factorisation scale variations in simulated signal samples and contribute up to 5\% at LQ masses above 1~\TeV.

Uncertainties in the modelling of the simulated SM background processes and in their theoretical
cross-sections are also taken into account.
 
The shape uncertainty in the modelling of the DY background is defined by taking the largest difference between data and MC
predictions in the control regions dominated by the DY process. The uncertainty is split into two parts,
$\sigma^{Z}=\pm 0.2\log(m_{\ell j}^\mathrm{max}/800~\GeV)$ and $\sigma^{Z}=\pm 0.4\log(m_{\ell j}^\mathrm{max}/200~\GeV)$,
to allow a shape difference between low and high $m_{{\ell j}}$.  When added in
quadrature these uncertainties approximately cover the observed disagreement. The $m_{{\ell j}}^\mathrm{max}$ variable is used instead of $m_{{\ell j}}^\mathrm{Av}$
as this leads to slightly larger, hence more conservative uncertainties. MC predictions were
found to describe the data within these errors in the SB region.
Since the DY shape modelling uncertainty is determined directly
from the difference between the data and the simulation, most of the
experimental uncertainties are not applied as this avoids double counting.
Simulated samples also exhibit differences with respect to data, for example due to
jet energy resolution, which might contribute to any disagreement.
The $b$- and $c$-tagging uncertainties are, however, applied as these
can change the normalisation between regions, and this is not taken
into account in the modelling studies.  The DY shape modelling
uncertainty is treated as uncorrelated among DY+light-jets, DY+$c$-jets, and
DY+$b$-jets processes.  In addition to the shape uncertainties, the DY+$c$-jets and
DY+$b$-jets processes are each assigned a 10\% normalisation uncertainty, where
this value represents the largest difference between data and MC simulation in
the $Z$ control region for any number of $b$-tags. 
 
The $t\bar{t}$  modelling is determined in a similar way to the DY
by comparing data and MC predictions in the Top control regions.
Reasonable agreement between data and simulation is found and an
error of $\sigma^{}_{t\bar{t}}=\pm 0.5 \log(m_{\ell j}^\mathrm{max}/200~\GeV)$ is assigned
to cover any possible differences. As in the DY estimates, the experimental
uncertainties, with the exception of $b$- and $c$-tagging ones,
are not applied to the $t\bar{t}$ simulation. The normalisation of the $t\bar{t}$ background
is left as a free parameter in all fits.

The extrapolation uncertainty for the $t\bar{t}$ background is evaluated
using a falling exponential function as
an alternative to the functional form described in Section~\ref{sec:background}. Differences from the nominal form are as large as
100\% at very high LQ candidate mass for all channels, but 
the impact on the results is minimal due to the low $t\bar{t}$ rate above 1.3~\TeV.
 
Finally, normalisation uncertainties are associated with the predictions of diboson and single-top-quark
processes, and non-prompt and misidentified leptons. A 30\% uncertainty is
assigned to the diboson predictions, dominated by theoretical
modelling uncertainties and estimated as in Ref.~\cite{Aaboud:2017opj}. The dominant uncertainty for
single-top-quark predictions also arise from theoretical and modelling uncertainties of the
$Wt$ process. They are found to be around 35\% and are computed using differences
between the predictions from the nominal sample and those from additional samples
differing in hard-scattering generator, modelling of the $t\bar{t}$ and $Wt$ interference
term, and other parameter settings. A 25\% uncertainty is assigned to the background from
non-prompt and misidentified leptons, computed using the difference between data and
MC $W$+jets predictions in the same-sign leptons control sample described in Section~\ref{sec:background}.
 
\section{Results}
\label{sec:result}
The data are compared with the expectation by performing simultaneous maximum-likelihood fits to the distribution of
$m_{{\ell j}}^{\mathrm{Av}}$ in the signal, sideband and Top control regions.
The Top CRs contain a negligible signal expectation and are used to constrain the top-quark background.
The SB regions are used to constrain the DY background. They have a low but non-negligible signal
expectation and therefore are treated in the fit in the same way as the SRs.
A separate fit is performed for each signal hypothesis.  Confidence intervals are
based on a profile-likelihood-ratio test statistic~\cite{Cowan:2010st}, assuming asymptotic distributions for
the test statistic.\footnote{Cross-checks with sampling distributions generated using pseudo-experiments
were performed to test the accuracy of the asymptotic approximation for the high-mass part of the lepton--jet spectrum.
The approximation is found to lead to limits that are slightly stronger than those obtained with pseudo-experiments,
i.e.\ about 15\% at 1.8~\TeV, independent of the channel.
The impact of this approximation on the mass limits is below 50~\GeV. }
The systematic uncertainties affecting the signal
and background normalisations and shapes across categories are
parameterised by making the likelihood function depend on dedicated
nuisance parameters, constrained by additional Gaussian or log-normal
probability terms.
 
For the $q\ell$ signals, the pretag SR, SB, and Top CR are used.
For the $c\ell$ channels, the SR and SB in untagged, $c$-tag and $b$-tag
categories are used together with the Top CR for $c$-tag and $b$-tag.
For the $b\ell$ channels, the SR and SB in 0-, 1-, and 2-tag categories
are used together with the Top CR in 1- and 2-tag.
In all fits the DY and $t\bar{t}$ normalisations are treated as a single free parameter while
different uncertainties in the shapes of distributions are assigned to the
events as described in Section~\ref{sec:sys}.
 
All other backgrounds 
are set to their MC expectations and are allowed to float within their respective uncertainties.
 
The event yields in the SR for all channels are listed in Tables~\ref{tab:qlqlSR} to \ref{tab:bmubmuSR}.
The SM background expectations resulting from the fits are reported showing statistical plus systematic uncertainties.
The largest background contribution in $q\ell$ channels arises from DY$+$ light-jets,
whilst the contribution from $t\bar{t}$ is largest for the
signal regions relevant for the $c\ell$- and $b\ell$-jet channels.
Single-top-quark and diboson processes as well as misidentified/non-prompt lepton contributions are subdominant in all
regions. No significant differences are observed between expected and observed yields
in all selections and channels considered. Since the SRs are not mutually exclusive, the same data are used
across the various channels.
 
\begin{table}[htbp]
 
\begin{center}
\small
\begin{tabular}{lcc}
\hline
& $\mathrm{LQ}\rightarrow q e$ & $\mathrm{LQ}\rightarrow q \mu$ \\
\hline
$t\bar{t}$ &  1790  $\pm$  220 &  1910  $\pm$  240\\
Single top &  ~~390  $\pm$  110 &  ~~430  $\pm$  120\\
DY+light-jets &  2820  $\pm$  180 &  3040  $\pm$  180\\
DY+$c$-jets &  521  $\pm$  93 &  528  $\pm$  90\\
DY+$b$-jets &  233  $\pm$  44 &  252  $\pm$  46\\
$W$+jets &  126  $\pm$  32 &  ~~8.5  $\pm$  2.2\\
Diboson &  31.8  $\pm$  9.6 &  12.4  $\pm$  3.7\\
\hline
Fitted SM background events &  5910  $\pm$  67~~ &  6185  $\pm$  77~~~\\
\hline
Observed events &  5881  &  6169 \\
\hline
Signal ($m_\mathrm{LQ} = 1$~\TeV) &  591  $\pm$  45 &  503  $\pm$  27\\
Signal ($m_\mathrm{LQ} = 1.5$~\TeV) &  22.1  $\pm$  1.7 &  15.4  $\pm$  1.0~~\\
 
\hline
\end{tabular}
\end{center}
\caption{Observed and expected numbers of events in pretag SRs for $\mathrm{LQ}\rightarrow q \ell$,
where SM predictions are the result of fits performed using 139~fb$^{-1}$ of data.
The uncertainties quoted for the fitted SM background include both the statistical and systematic components.
Yields for two LQ scenarios are also shown for comparison.}
\label{tab:qlqlSR}
 
\begin{center}
\footnotesize
\begin{tabular}{lcccccc}
\hline
& \multicolumn{3}{c}{$\mathrm{LQ}\rightarrow c e$} & \multicolumn{3}{c}{$\mathrm{LQ}\rightarrow c \mu$} \\
& untagged & $b$-tag & $c$-tag & untagged & $b$-tag & $c$-tag \\
\hline
$t\bar{t}$ & 291 $\pm$ 18 & 964 $\pm$ 51 & 227 $\pm$ 14 & 293 $\pm$ 16 & 1049 $\pm$ 50~~ & 237 $\pm$ 14\\
Single top & ~~35 $\pm$ 11 & 129 $\pm$ 39 & 28.7 $\pm$ 9.0 & ~~37 $\pm$ 10 & 166 $\pm$ 46 & ~~38 $\pm$ 11\\
DY+light-jets & 2872 $\pm$ 74~~ & 32.3 $\pm$ 8.6 & 101 $\pm$ 11 & 3120 $\pm$ 71~~ & 29.0 $\pm$ 9.4 & 123 $\pm$ 13\\
DY+$c$-jets & 367 $\pm$ 49 & ~~80 $\pm$ 12 & 135 $\pm$ 17 & 340 $\pm$ 46 & ~~67 $\pm$ 10 & 155 $\pm$ 20\\
DY+$b$-jets & 39.4 $\pm$ 5.7 & 166 $\pm$ 24 & 31.5 $\pm$ 4.8 & 40.4 $\pm$ 5.7 & 165 $\pm$ 23 & 35.1 $\pm$ 5.2\\
$W$+jets & 101 $\pm$ 26 & 10.2 $\pm$ 2.7 & ~~7.5 $\pm$ 2.0 & ~~6.3 $\pm$ 1.6 & ~~1.39 $\pm$ 0.36 & ~~0.81 $\pm$ 0.21\\
Diboson & 23.5 $\pm$ 7.2 & ~~2.58 $\pm$ 0.79 & ~~3.6 $\pm$ 1.1 & ~~9.0 $\pm$ 2.7 & ~~1.21 $\pm$ 0.37 & ~~1.45 $\pm$ 0.44\\
\hline
Fitted SM events & 3728 $\pm$ 53~~ & 1384 $\pm$ 26~~ & 534 $\pm$ 17 & 3846 $\pm$ 55~~ & 1478 $\pm$ 26~~ & 591 $\pm$ 18\\
\hline
Observed events & 3714 & 1366 & 535 & 3824 & 1484 & 591\\
\hline
Signal ($m_\mathrm{LQ} = 1$~\TeV) & 312 $\pm$ 26 & ~~71 $\pm$ 12 & 129 $\pm$ 13 & 265 $\pm$ 17 & 58.0 $\pm$ 9.1 & 111.5 $\pm$ 9.5~~~\\
Signal ($m_\mathrm{LQ} = 1.5$~\TeV) & 13.7 $\pm$ 1.2 & ~~2.33 $\pm$ 0.38 & ~~3.10 $\pm$ 0.30 & ~~9.72 $\pm$ 0.69 & ~~1.49 $\pm$ 0.28 & ~~1.99 $\pm$ 0.20\\
 
\hline
\end{tabular}
\end{center}
\caption{Observed and expected numbers of events in untagged, $c$- and $b$-tag SRs for $\mathrm{LQ}\rightarrow c \ell$, where SM predictions are
the result of fits performed using 139~fb$^{-1}$ of data.
The uncertainties quoted for the fitted SM background include both the statistical and systematic components. Yields for two LQ
scenarios are also shown for comparison.}
\label{tab:cmucmuSR}
 
\begin{center}
\footnotesize
\begin{tabular}{lcccccc}
\hline
& \multicolumn{3}{c}{$\mathrm{LQ}\rightarrow b e$} & \multicolumn{3}{c}{$\mathrm{LQ}\rightarrow b \mu$} \\
& 0-tag & 1-tag & 2-tag & 0-tag & 1-tag & 2-tag \\
\hline
$t\bar{t}$ & 469 $\pm$ 22 & 919 $\pm$ 33 & 255 $\pm$ 11 & 487 $\pm$ 22 & 1001 $\pm$ 35~~ & 295 $\pm$ 12\\
Single top & ~~51 $\pm$ 11 & 109 $\pm$ 24 & ~~48 $\pm$ 10 & ~~48 $\pm$ 10 & 122 $\pm$ 25 & ~~49 $\pm$ 10\\
DY+light-jets & 3035 $\pm$ 95~~ & 29.2 $\pm$ 8.0 & ~~0.105 $\pm$ 0.057 & 3318 $\pm$ 93~~ & ~~36 $\pm$ 11 & ~~0.099 $\pm$ 0.059\\
DY+$c$-jets & 479 $\pm$ 77 & ~~92 $\pm$ 15 & ~~1.68 $\pm$ 0.34 & 464 $\pm$ 75 & ~~75 $\pm$ 13 & ~~1.61 $\pm$ 0.33\\
DY+$b$-jets & 54.2 $\pm$ 7.7 & 165 $\pm$ 23 & 25.9 $\pm$ 3.6 & 52.5 $\pm$ 7.6 & 151 $\pm$ 22 & 21.1 $\pm$ 3.0\\
$W$+jets & 113 $\pm$ 29 & ~~9.4 $\pm$ 2.4 & ~~1.02 $\pm$ 0.27 & ~~7.5 $\pm$ 1.9 & ~~0.97 $\pm$ 0.25 & ~~0.110 $\pm$ 0.028\\
Diboson & 27.8 $\pm$ 8.5 & ~~2.63 $\pm$ 0.81 & ~~0.33 $\pm$ 0.10 & 10.8 $\pm$ 3.2 & ~~1.21 $\pm$ 0.37 & ~~0.141 $\pm$ 0.043\\
\hline
Fitted SM events & 4229 $\pm$ 57~~ & 1326 $\pm$ 25~~ & 332.4 $\pm$ 9.0~~ & 4389 $\pm$ 59~~ & 1387 $\pm$ 25~~ & 367.1 $\pm$ 9.3~~~\\
\hline
Observed events & 4214 & 1314 & 316 & 4367 & 1408 & 340\\
\hline
Signal ($m_\mathrm{LQ} = 1$~\TeV) & 102 $\pm$ 13 & 237 $\pm$ 19 & 149 $\pm$ 13 & ~~87 $\pm$ 11 & 200 $\pm$ 12 & 124.1 $\pm$ 8.7~~~\\
Signal ($m_\mathrm{LQ} = 1.5$~\TeV) & ~~5.69 $\pm$ 0.90 & ~~8.72 $\pm$ 0.76 & ~~3.57 $\pm$ 0.33 & ~~3.89 $\pm$ 0.61 & ~~6.11 $\pm$ 0.50 & ~~2.38 $\pm$ 0.20\\
 
\hline
\end{tabular}
\end{center}
\caption{Observed and expected numbers of events in 0-, 1- and 2-tag SRs for $\mathrm{LQ}\rightarrow b \ell$, where SM predictions are the result of
fits performed using 139~fb$^{-1}$ of data. The uncertainties quoted for the fitted SM background include both the statistical and systematic components.
Yields for two LQ scenarios are also shown for comparison.}
\label{tab:bmubmuSR}
\end{table}
 
\begin{figure}[htbp]
\centering
\includegraphics[width=0.32\textwidth]{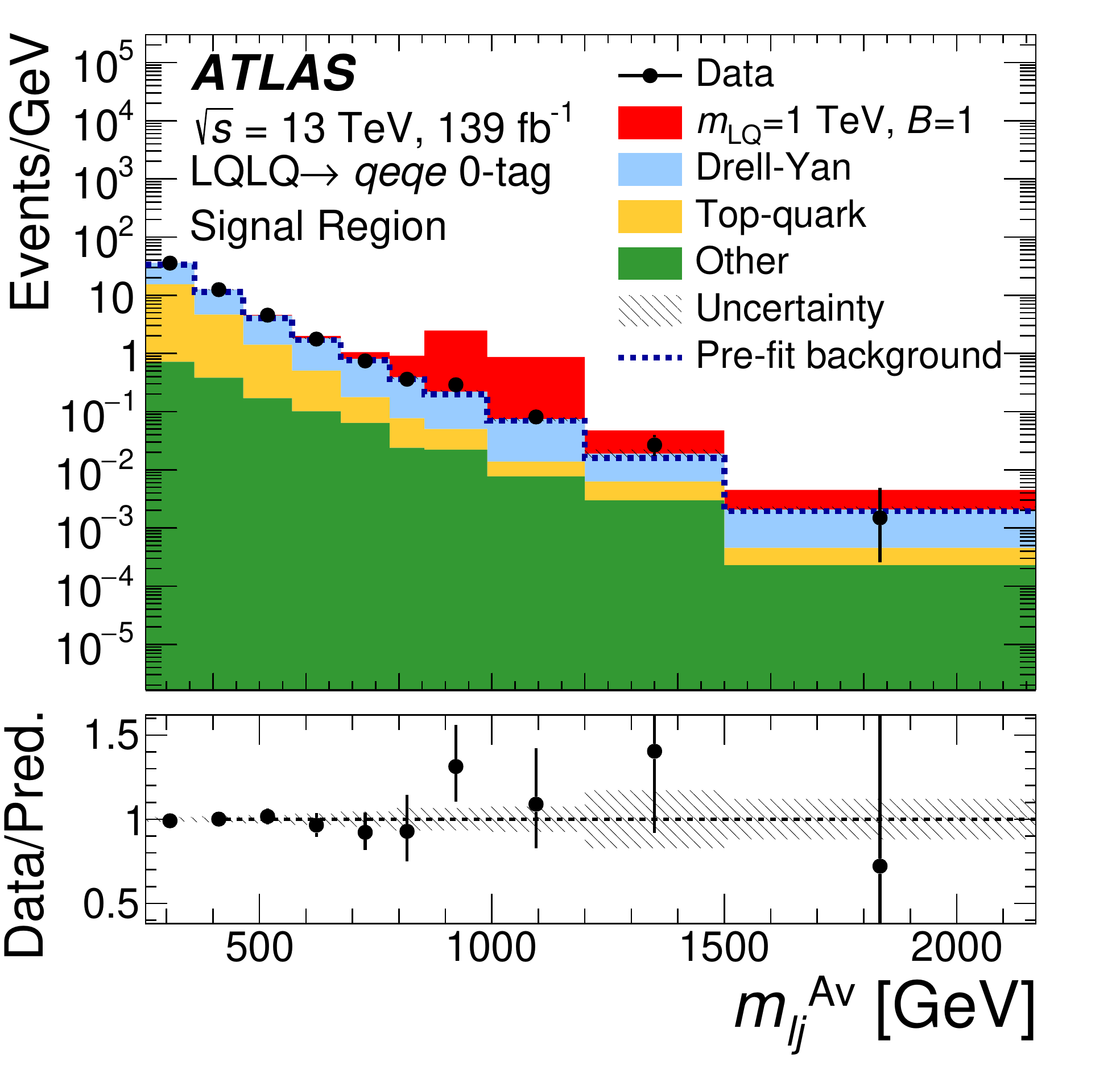}
\includegraphics[width=0.32\textwidth]{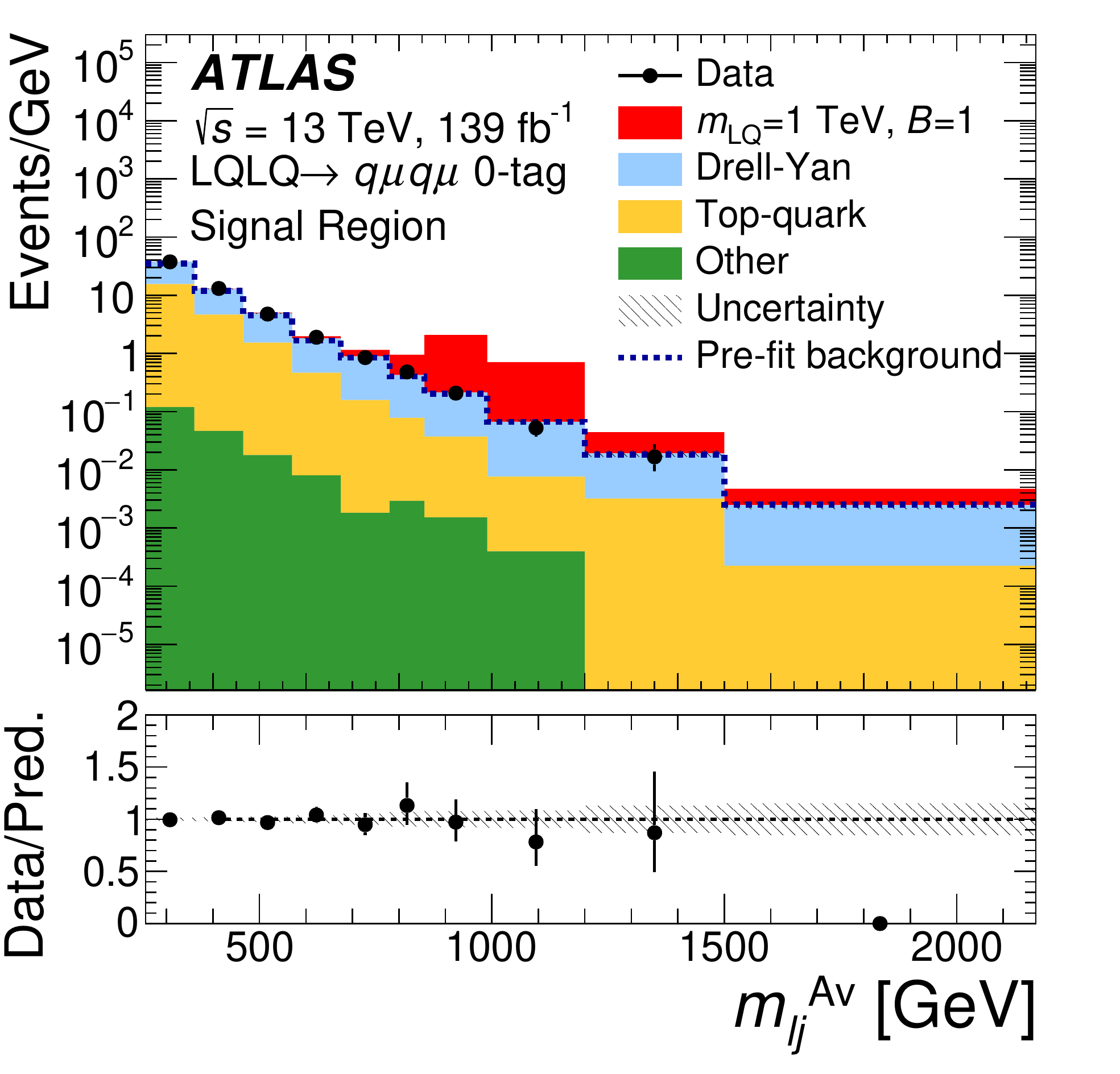}
\caption{Post-fit distributions of $m_{{\ell j}}^\mathrm{Av}$ in the pretag signal regions for the $q e$ (left) and $q\mu$
(right) channels. The expected signals, shown for $m_{\mathrm{LQ}}=1$~\TeV\ and $\mathcal{B}(\mathrm{LQ} \to qe / q\mu) = 1 $,
are shown for illustrative purposes. The category `Top quark' refers to $t\bar{t}$ and single-top-quark processes. The category `Other' refers to  diboson and $W$+jet production.
The hatched band represents the total uncertainty in the background predictions.
}
\label{fig:Results:postfitqlql}
 
\centering
\includegraphics[width=0.32\textwidth]{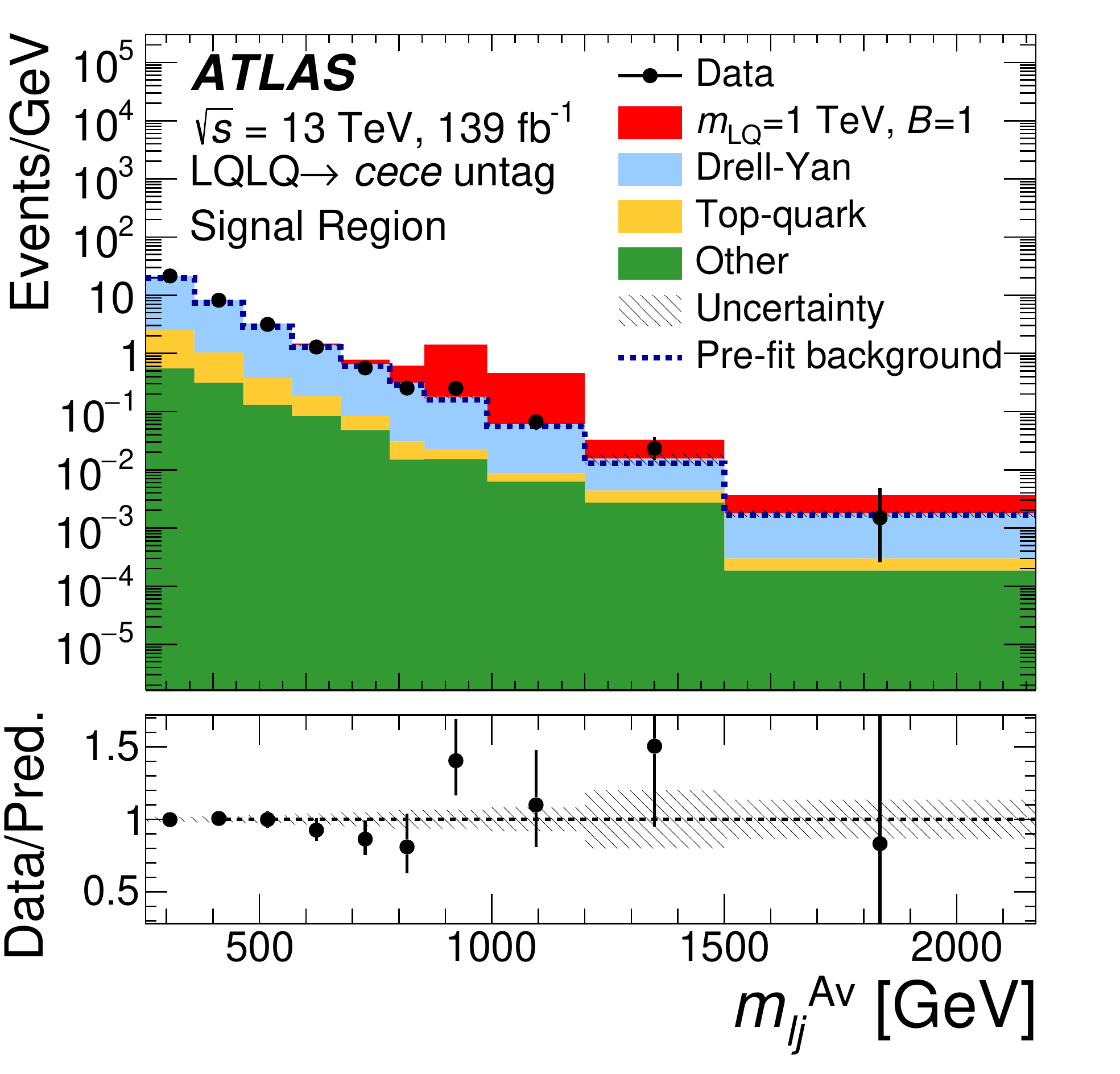}
\includegraphics[width=0.32\textwidth]{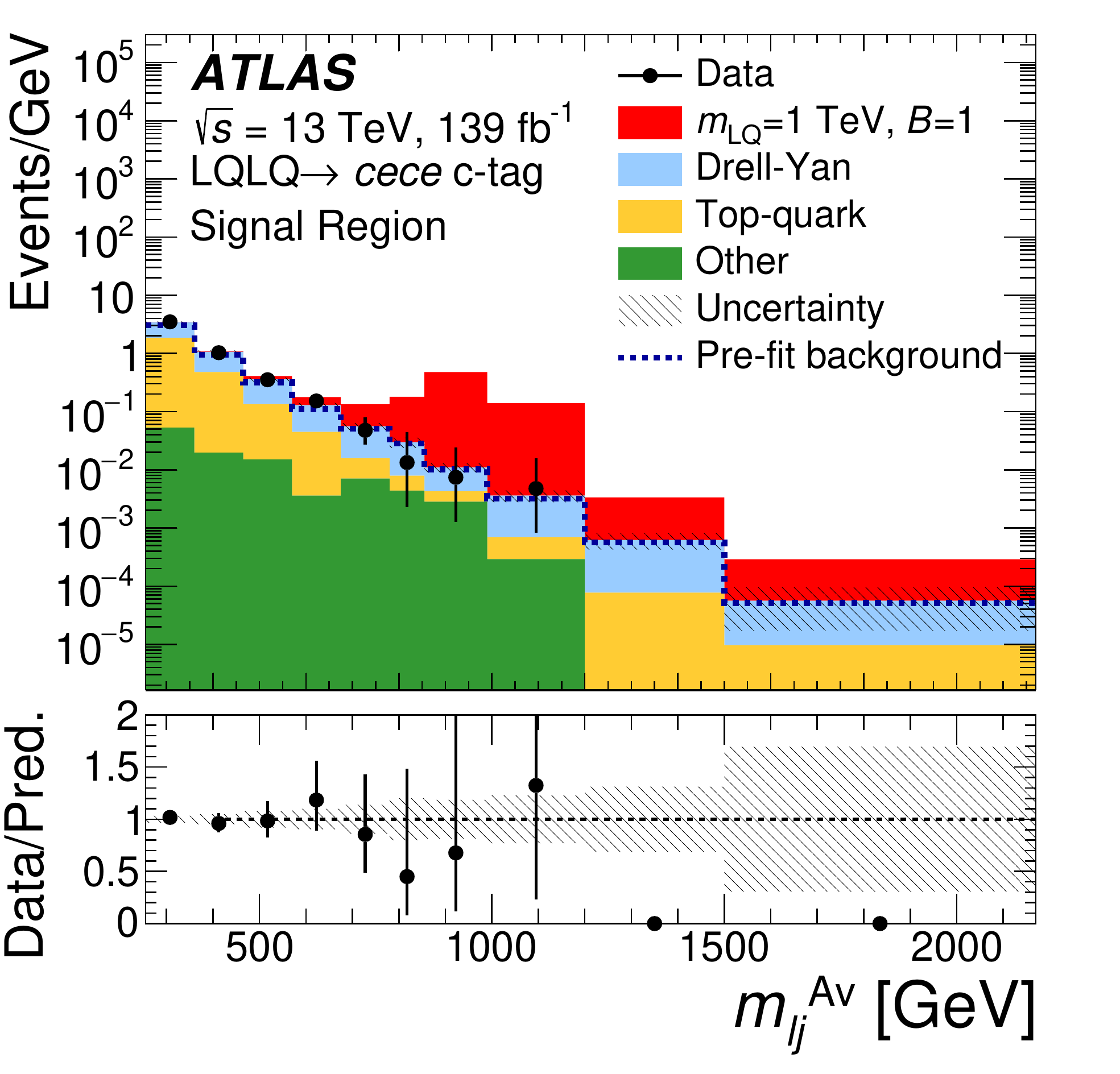}
\includegraphics[width=0.32\textwidth]{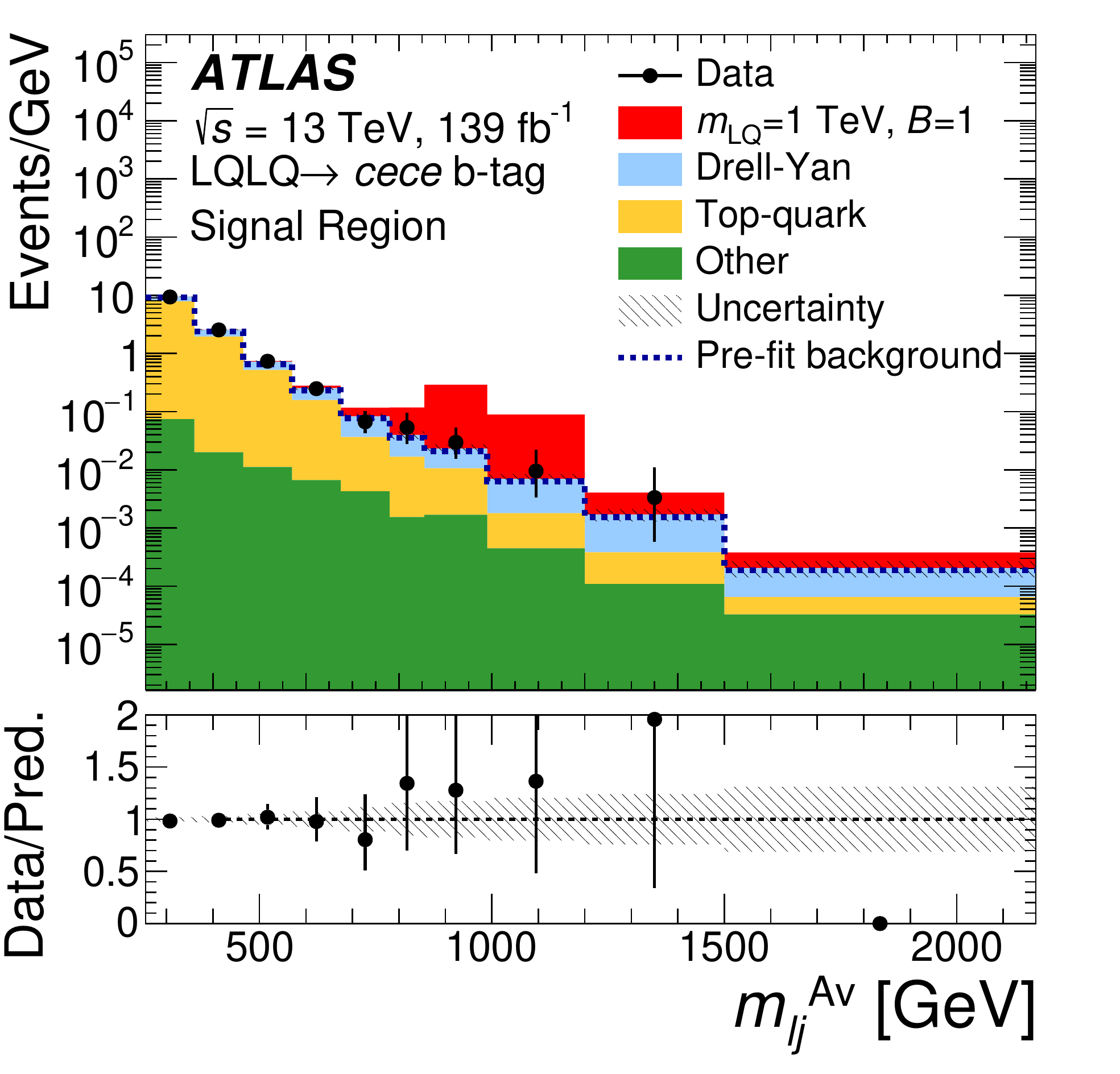}
\caption{Post-fit distributions of $m_{{\ell j}}^\mathrm{Av}$ in the $ce$ signal regions: untagged (left), $c$-tag (middle), and $b$-tag (right). The expected signals,
shown for $m_{\mathrm{LQ}}=1$~\TeV\ and $\mathcal{B}(\mathrm{LQ} \to ce)=1$, are shown for illustrative purposes.
The category `Top quark' refers to $t\bar{t}$ and single-top-quark processes.
The category `Other' refers to diboson and $W$+jet production.
The hatched band represents the total uncertainty in the background predictions.
}
\label{fig:Results:postfitcece}
\centering
\includegraphics[width=0.32\textwidth]{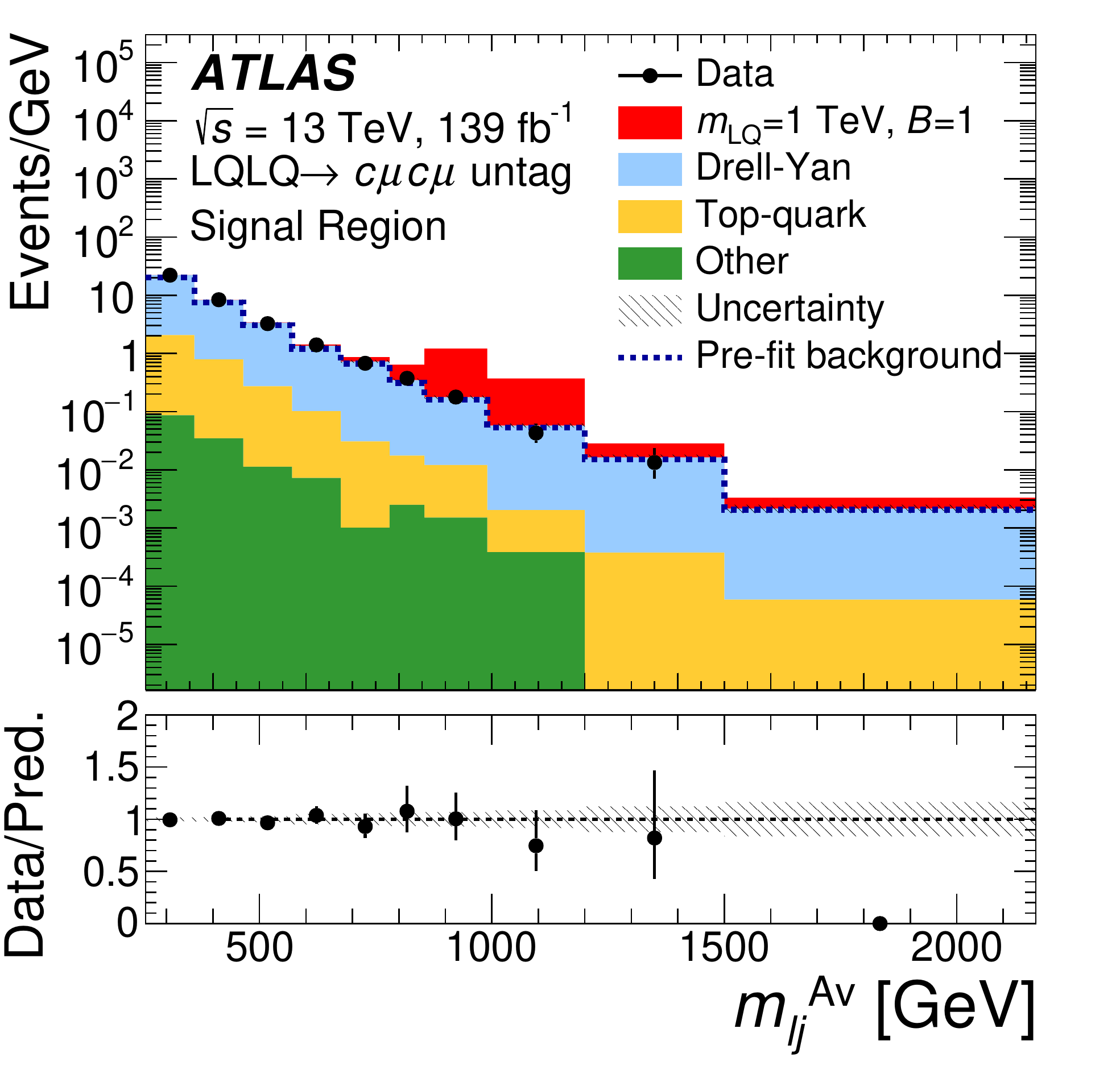}
\includegraphics[width=0.32\textwidth]{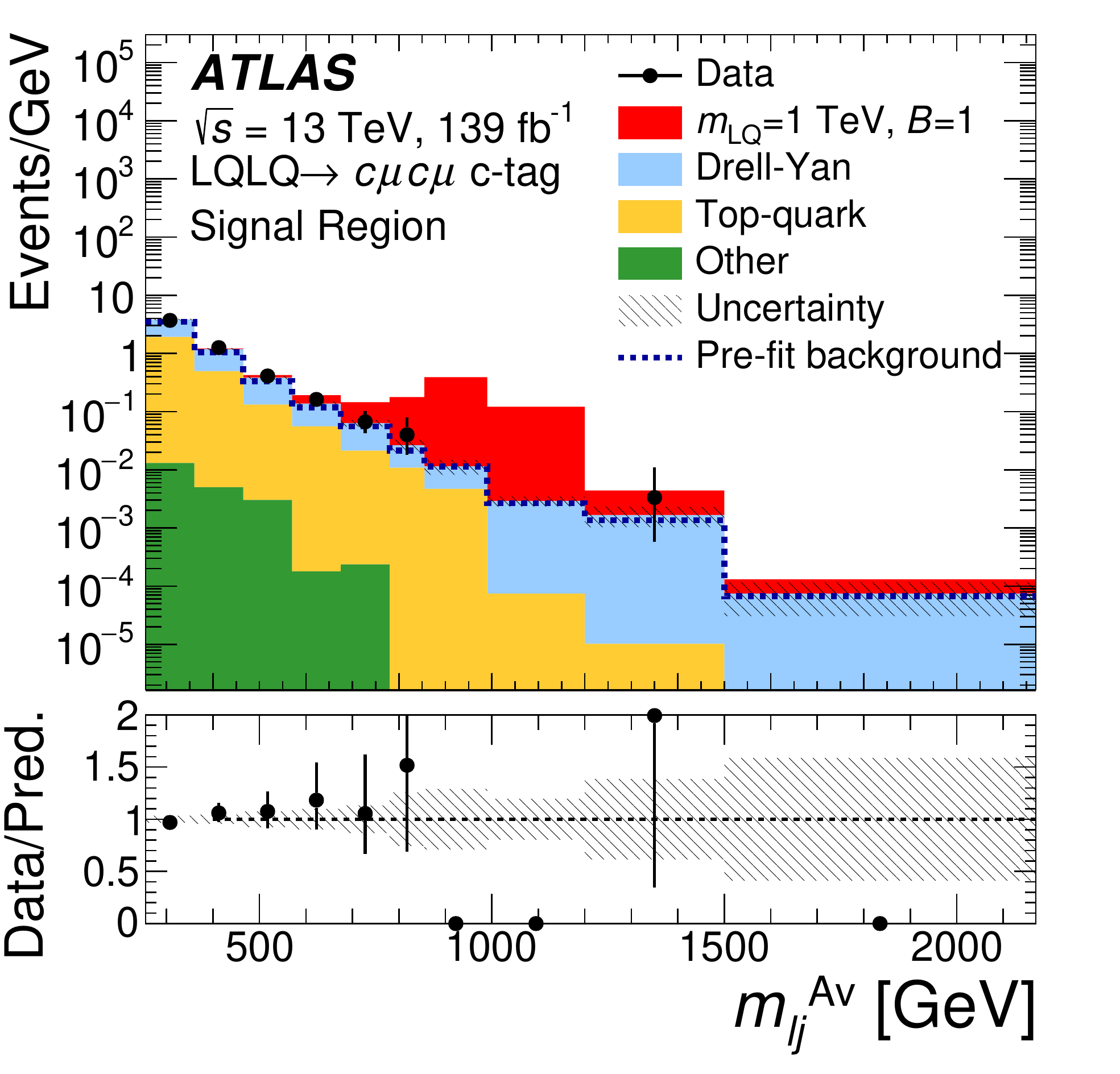}
\includegraphics[width=0.32\textwidth]{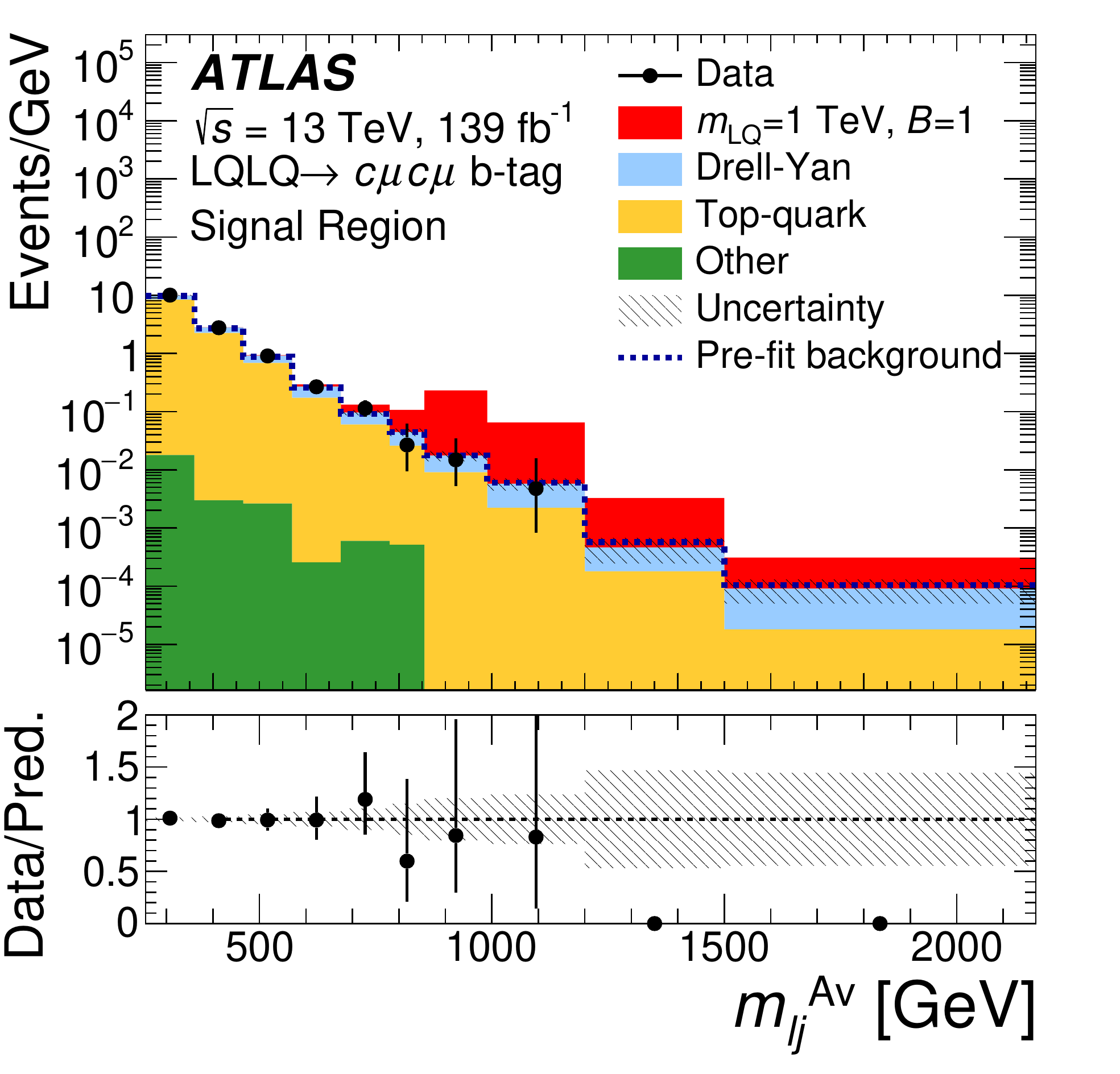}
\caption{Post-fit distributions of $m_{{\ell j}}^\mathrm{Av}$ in the $c\mu$ signal regions: untagged (left), $c$-tag (middle), and $b$-tag (right). The expected signals,
shown for $m_{\mathrm{LQ}}=1$~\TeV\ and $\mathcal{B}(\mathrm{LQ} \to c\mu)=1$, are shown for illustrative purposes.
The category `Top quark' refers to $t\bar{t}$ and single-top-quark processes.
The category `Other' refers to diboson and $W$+jet production.
The hatched band represents the total uncertainty in the background predictions.
}
\label{fig:Results:postfitcmucmu}
\end{figure}

\begin{figure}[htbp]
\centering
\includegraphics[width=0.32\textwidth]{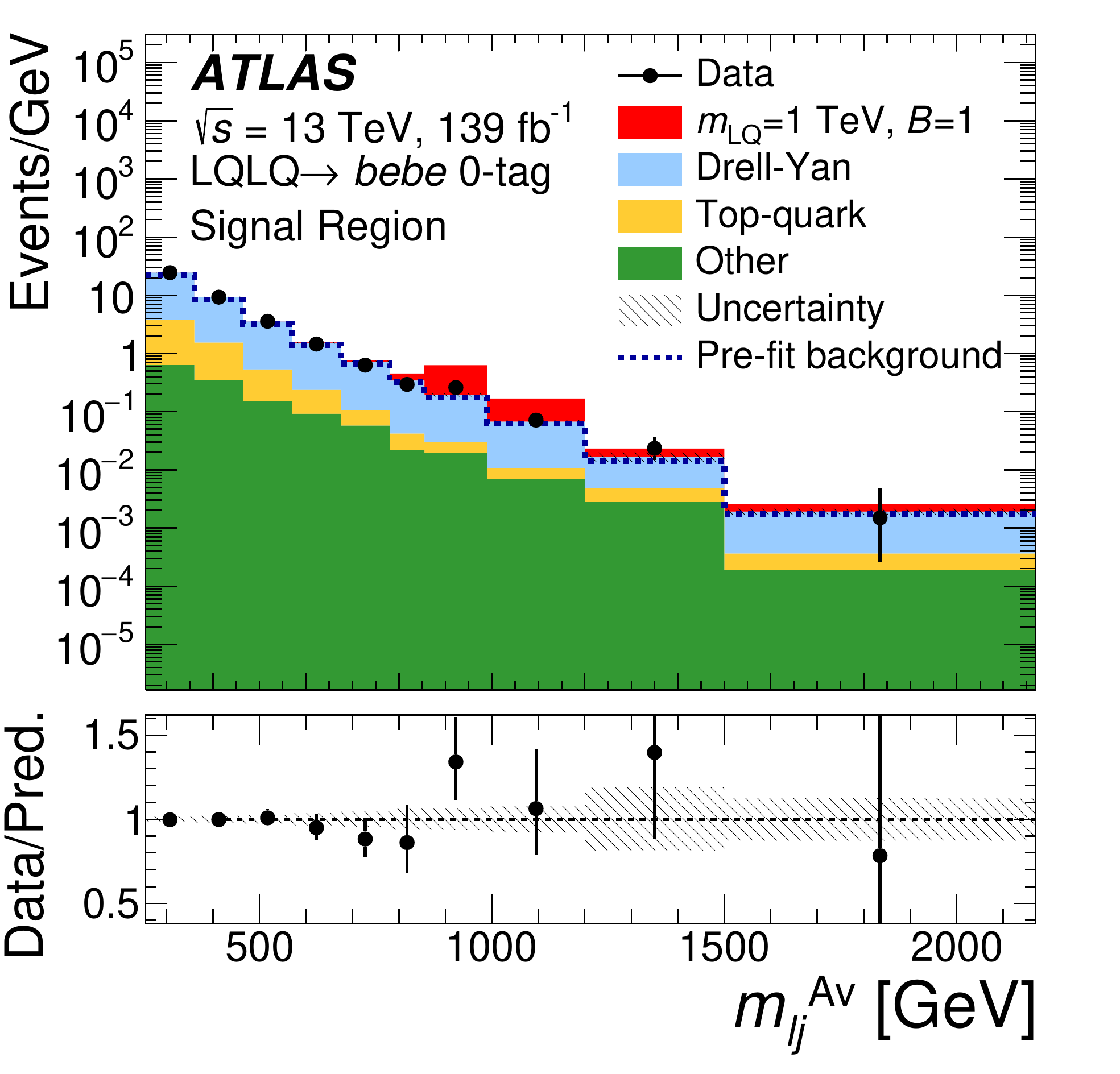}
\includegraphics[width=0.32\textwidth]{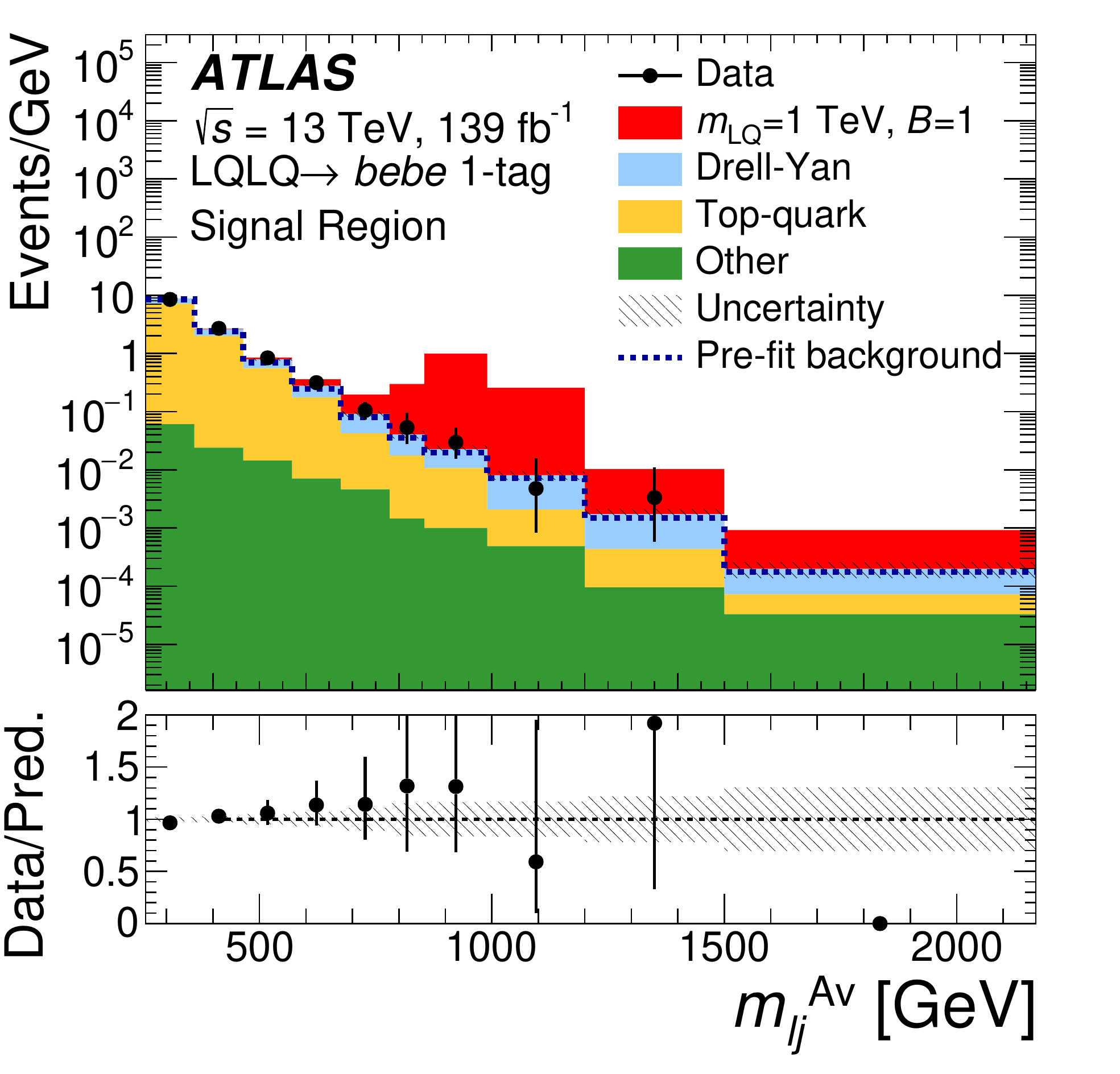}
\includegraphics[width=0.32\textwidth]{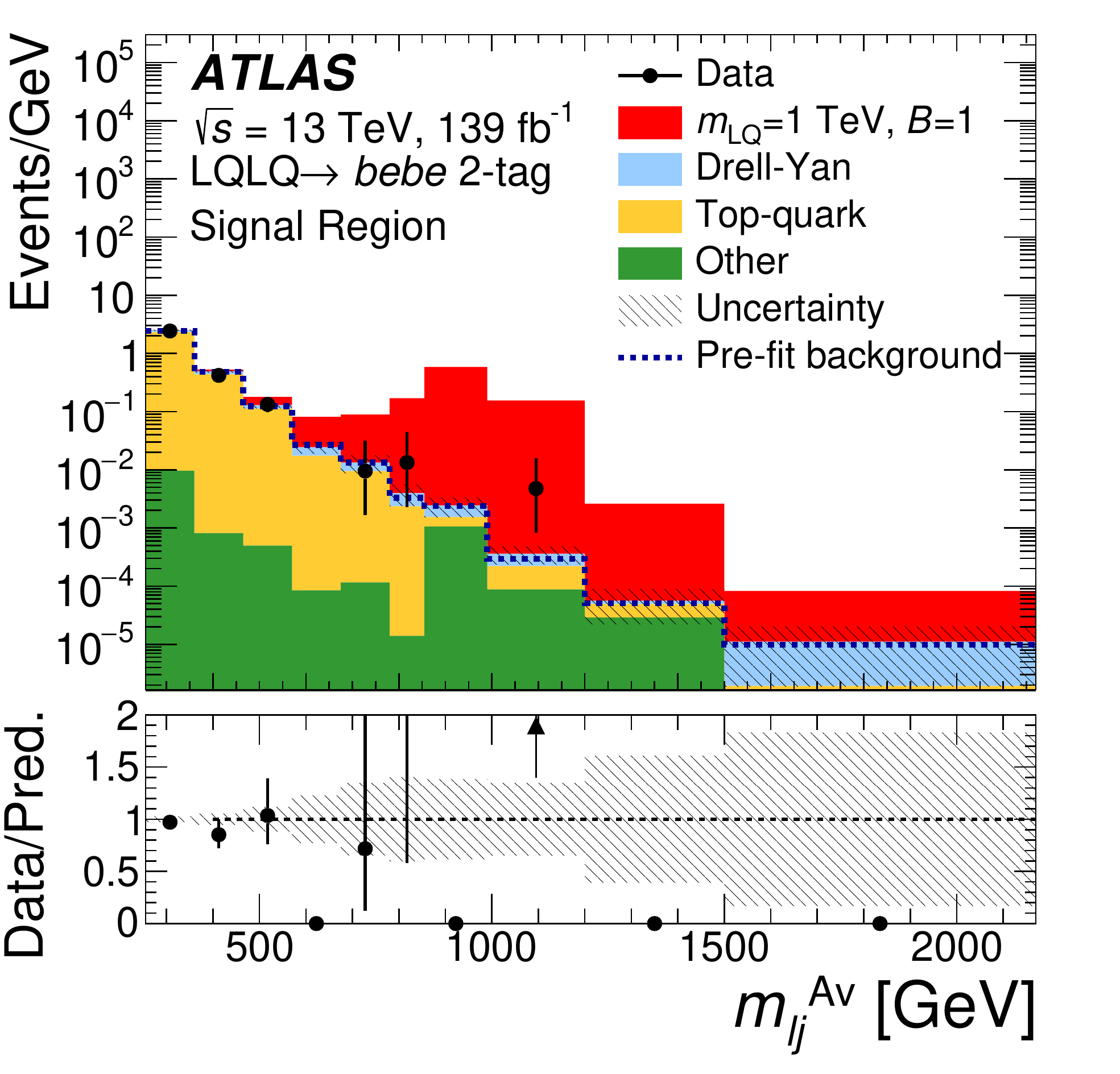}
\caption{Post-fit distributions of $m_{{\ell j}}^\mathrm{Av}$ in the $be$ signal regions: 0-tag (left), 1-tag (middle), and 2-tag (right). The expected signals,
shown for $m_{\mathrm{LQ}}=1$~\TeV\ and $\mathcal{B}(\mathrm{LQ} \to be)=1$, are shown for illustrative purposes.
The category `Top quark' refers to $t\bar{t}$ and single-top-quark processes.
The category `Other' refers to diboson and $W$+jet production.
The hatched band represents the total uncertainty in the background predictions.
}
\label{fig:Results:postfitbebe}
\centering
\includegraphics[width=0.32\textwidth]{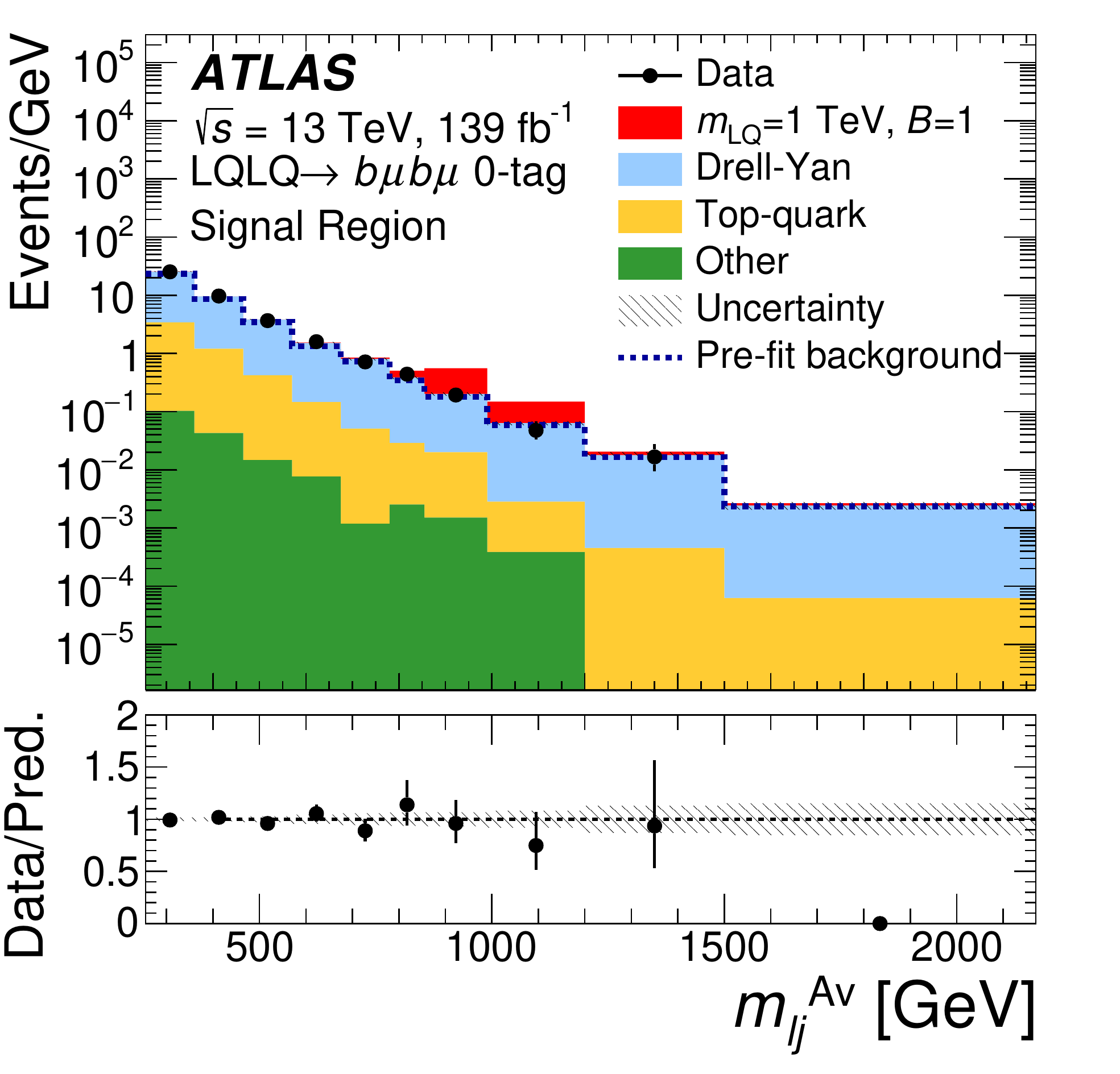}
\includegraphics[width=0.32\textwidth]{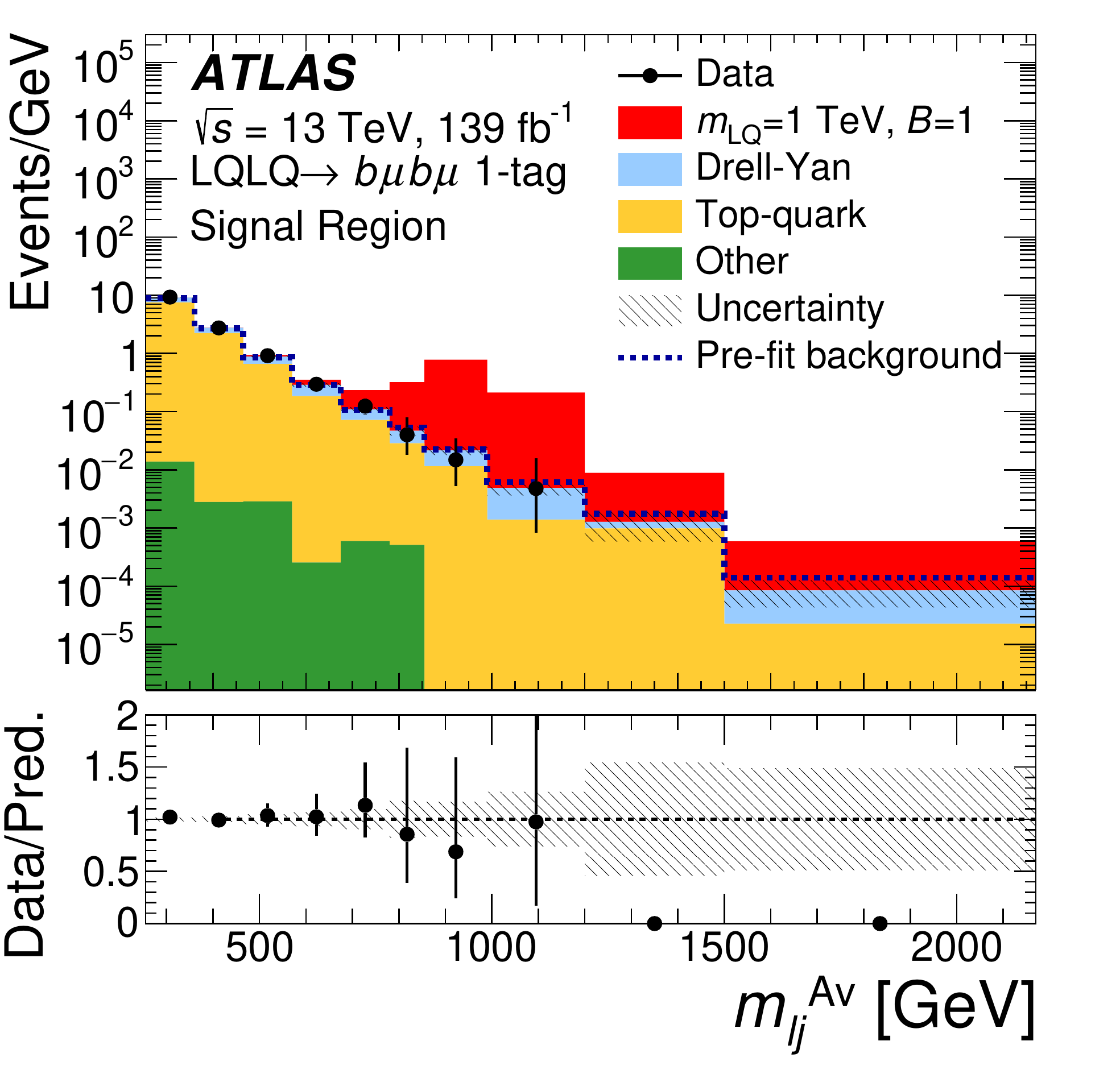}
\includegraphics[width=0.32\textwidth]{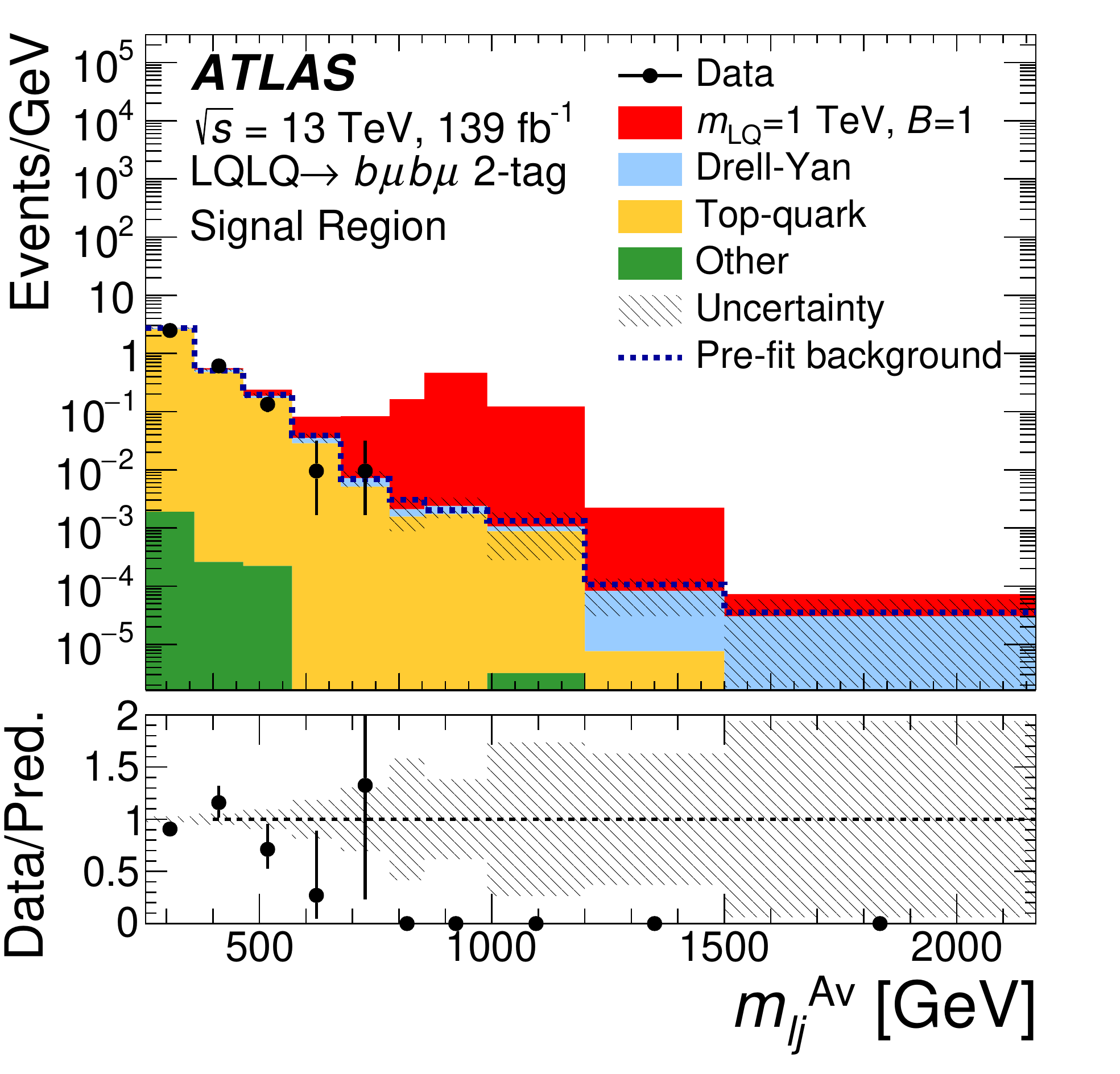}
\caption{Post-fit distributions of $m_{{\ell j}}^\mathrm{Av}$ in the $b\mu$ signal regions: 0-tag (left), 1-tag (middle), and 2-tag (right). The expected signals,
shown for $m_{\mathrm{LQ}}=1$~\TeV\ and $\mathcal{B}(\mathrm{LQ} \to b\mu)=1$, are shown for illustrative purposes.
The category `Top quark' refers to $t\bar{t}$ and single-top-quark processes.
The category `Other' refers to diboson and $W$+jet production.
The hatched band represents the total uncertainty in the background predictions.
}
\label{fig:Results:postfitbmubmu}
\end{figure}

 
Figures~\ref{fig:Results:postfitqlql} to~\ref{fig:Results:postfitbmubmu} show comparisons between the
observed data and the post-fit SM predictions for $m_{\ell j}^\mathrm{Av}$ for all signal regions
in the $q\ell$, $c\ell$, and $b\ell$ channels. In each case, the expected distribution for one scenario with LQ mass
of 1~\TeV\ is shown for illustrative purposes, considering $\mathrm{LQ} \to qe / q\mu$ for the pretag channels
(with $q=u$-, $d$- or $s$-quark),
$\mathrm{LQ} \to ce / c\mu$ for the $c\ell$ channels, and $\mathrm{LQ} \to be / b\mu$ for the $b\ell$ channels.
Only the data and predictions within the  mass range shown in the Figures are used in the fit, although it should
be noted that no data events are recorded above the range.
 
\begin{figure}[htbp]
\centering
\includegraphics[width=0.45\textwidth]{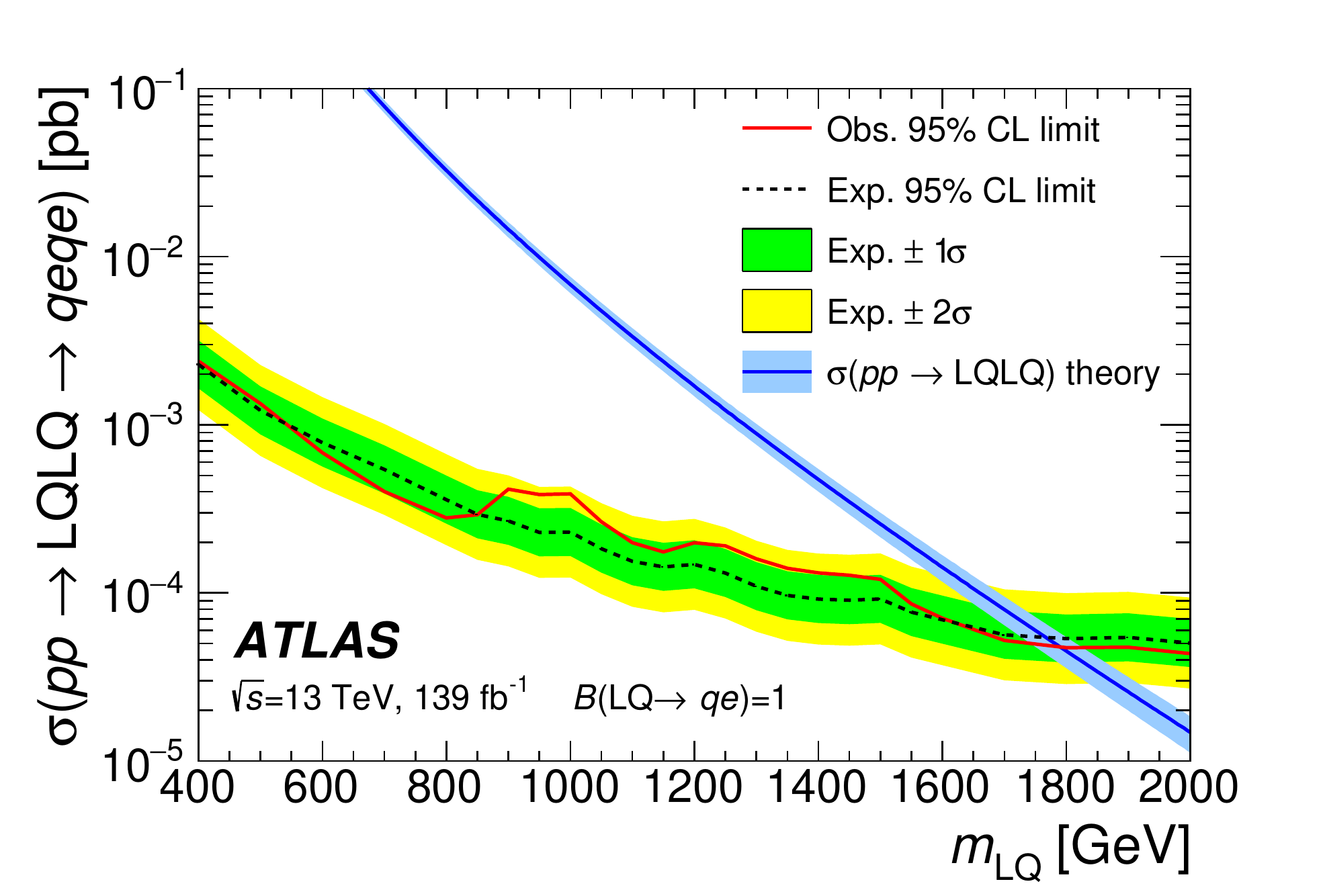}
\includegraphics[width=0.45\textwidth]{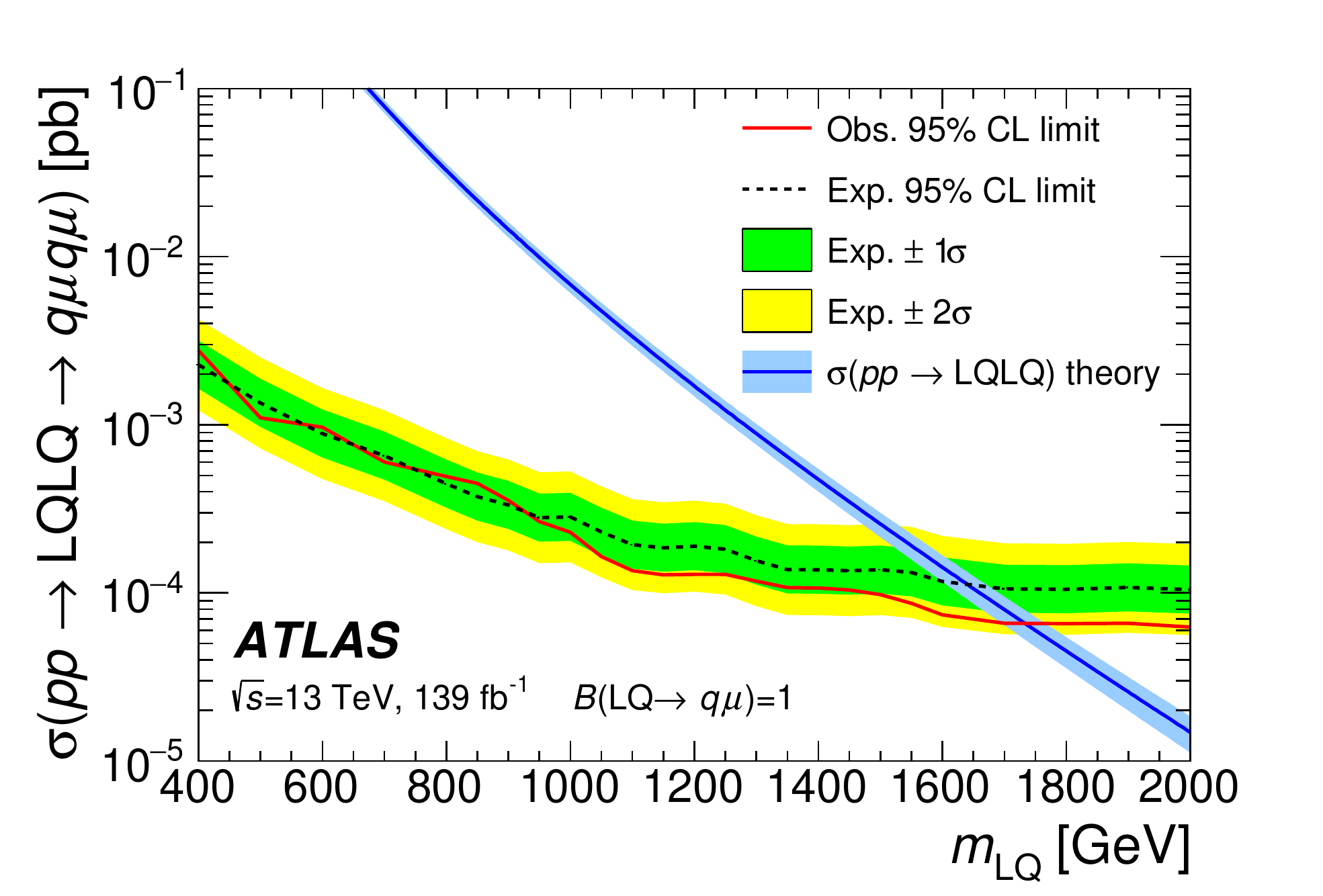}
\includegraphics[width=0.45\textwidth]{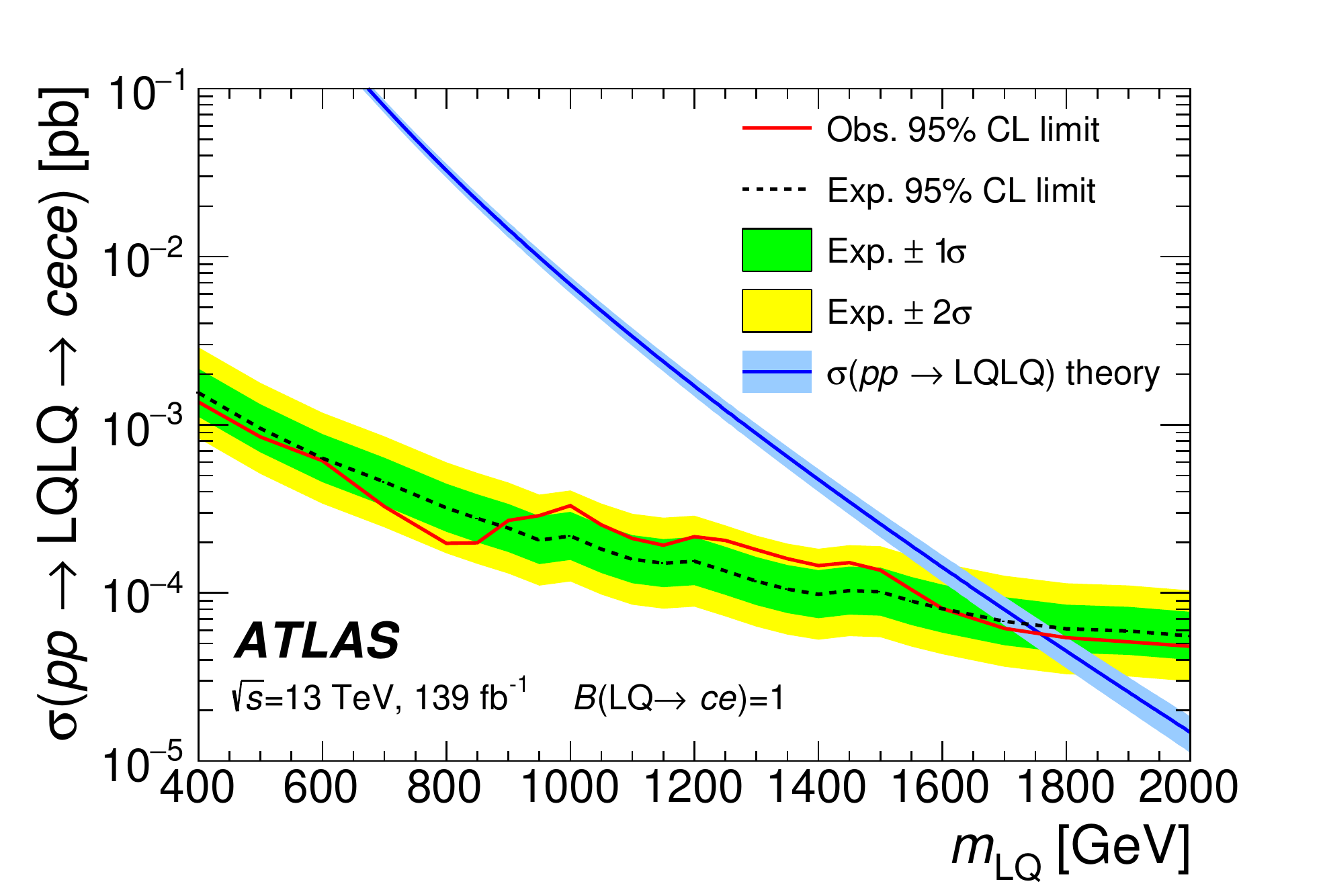}
\includegraphics[width=0.45\textwidth]{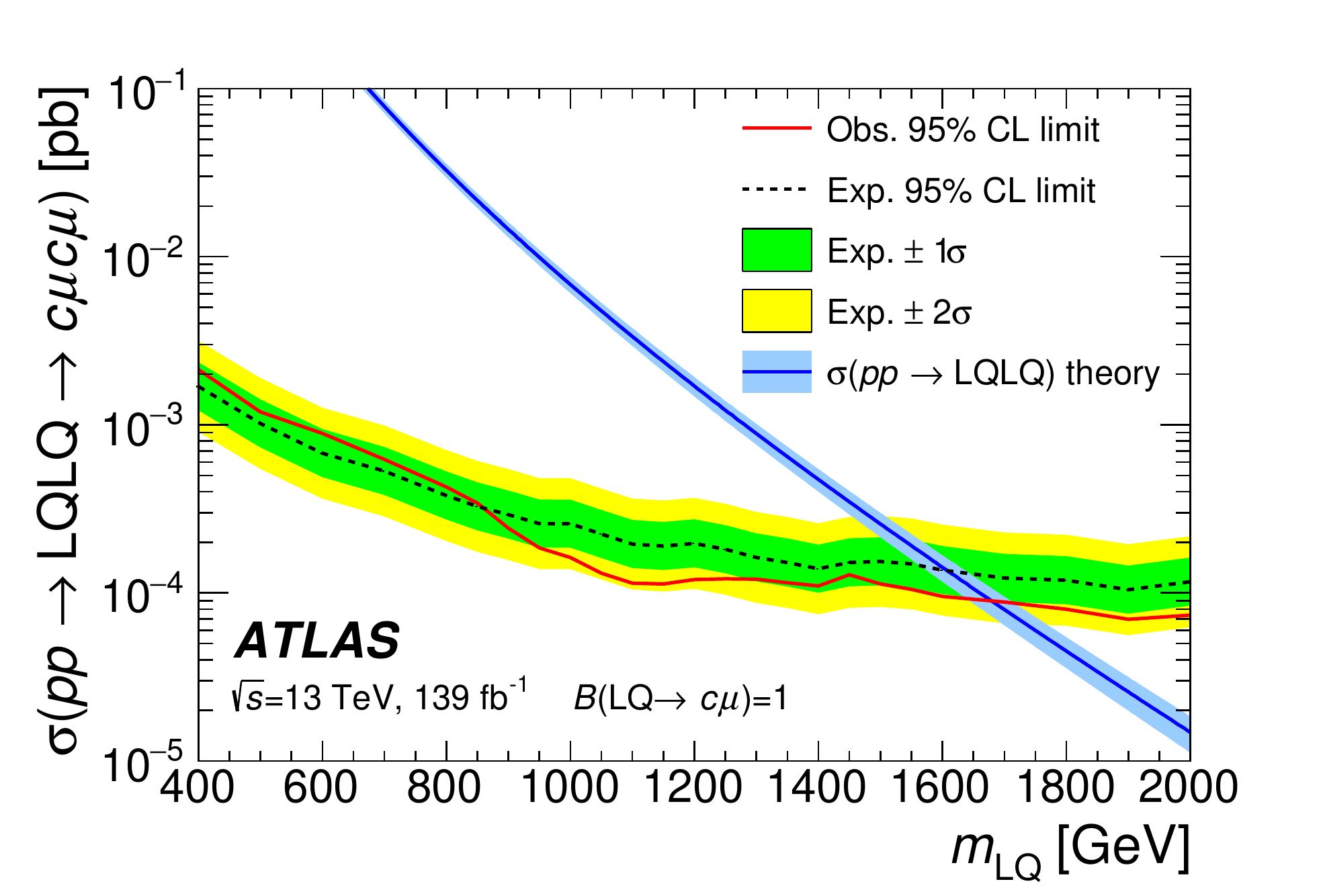}
\includegraphics[width=0.45\textwidth]{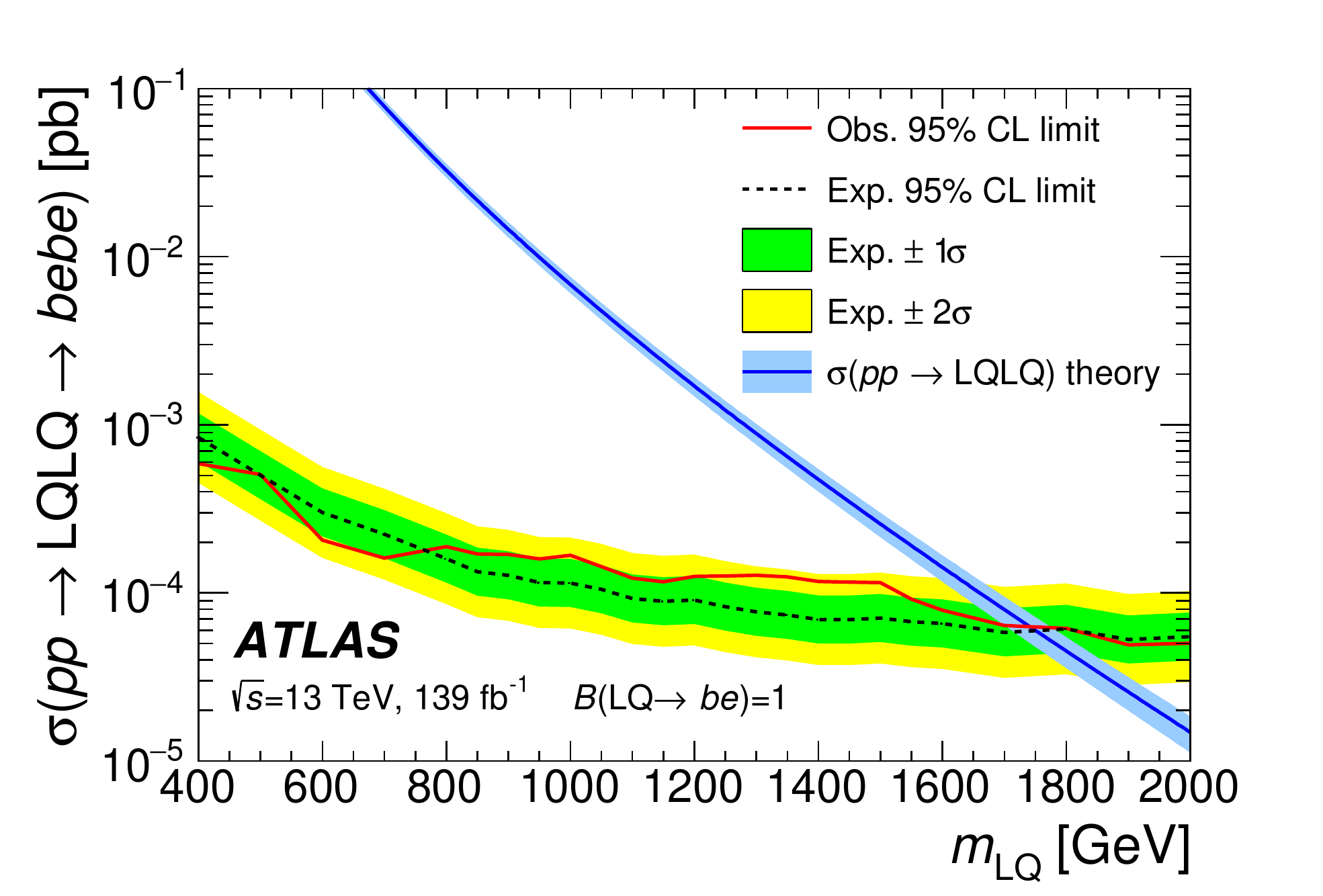}
\includegraphics[width=0.45\textwidth]{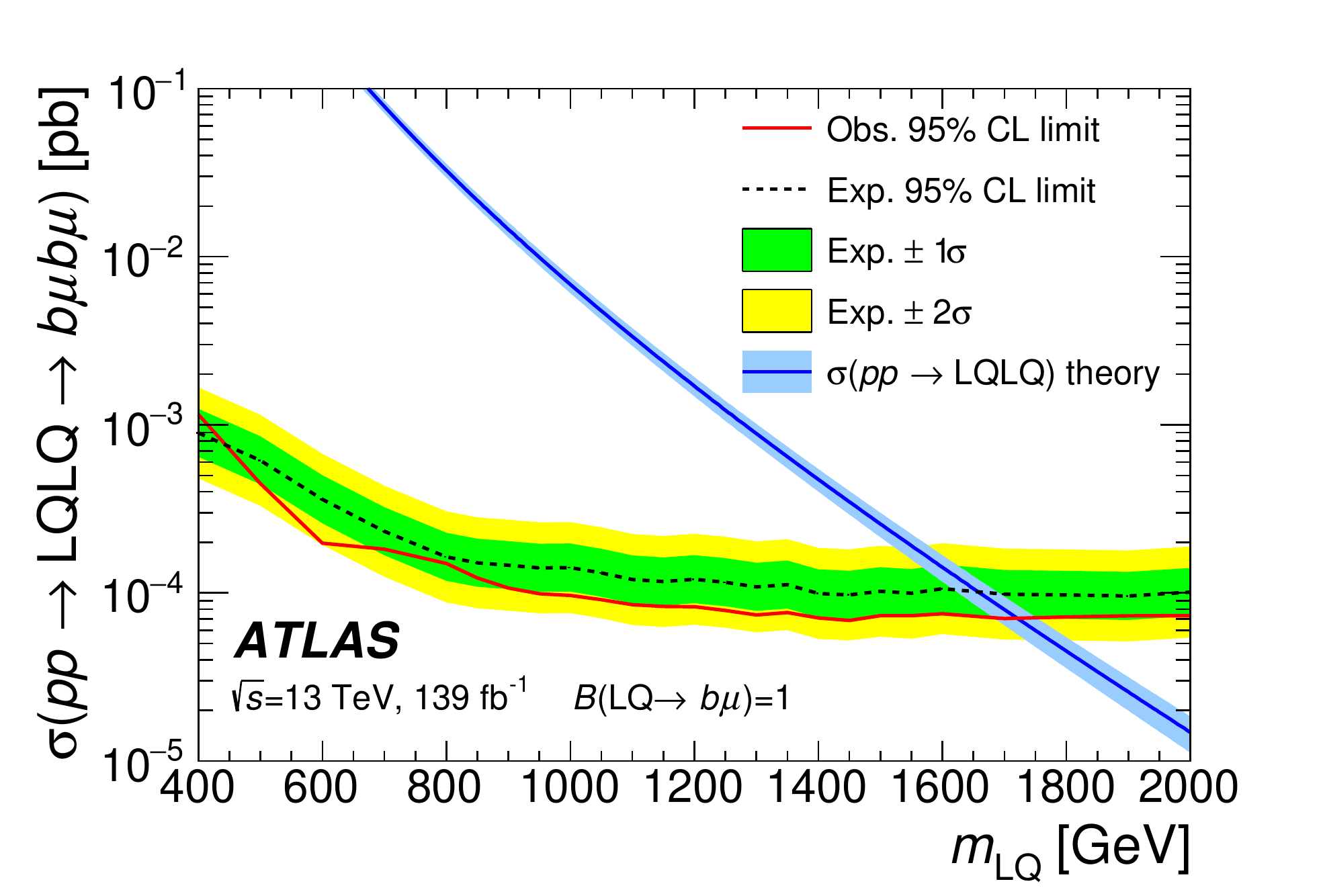}
 
\caption{The observed (solid line) and expected (dashed line) limits on the leptoquark pair production cross-section at  95\% CL for $\mathcal{B}=1$ into electrons or muons, shown as a function of $m_{\mathrm{LQ}}$ for the different leptoquark channels. The green and yellow
bands show the $\pm 1\sigma$ and $\pm 2\sigma$ ranges of the expected limit.
Also included on the plots is the expected theoretical cross-section.
The thickness of the theory curve represents the theoretical uncertainty from PDFs,
renormalisation and factorisation scales, and the strong coupling constant $\alpha^{}_\mathrm{S}$.}
\label{fig:Results:limits}
\end{figure}

As no evidence of an excess at any mass in any of the channels was
found, upper limits on the leptoquark production cross-section are
computed at the 95\% confidence level using a modified frequentist CL$_\text{s}$
method~\cite{Cowan:2010st,Read:2002hq}.
The limits are shown in Figure~\ref{fig:Results:limits} as a function
of $m_{\mathrm{LQ}}$ for a 100\% branching ratio into charged leptons.
They were calculated for LQ masses of the generated samples, and a linear
interpolation has been made between mass points.
The theoretical prediction for the cross-section of scalar leptoquark
pair production is shown by the solid line along with the uncertainties.
Exclusion limits are driven by the small number of data events populating the
high-mass part of the lepton--jet spectrum.
The limits at large $m_{\mathrm{LQ}}$ are more stringent for decays with electrons
than for decays with muons, due to the better electron resolution at high \pt.
The decays involving $c$- and $b$-quarks have lower cross-section limits
at low mass, due to the lower rate of SM background contributions in
the tagged categories.
 
\begin{figure}[htbp]
\centering
\includegraphics[width=0.45\textwidth]{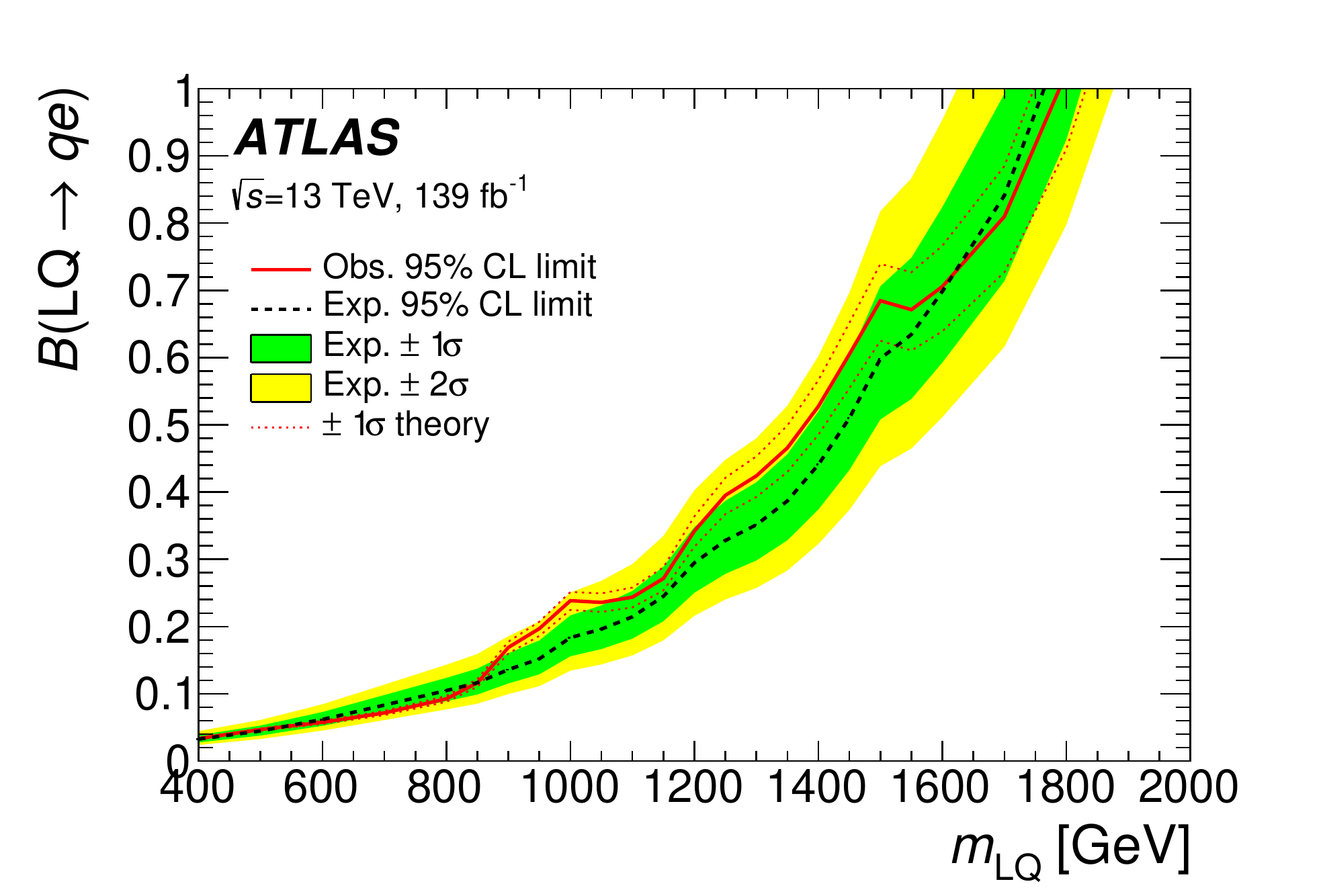}
\includegraphics[width=0.45\textwidth]{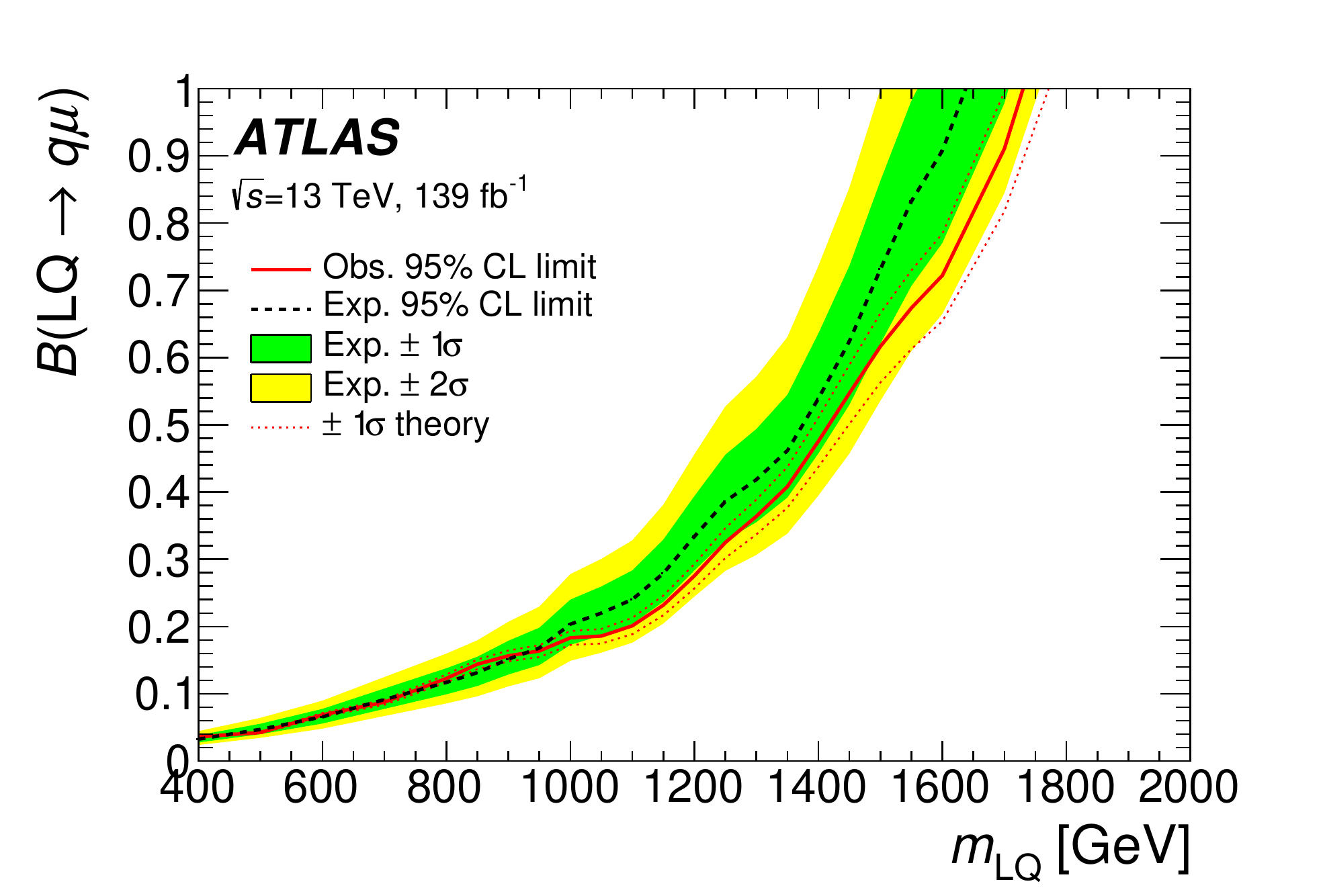}
\includegraphics[width=0.45\textwidth]{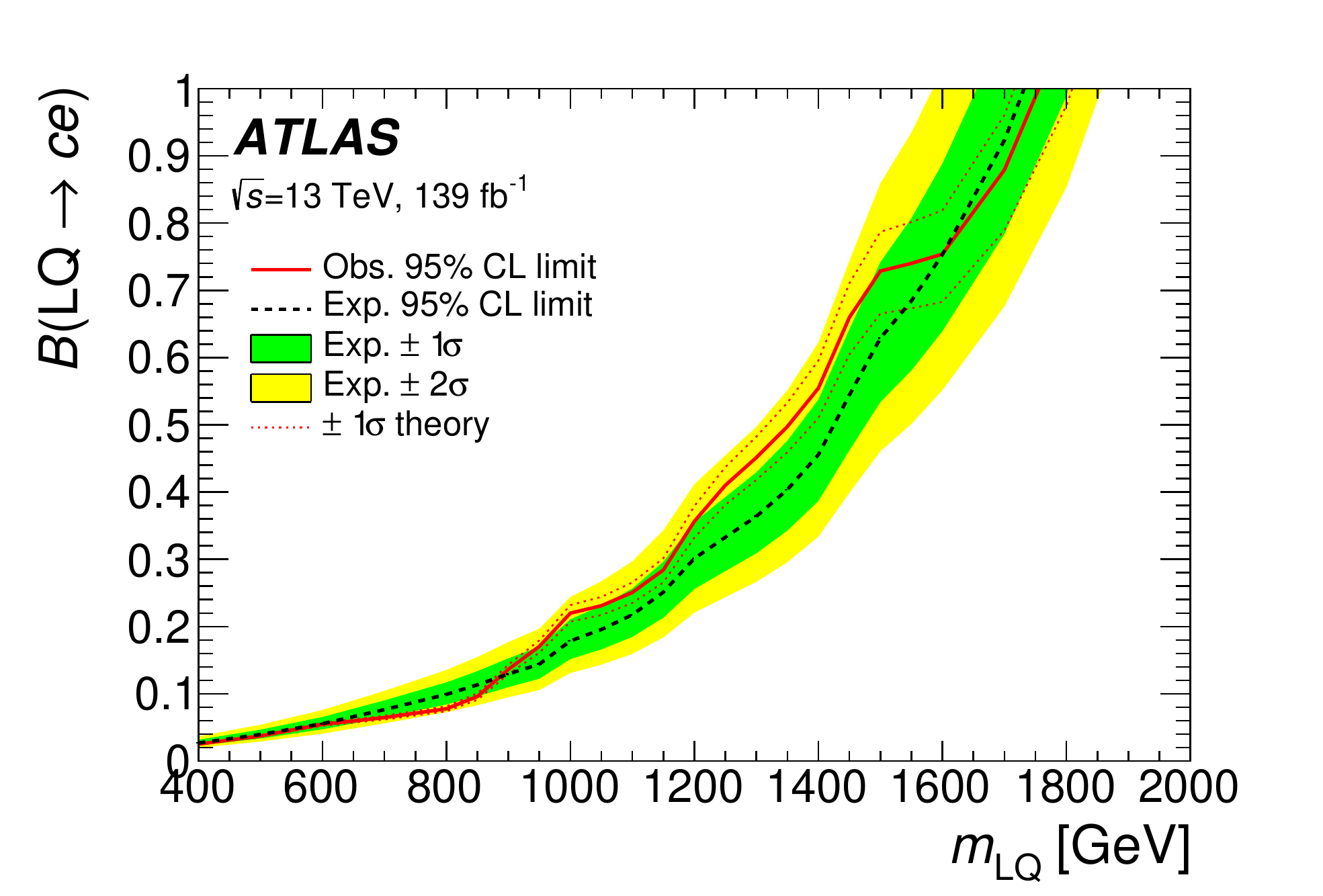}
\includegraphics[width=0.45\textwidth]{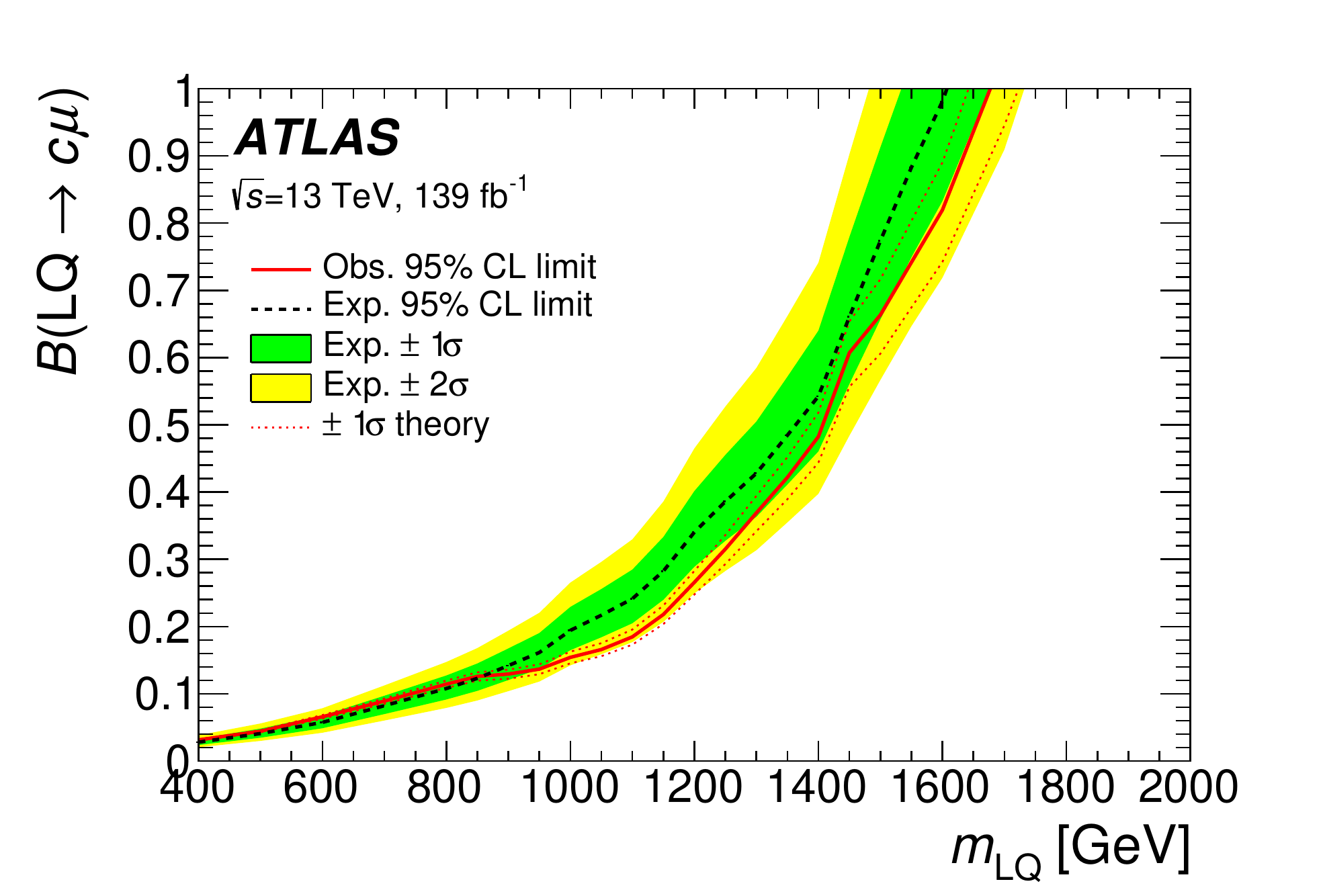}
\includegraphics[width=0.45\textwidth]{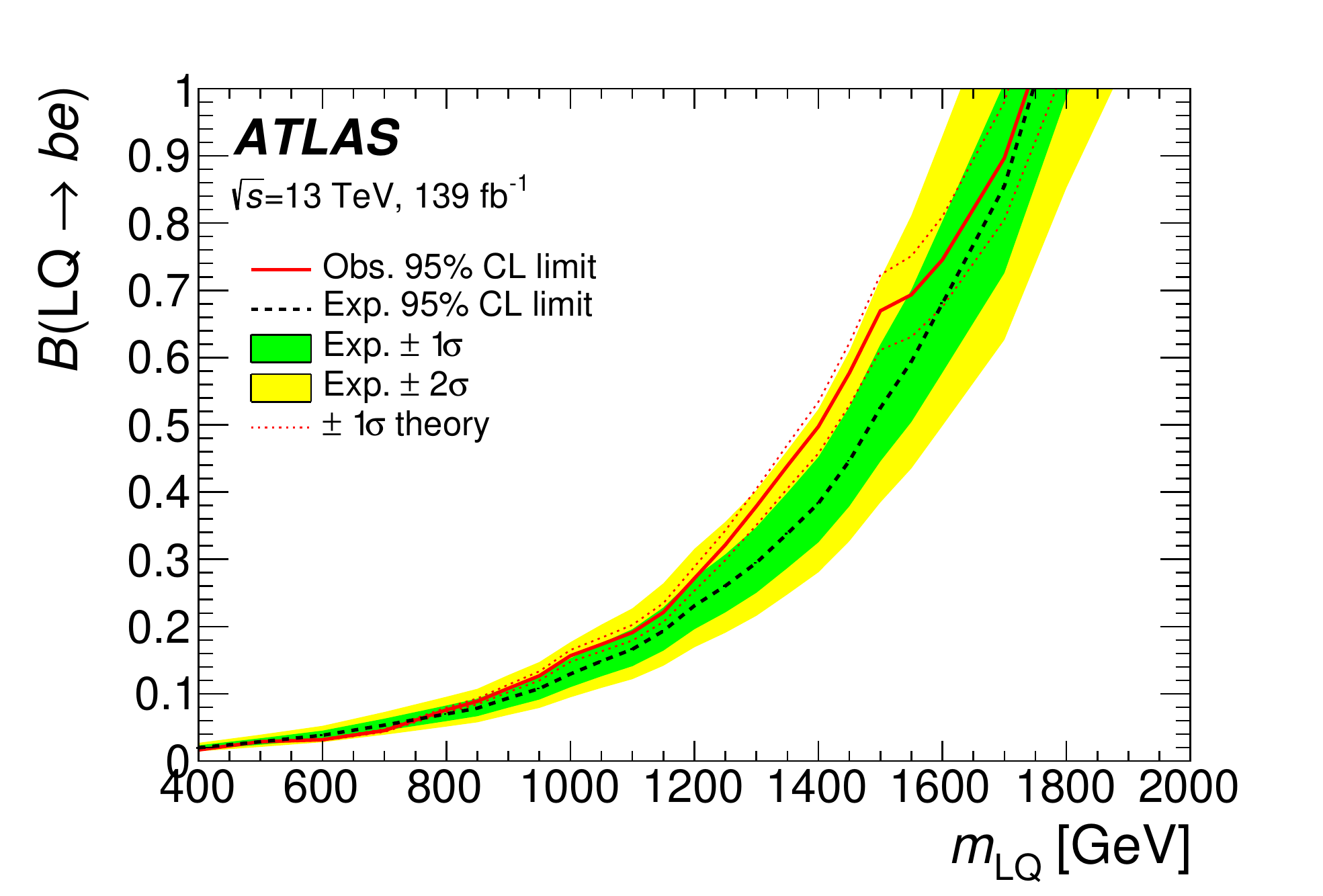}
\includegraphics[width=0.45\textwidth]{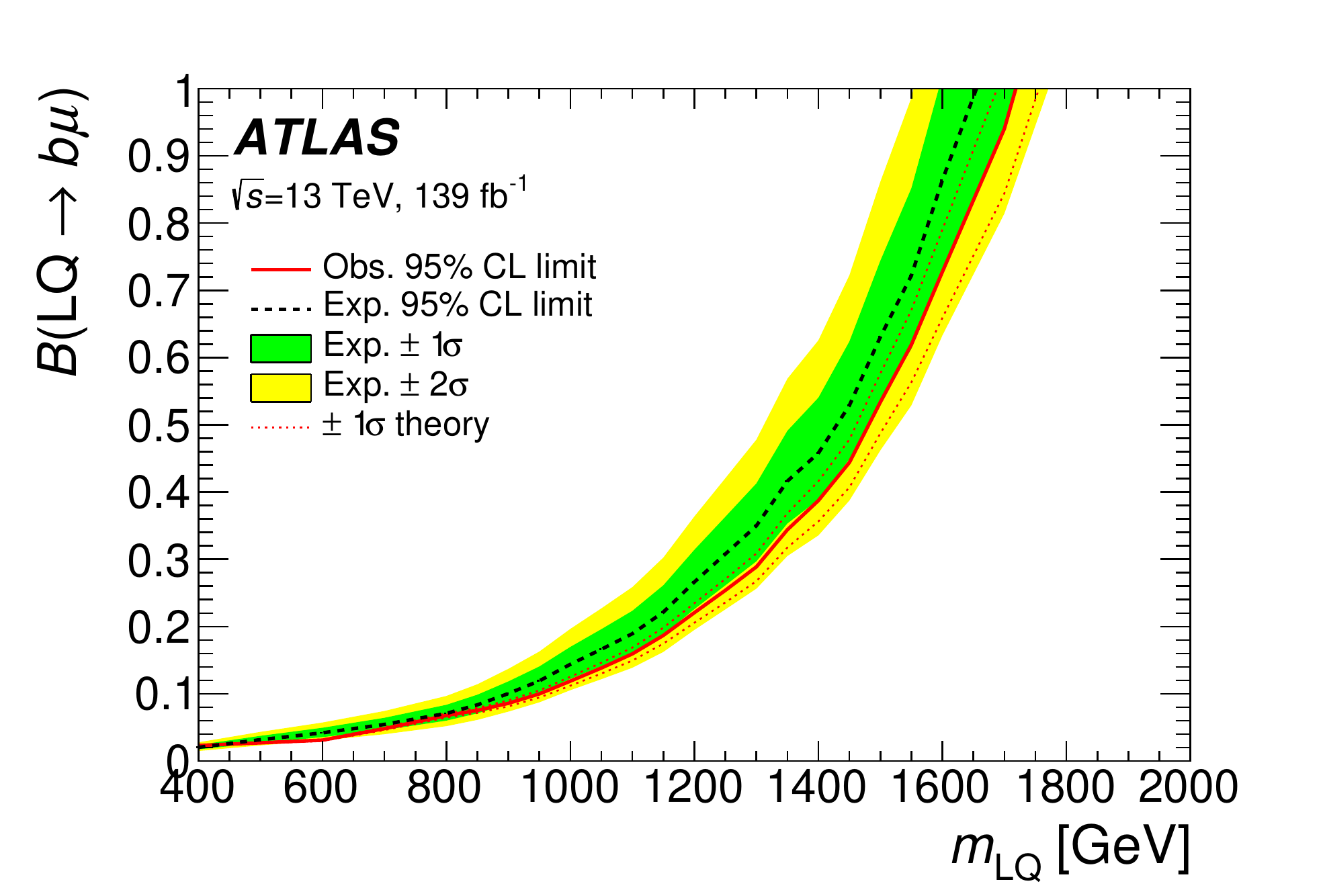}
 
\caption{The observed (solid line) and expected (dashed line) limits on the leptoquark branching ratio $\mathcal{B}$ into a quark and an electron or a muon at  95\% CL, shown as a function of $m_{\mathrm{LQ}}$ for the different leptoquark channels. The green and yellow
bands show the $\pm 1\sigma$ and $\pm 2\sigma$ ranges of the expected limit. The error band on the observed  curve (dotted lines) represents the uncertainty in the theoretical cross-section from PDFs,
renormalisation and factorisation scales, and the strong coupling constant $\alpha^{}_\mathrm{S}$.}
\label{fig:Results:bf}
\end{figure}

The results of the fit may also be expressed as limits on the branching
ratio into charged leptons as shown in Figure~\ref{fig:Results:bf}.
In this case, it is assumed that there is zero acceptance for
LQ decays involving neutrinos or top quarks. Furthermore, it is assumed that the
LQs can decay into only one specific combination of lepton flavour and quark flavour.
The $\mathcal{B}$ limit is computed as
$\sqrt{\sigma_\mathrm{obs}/\sigma_\mathrm{theory}}$, where $\sigma_\mathrm{obs}$ is the observed
LQ pair production cross-section exclusion limit with $\mathcal{B}=1$ into charged leptons
and $\sigma_\mathrm{theory}$ is the theoretical cross-section. Constraints on the LQ masses are
reduced by no more than 20\% for $\mathcal{B}=0.5$, and LQs with mass around 800~\GeV\ can be excluded
for branching ratios into charged leptons as low as $0.1$ (up to 900~\GeV\ for $b\ell$ channels).
This result improves upon the sensitivity of previous scalar
LQ searches by about 300--400~\GeV\ in LQ mass depending on the lepton flavour, and it establishes
for the first time limits on cross-generational LQ decays using dedicated $c$- and $b$-tagging algorithms.
 
\FloatBarrier

\section{Conclusion}
\label{sec:conclusion}
 
A search for a new-physics resonances decaying into a lepton and a jet performed by the ATLAS experiment is presented.
Scalar leptoquarks, pair produced in \(pp\) collisions at $\sqrt{s}=13$~\TeV\ at the LHC,
are considered using an integrated luminosity of 139~fb$^{-1}$, corresponding to the full Run 2 dataset.
Leptoquarks are searched for in events with two electrons or muons and
two or more jets. Tagging algorithms are used to  identify jets arising from the fragmentation of $b$-quarks ($b$-jets)
and, for the first time, of $c$-quarks ($c$-jets).
The observed yield in each channel is consistent with SM background expectations.
Leptoquarks with masses below 1.8~\TeV\ and 1.7~\TeV\ are excluded in the electron and muon channels, respectively,
assuming a branching ratio into a charged lepton and a quark of 100\%, with minimal dependency on the quark flavour.
Upper limits on the aforementioned branching ratio are also presented. LQs with masses up to around 800~\GeV\ can be
excluded for branching ratios into charged leptons as low as $0.1$, assuming that there is zero acceptance for
LQ decays involving neutrinos or top quarks, and that only one charged lepton plus quark decay mode at the time is possible.
This result improves upon the sensitivity of previous scalar
LQ searches by about 300--400~\GeV\ in LQ mass depending on the lepton flavour, and it establishes
for the first time limits on cross-generational LQ decays using dedicated $c$- and $b$-jet identification algorithms.
 
\section*{Acknowledgements}
 

We thank CERN for the very successful operation of the LHC, as well as the
support staff from our institutions without whom ATLAS could not be
operated efficiently.
 
We acknowledge the support of ANPCyT, Argentina; YerPhI, Armenia; ARC, Australia; BMWFW and FWF, Austria; ANAS, Azerbaijan; SSTC, Belarus; CNPq and FAPESP, Brazil; NSERC, NRC and CFI, Canada; CERN; CONICYT, Chile; CAS, MOST and NSFC, China; COLCIENCIAS, Colombia; MSMT CR, MPO CR and VSC CR, Czech Republic; DNRF and DNSRC, Denmark; IN2P3-CNRS and CEA-DRF/IRFU, France; SRNSFG, Georgia; BMBF, HGF and MPG, Germany; GSRT, Greece; RGC and Hong Kong SAR, China; ISF and Benoziyo Center, Israel; INFN, Italy; MEXT and JSPS, Japan; CNRST, Morocco; NWO, Netherlands; RCN, Norway; MNiSW and NCN, Poland; FCT, Portugal; MNE/IFA, Romania; MES of Russia and NRC KI, Russia Federation; JINR; MESTD, Serbia; MSSR, Slovakia; ARRS and MIZ\v{S}, Slovenia; DST/NRF, South Africa; MICINN, Spain; SRC and Wallenberg Foundation, Sweden; SERI, SNSF and Cantons of Bern and Geneva, Switzerland; MOST, Taiwan; TAEK, Turkey; STFC, United Kingdom; DOE and NSF, United States of America. In addition, individual groups and members have received support from BCKDF, CANARIE, Compute Canada and CRC, Canada; ERC, ERDF, Horizon 2020, Marie Sk{\l}odowska-Curie Actions and COST, European Union; Investissements d'Avenir Labex, Investissements d'Avenir Idex and ANR, France; DFG and AvH Foundation, Germany; Herakleitos, Thales and Aristeia programmes co-financed by EU-ESF and the Greek NSRF, Greece; BSF-NSF and GIF, Israel; La Caixa Banking Foundation, CERCA Programme Generalitat de Catalunya and PROMETEO and GenT Programmes Generalitat Valenciana, Spain; G\"{o}ran Gustafssons Stiftelse, Sweden; The Royal Society and Leverhulme Trust, United Kingdom.
 
The crucial computing support from all WLCG partners is acknowledged gratefully, in particular from CERN, the ATLAS Tier-1 facilities at TRIUMF (Canada), NDGF (Denmark, Norway, Sweden), CC-IN2P3 (France), KIT/GridKA (Germany), INFN-CNAF (Italy), NL-T1 (Netherlands), PIC (Spain), ASGC (Taiwan), RAL (UK) and BNL (USA), the Tier-2 facilities worldwide and large non-WLCG resource providers. Major contributors of computing resources are listed in Ref.~\cite{comp_Tier2s}.
 

\printbibliography

@article{PhysRevD.68.116004,
  title = {CP-violating electron-nucleon interactions from leptoquark exchange},
  author = {Herczeg, Peter},
  journal = {Phys. Rev. D},
  volume = {68},
  issue = {11},
  pages = {116004},
  numpages = {10},
  year = {2003},
  month = {Dec},
  publisher = {American Physical Society},
  doi = {10.1103/PhysRevD.68.116004},
  url = {https://link.aps.org/doi/10.1103/PhysRevD.68.116004}
}

@article{FRITZSCH1975193,
title = "Unified interactions of leptons and hadrons",
journal = "Annals of Physics",
volume = "93",
number = "1",
pages = "193 - 266",
year = "1975",
%issn = "0003-4916",
doi = "https://doi.org/10.1016/0003-4916(75)90211-0",
%url = "http://www.sciencedirect.com/science/article/pii/0003491675902110",
author = "Harald Fritzsch and Peter Minkowski",
}

@article{PhysRevLett.32.438,
  title = {Unity of All Elementary-Particle Forces},
  author = {Georgi, Howard and Glashow, S. L.},
  journal = {Phys. Rev. Lett.},
  volume = {32},
  pages = {438--441},
  numpages = {0},
  year = {1974},
  month = {Feb},
  publisher = {American Physical Society},
  doi = {10.1103/PhysRevLett.32.438},
%  url = {https://link.aps.org/doi/10.1103/PhysRevLett.32.438}
}

@article{Borschensky:2020hot,
      author         = "Borschensky, Christoph and Fuks, Benjamin and Kulesza,
                        Anna and Schwartländer, Daniel",
      title          = "{Precision predictions for scalar leptoquark
                        pair-production at hadron colliders}",
      year           = "2020",
      eprint         = "2002.08971",
      archivePrefix  = "arXiv",
      primaryClass   = "hep-ph",
      reportNumber   = "MS-TP-20-09",
      SLACcitation   = "%%CITATION = ARXIV:2002.08971;%%"
}

@article{BUCHMULLER1987442,
title = "Leptoquarks in lepton-quark collisions",
journal = "Phys. Lett. B",
volume = "191",
number = "4",
pages = "442 - 448",
year = "1987",
%issn = "0370-2693",
doi = "https://doi.org/10.1016/0370-2693(87)90637-X",
%url = "http://www.sciencedirect.com/science/article/pii/037026938790637X",
author = "W. Buchmüller and R. Rückl and D. Wyler",
abstract = "Low-energy experiments permit the existence of leptoquarks with masses of order 100 GeV and couplings to quark-lepton pairs as large as gauge couplings. We study systematically the signatures of all possible scalar and vector leptoquarks in electron (positron)-proton collisions. Clear evidence for leptoquarks would be narrow peaks in the x-distributions of inclusive neutral and charged current processes. At HERA one will be able to explore the mass range up to 300 GeV through direct production, and even somewhat beyond the CM energy of 314 GeV through virtual effects. Conversely, leptoquarks with masses of 200 GeV can be discovered for couplings as small as 10−3 αem."
}

@article{Aaij:2019kcg,
    author = "{LHCb Collaboration}",
    title = "{Observation of CP Violation in Charm Decays}",
    eprint = "1903.08726",
    archivePrefix = "arXiv",
    primaryClass = "hep-ex",
    reportNumber = "LHCb-PAPER-2019-006, CERN-EP-2019-042",
    doi = "10.1103/PhysRevLett.122.211803",
    journal = "Phys. Rev. Lett.",
    volume = "122",
    number = "21",
    pages = "211803",
    year = "2019"
}

@article{Barger:2000gv,
      author         = "Barger, Vernon D. and Cheung, King-man",
      title          = "{Atomic parity violation, leptoquarks, and contact
                        interactions}",
      journal        = "Phys. Lett. B",
      volume         = "480",
      year           = "2000",
      pages          = "149-154",
      doi            = "10.1016/S0370-2693(00)00401-9",
      eprint         = "hep-ph/0002259",
      archivePrefix  = "arXiv",
      primaryClass   = "hep-ph",
      reportNumber   = "UCD-2000-8, MAD-00-1163",
      SLACcitation   = "%%CITATION = HEP-PH/0002259;%%"
}

@article{Catani:2009sm,
        author         = "Catani, Stefano and Cieri, Leandro and Ferrera, Giancarlo
                                  and de Florian, Daniel and Grazzini, Massimiliano",
        title          = "{Vector Boson Production at Hadron Colliders: A Fully Exclusive QCD Calculation at Next-to-Next-to-Leading Order}",
        journal        = "Phys. Rev. Lett.",
        volume         = "103",
        pages          = "082001",
        doi            = "10.1103/PhysRevLett.103.082001",
        year           = "2009",
        eprint         = "0903.2120",
        archivePrefix  = "arXiv",
        primaryClass   = "hep-ph",
        SLACcitation   = "%%CITATION = ARXIV:0903.2120;%%",
}

@article{Czakon:2011xx,
      author         = "Czakon, Michal and Mitov, Alexander",
      title          = "{Top++: A Program for the Calculation of the Top-Pair
                       Cross-Section at Hadron Colliders}",
      year           = "2014",
      journal       = "Compt/~Phys.~Commun.",
      vol             = 185,
      pages        = 2930,
      eprint         = "1112.5675",
      archivePrefix  = "arXiv",
      primaryClass   = "hep-ph",
      SLACcitation   = "%%CITATION = ARXIV:1112.5675;%%",
}

@article{Kidonakis:2011wy,
      author         = "Kidonakis, Nikolaos",
      title          = "{Next-to-next-to-leading logarithm resummation for 
s-channel single top quark production}",
      journal        = "Phys. Rev. D",
      volume         = "83",
      year           = "2011",
      pages          = "091503",
      doi            = "10.1103/PhysRevD.83.091503",
      eprint         = "1103.2792",
      archivePrefix  = "arXiv",
      primaryClass   = "hep-ph",
      SLACcitation   = "%%CITATION = ARXIV:1103.2792;%%"
}

@article{Kidonakis:2010ux,
      author         = "Kidonakis, Nikolaos",
      title          = "{Two-loop soft anomalous dimensions for single top quark
                        associated production with a W- or H-}",
      journal        = "Phys. Rev. D",
      volume         = "82",
      year           = "2010",
      pages          = "054018",
      doi            = "10.1103/PhysRevD.82.054018",
      eprint         = "1005.4451",
      archivePrefix  = "arXiv",
      primaryClass   = "hep-ph",
      SLACcitation   = "%%CITATION = ARXIV:1005.4451;%%"
}

@article{Kidonakis:2010tc,
      author         = "Kidonakis, Nikolaos",
      title          = "{NNLL resummation for s-channel single top quark
                        production}",
      journal        = "Phys. Rev. D",
      volume         = "81",
      pages          = "054028",
      doi            = "10.1103/PhysRevD.81.054028",
      year           = "2010",
      eprint         = "1001.5034",
      archivePrefix  = "arXiv",
      primaryClass   = "hep-ph",
      SLACcitation   = "%%CITATION = ARXIV:1001.5034;%%",
}

@article{Lonnblad:2012ix,
      author         = "L{\"o}nnblad, Leif and Prestel, Stefan",
      title          = "{Merging Multi-leg NLO Matrix Elements with Parton
                        Showers}",
      journal        = "JHEP",
      volume         = "03",
      year           = "2013",
      pages          = "166",
      doi            = "10.1007/JHEP03(2013)166",
      eprint         = "1211.7278",
      archivePrefix  = "arXiv",
      primaryClass   = "hep-ph",
      reportNumber   = "MCNET-12-17, LU-TP-12-43",
      SLACcitation   = "%%CITATION = ARXIV:1211.7278;%%"
}

@article{Beenakker:2016lwe,
      author         = "Beenakker, Wim and Borschensky, Christoph and Kr{\"a}mer, Michael and Kulesza, Anna and Laenen, Eric",
      title          = "{NNLL-fast: predictions for coloured supersymmetric particle production at the LHC with threshold and Coulomb resummation}",
      journal        = "JHEP",
      volume         = "12",
      year           = "2016",
      pages          = "133",
      doi            = "10.1007/JHEP12(2016)133",
      eprint         = "1607.07741",
      archivePrefix  = "arXiv",
      primaryClass   = "hep-ph",
      reportNumber   = "MS-TP-16-17, NIKHEF-2016-037, TTK-16-28",
      SLACcitation   = "%%CITATION = ARXIV:1607.07741;%%"
}

@article{Beenakker:2016gmf,
      author         = "Beenakker, Wim and Borschensky, Christoph and Heger, Raphael and Kr{\"a}mer, Michael and Kulesza, Anna and Laenen, Eric",
      title          = "{NNLL resummation for stop pair-production at the LHC}",
      journal        = "JHEP",
      volume         = "05",
      year           = "2016",
      pages          = "153",
      doi            = "10.1007/JHEP05(2016)153",
      eprint         = "1601.02954",
      archivePrefix  = "arXiv",
      primaryClass   = "hep-ph",
      reportNumber   = "MS-TP-16-01, TTK-16-01, NIKHEF-2016-002",
      SLACcitation   = "%%CITATION = ARXIV:1601.02954;%%"
}

@Article{Beenakker:1997ut,
     author    = "Beenakker, W. and Kr{\"a}mer, M. and Plehn, T. and Spira, M. and Zerwas, P. M.",
     title     = "Stop production at hadron colliders",
     journal   = "Nucl. Phys. B",
     volume    = "515",
     year      = "1998",
     pages     = "3",
     doi       = "10.1016/s0550-3213(98)00014-5",
     eprint    = "hep-ph/9710451",
     archivePrefix  = "arXiv",
     primaryClass   = "hep-ph",
     SLACcitation  = "%%CITATION = HEP-PH/9710451;%%"
}

@article{Beenakker:2010nq,
      author         = "Beenakker, Wim and Brensing, Silja and Kr{\"a}mer, Michael and Kulesza, Anna and Laenen, Eric and Niessen, Irene",
      title          = "{Supersymmetric top and bottom squark production at hadron colliders}",
      journal        = "JHEP",
      volume         = "08",
      year           = "2010",
      pages          = "098",
      doi            = "10.1007/JHEP08(2010)098",
      eprint         = "1006.4771",
      archivePrefix  = "arXiv",
      primaryClass   = "hep-ph",
      reportNumber   = "CERN-PH-TH-2010-142, ITFA-2010-015, ITP-UU-10-17,
                        NIKHEF-2010-016, TTK-10-33",
      SLACcitation   = "%%CITATION = ARXIV:1006.4771;%%"
}

@article{Aaboud:2019bye,
      author         = "{ATLAS Collaboration}",
      title          = "{Searches for third-generation scalar leptoquarks in
                        $\sqrt{s}$ = 13 TeV pp collisions with the ATLAS
                        detector}",
      collaboration  = "ATLAS",
      journal        = "JHEP",
      volume         = "06",
      year           = "2019",
      pages          = "144",
      doi            = "10.1007/JHEP06(2019)144",
      eprint         = "1902.08103",
      archivePrefix  = "arXiv",
      primaryClass   = "hep-ex",
      reportNumber   = "CERN-EP-2019-026",
      SLACcitation   = "%%CITATION = ARXIV:1902.08103;%%"
}

@article{Aaboud:2017opj,
    author         = "{ATLAS Collaboration}",
    title          = "{Search for \(B-L\) \(R\)-parity-violating top squarks in \(\sqrt{s} = 13\,\text{TeV}\) \(pp\) collisions with the ATLAS experiment}",
    journal        = "Phys. Rev. D",
    volume         = "97",
    year           = "2018",
    pages          = "032003",
    doi            = "10.1103/PhysRevD.97.032003",
    reportNumber   = "CERN-EP-2017-171",
    eprint         = "1710.05544",
    archivePrefix  = "arXiv",
    primaryClass   = "hep-ex",

      author         = "{ATLAS Collaboration}",
      title          = "{Search for B-L R-parity-violating top squarks in $\sqrt
                        s$ =13  TeV pp collisions with the ATLAS experiment}",
      collaboration  = "ATLAS",
      journal        = "Phys. Rev. D",
      volume         = "97",
      year           = "2018",
      number         = "3",
      pages          = "032003",
      doi            = "10.1103/PhysRevD.97.032003",
      eprint         = "1710.05544",
      archivePrefix  = "arXiv",
      primaryClass   = "hep-ex",
      reportNumber   = "CERN-EP-2017-171",
      SLACcitation   = "%%CITATION = ARXIV:1710.05544;%%"
}

@article{Dorsner:2018ynv,
      author         = "Doršner, Ilja and Greljo, Admir",
      title          = "{Leptoquark toolbox for precision collider studies}",
      journal        = "JHEP",
      volume         = "05",
      year           = "2018",
      pages          = "126",
      doi            = "10.1007/JHEP05(2018)126",
      eprint         = "1801.07641",
      archivePrefix  = "arXiv",
      primaryClass   = "hep-ph",
      reportNumber   = "MITP-18-005",
      SLACcitation   = "%%CITATION = ARXIV:1801.07641;%%"
}

@article{Buttazzo:2017ixm,
      author         = "Buttazzo, Dario and Greljo, Admir and Isidori, Gino and
                        Marzocca, David",
      title          = "{B-physics anomalies: a guide to combined explanations}",
      journal        = "JHEP",
      volume         = "11",
      year           = "2017",
      pages          = "044",
      doi            = "10.1007/JHEP11(2017)044",
      eprint         = "1706.07808",
      archivePrefix  = "arXiv",
      primaryClass   = "hep-ph",
      reportNumber   = "ZU-TH-18-17",
      SLACcitation   = "%%CITATION = ARXIV:1706.07808;%%"
}

@article{DiLuzio:2017chi,
      author         = "Di Luzio, Luca and Nardecchia, Marco",
      title          = "{What is the scale of new physics behind the $B$-flavour
                        anomalies?}",
      journal        = "Eur. Phys. J. C",
      volume         = "77",
      year           = "2017",
      number         = "8",
      pages          = "536",
      doi            = "10.1140/epjc/s10052-017-5118-9",
      eprint         = "1706.01868",
      archivePrefix  = "arXiv",
      primaryClass   = "hep-ph",
      reportNumber   = "IPPP-17-50, CERN-TH-2017-126",
      SLACcitation   = "%%CITATION = ARXIV:1706.01868;%%"
}

@article{Bauer:2015knc,
      author         = "Bauer, Martin and Neubert, Matthias",
      title          = "{Minimal Leptoquark Explanation for the R$_{D^{(*)}}$ ,
                        R$_K$ , and $(g-2)_\mu$ Anomalies}",
      journal        = "Phys. Rev. Lett.",
      volume         = "116",
      year           = "2016",
      number         = "14",
      pages          = "141802",
      doi            = "10.1103/PhysRevLett.116.141802",
      eprint         = "1511.01900",
      archivePrefix  = "arXiv",
      primaryClass   = "hep-ph",
      reportNumber   = "MITP-15-100",
      SLACcitation   = "%%CITATION = ARXIV:1511.01900;%%"
}

@article{Freytsis:2015qca,
      author         = "Freytsis, Marat and Ligeti, Zoltan and Ruderman, Joshua
                        T.",
      title          = "{Flavor models for $\bar{B} \to D^{(*)} \tau \bar{\nu}$}",
      journal        = "Phys. Rev. D",
      volume         = "92",
      year           = "2015",
      number         = "5",
      pages          = "054018",
      doi            = "10.1103/PhysRevD.92.054018",
      eprint         = "1506.08896",
      archivePrefix  = "arXiv",
      primaryClass   = "hep-ph",
      SLACcitation   = "%%CITATION = ARXIV:1506.08896;%%"
}

@article{Gripaios:2014tna,
      author         = "Gripaios, Ben and Nardecchia, Marco and Renner, S. A.",
      title          = "{Composite leptoquarks and anomalies in $B$-meson
                        decays}",
      journal        = "JHEP",
      volume         = "05",
      year           = "2015",
      pages          = "006",
      doi            = "10.1007/JHEP05(2015)006",
      eprint         = "1412.1791",
      archivePrefix  = "arXiv",
      primaryClass   = "hep-ph",
      SLACcitation   = "%%CITATION = ARXIV:1412.1791;%%"
}

@article{Hiller:2014yaa,
      author         = "Hiller, Gudrun and Schmaltz, Martin",
      title          = "{$R_K$ and future $b \to s \ell \ell$ physics beyond the
                        standard model opportunities}",
      journal        = "Phys. Rev. D",
      volume         = "90",
      year           = "2014",
      pages          = "054014",
      doi            = "10.1103/PhysRevD.90.054014",
      eprint         = "1408.1627",
      archivePrefix  = "arXiv",
      primaryClass   = "hep-ph",
      reportNumber   = "DO-TH-14-17",
      SLACcitation   = "%%CITATION = ARXIV:1408.1627;%%"
}

@article{Aaij:2014ora,
      author = "{LHCb Collaboration}",
      title          = "{Test of lepton universality using $B^{+}\rightarrow
                        K^{+}\ell^{+}\ell^{-}$ decays}",
      collaboration  = "LHCb Collaboration",
      journal        = "Phys. Rev. Lett.",
      volume         = "113",
      year           = "2014",
      pages          = "151601",
      doi            = "10.1103/PhysRevLett.113.151601",
      eprint         = "1406.6482",
      archivePrefix  = "arXiv",
      primaryClass   = "hep-ex",
      reportNumber   = "CERN-PH-EP-2014-140, LHCB-PAPER-2014-024",
      SLACcitation   = "%%CITATION = ARXIV:1406.6482;%%"
}

@article{Pas:2015hca,
      author         = "Päs, Heinrich and Schumacher, Erik",
      title          = "{Common origin of $R_K$ and neutrino masses}",
      journal        = "Phys. Rev. D",
      volume         = "92",
      year           = "2015",
      number         = "11",
      pages          = "114025",
      doi            = "10.1103/PhysRevD.92.114025",
      eprint         = "1510.08757",
      archivePrefix  = "arXiv",
      primaryClass   = "hep-ph",
      SLACcitation   = "%%CITATION = ARXIV:1510.08757;%%"
}

@article{Dorsner:2017wwn,
      author         = "Doršner, Ilja and Fajfer, Svjetlana and Košnik, Nejc",
      title          = "{Leptoquark mechanism of neutrino masses within the grand
                        unification framework}",
      journal        = "Eur. Phys. J. C",
      volume         = "77",
      year           = "2017",
      number         = "6",
      pages          = "417",
      doi            = "10.1140/epjc/s10052-017-4987-2",
      eprint         = "1701.08322",
      archivePrefix  = "arXiv",
      primaryClass   = "hep-ph",
      SLACcitation   = "%%CITATION = ARXIV:1701.08322;%%"
}

@article{Popov:2016fzr,
      author         = "Popov, Oleg and White, Graham A",
      title          = "{One Leptoquark to unify them? Neutrino masses and
                        unification in the light of $(g-2)_\mu$, $R_{D^{(\star)}}$
                        and $R_K$ anomalies}",
      journal        = "Nucl. Phys. B",
      volume         = "923",
      year           = "2017",
      pages          = "324-338",
      doi            = "10.1016/j.nuclphysb.2017.08.007",
      eprint         = "1611.04566",
      archivePrefix  = "arXiv",
      primaryClass   = "hep-ph",
      reportNumber   = "COEPP-MN-17-3, UCRHEP-T57",
      SLACcitation   = "%%CITATION = ARXIV:1611.04566;%%"
}

@article{Babu:2010vp,
      author         = "Babu, K. S. and Julio, J.",
      title          = "{Two-Loop Neutrino Mass Generation through Leptoquarks}",
      journal        = "Nucl. Phys. B",
      volume         = "841",
      year           = "2010",
      pages          = "130-156",
      doi            = "10.1016/j.nuclphysb.2010.07.022",
      eprint         = "1006.1092",
      archivePrefix  = "arXiv",
      primaryClass   = "hep-ph",
      reportNumber   = "OSU-HEP-10-05",
      SLACcitation   = "%%CITATION = ARXIV:1006.1092;%%"
}

@article{PERF-2018-01,
      author         = "{ATLAS Collaboration}",
      title          = "{Electron and photon performance measurements with the
                        ATLAS detector using the 2015-2017 LHC proton-proton
                        collision data}",
      collaboration  = "ATLAS",
      doi            = "10.1088/1748-0221/14/12/P12006",
      year           = "2019",
      eprint         = "1908.00005",
      archivePrefix  = "arXiv",
      primaryClass   = "hep-ex",
      reportNumber   = "CERN-EP-2019-145",
      journal        = "JINST",
      volume         = "14",
      pages          = "P12006",
      SLACcitation   = "%%CITATION = ARXIV:1908.00005;%%"
}

@article{Aad:2019aic,
    author         = "{ATLAS Collaboration}",
    title          = "{ATLAS \(b\)-jet identification performance and efficiency measurement with \(t\bar{t}\) events in \(pp\) collisions at \(\sqrt{s} = 13\,\text{TeV}\)}",
    year           = "2019",
    reportNumber   = "CERN-EP-2019-132",
    eprint         = "1907.05120",
    archivePrefix  = "arXiv",
    primaryClass   = "hep-ex",
    journal        = "Eur. Phys. J. C",
    volume         = "79",
    pages          = "970",
    doi            = "10.1140/epjc/s10052-019-7450-8",
    SLACcitation   = "%%CITATION = ARXIV:1907.05120;%%"
}

@Article{Cowan:2010st,
    Author         = {Cowan, Glen and Cranmer, Kyle and Gross, Eilam and Vitells, Ofer},
    Eprint         = {1007.1727},
    Archiveprefix  = {arXiv},
    Primaryclass   = {physics.data-an},
    Journal        = {Eur. Phys. J. C},
    Pages          = {1554},
    doi            = "10.1140/epjc/s10052-011-1554-0",
    Title          = {{Asymptotic formulae for likelihood-based tests of new physics}},
    Volume         = {71},
    Year           = {2011},
    note       = "[Erratum: \href{http://dx.doi.org/10.1140/epjc/s10052-013-2501-z}{Eur. Phys. J. C {\textbf{73}} (2013) 2501}]"
}

@article{LUCID2,
  author={G. Avoni and others},
  title={The new LUCID-2 detector for luminosity measurement and monitoring in ATLAS},
  journal={JINST},
  volume={13},
  number={07},
  pages={P07017},
  doi="10.1088/1748-0221/13/07/P07017",
  year={2018},
}

@Article{Cacciari:2008gp,
     author    = "Cacciari, Matteo and Salam, Gavin P. and Soyez, Gregory",
     title     = "{The anti-\(k_t\) jet clustering algorithm}",
     journal   = "JHEP",
     volume    = "04",
     year      = "2008",
     pages     = "063",
     eprint    = "0802.1189",
     archivePrefix = "arXiv",
     primaryClass  =  "hep-ph",
     doi       = "10.1088/1126-6708/2008/04/063",
     SLACcitation  = "%%CITATION = 0802.1189;%%"
}

@Article{Fastjet,
      author         = "Cacciari, Matteo and Salam, Gavin P. and Soyez, Gregory",
      title          = "{FastJet user manual}",
      journal        = "Eur. Phys. J. C",
      volume         = "72",
      year           = "2012",
      pages          = "1896",
      doi            = "10.1140/epjc/s10052-012-1896-2",
      eprint         = "1111.6097",
      archivePrefix  = "arXiv",
      primaryClass   = "hep-ph",
      reportNumber   = "CERN-PH-TH-2011-297",
      SLACcitation   = "%%CITATION = ARXIV:1111.6097;%%"
}

@Article{Butterworth:2015oua,
      author         = "Butterworth, Jon and others",
      title          = "{PDF4LHC recommendations for LHC Run II}",
      journal        = "J. Phys. G",
      volume         = "43",
      year           = "2016",
      pages          = "023001",
      doi            = "10.1088/0954-3899/43/2/023001",
      eprint         = "1510.03865",
      archivePrefix  = "arXiv",
      primaryClass   = "hep-ph",
      reportNumber   = "OUTP-15-17P, SMU-HEP-15-12, TIF-UNIMI-2015-14,
                        LCTS-2015-27, CERN-PH-TH-2015-249",
      SLACcitation   = "%%CITATION = ARXIV:1510.03865;%%"
}

@article{Ball:2012cx,
      author         = "Ball, Richard D. and others",
      title          = "{Parton distributions with LHC data}",
      journal        = "Nucl. Phys. B",
      volume         = "867",
      year           = "2013",
      pages          = "244",
      doi            = "10.1016/j.nuclphysb.2012.10.003",
      eprint         = "1207.1303",
      archivePrefix  = "arXiv",
      primaryClass   = "hep-ph",
      reportNumber   = "EDINBURGH-2012-08, IFUM-FT-997, FR-PHENO-2012-014,
                        RWTH-TTK-12-25, CERN-PH-TH-2012-037, SFB-CPP-12-47\,
                        --CERN-PH-TH-2012-037",
      SLACcitation   = "%%CITATION = ARXIV:1207.1303;%%"
}

@article{Ball:2014uwa,
      author         = "Ball, Richard D. and others",
      title          = "{Parton distributions for the LHC Run II}",
      collaboration  = "NNPDF",
      journal        = "JHEP",
      volume         = "04",
      year           = "2015",
      pages          = "040",
      doi            = "10.1007/JHEP04(2015)040",
      eprint         = "1410.8849",
      archivePrefix  = "arXiv",
      primaryClass   = "hep-ph",
      reportNumber   = "EDINBURGH-2014-15, IFUM-1034-FT, CERN-PH-TH-2013-253,
                        OUTP-14-11P, CAVENDISH-HEP-14-11",
      SLACcitation   = "%%CITATION = ARXIV:1410.8849;%%"
}

@article{Sjostrand:2014zea,
      author         = "Sj{\"o}strand, Torbj{\"o}rn and Ask, Stefan and Christiansen,
                        Jesper R. and Corke, Richard and Desai, Nishita and Ilten,
                        Philip and Mrenna, Stephen and Prestel, Stefan and
                        Rasmussen, Christine O. and Skands, Peter Z.",
      title          = "{An introduction to PYTHIA 8.2}",
      journal        = "Comput. Phys. Commun.",
      volume         = "191",
      year           = "2015",
      pages          = "159",
      doi            = "10.1016/j.cpc.2015.01.024",
      eprint         = "1410.3012",
      archivePrefix  = "arXiv",
      primaryClass   = "hep-ph",
      reportNumber   = "LU-TP-14-36, MCNET-14-22, CERN-PH-TH-2014-190,
                        FERMILAB-PUB-14-316-CD, DESY-14-178, SLAC-PUB-16122,
                        --FERMILAB-PUB-14-316-CD",
      SLACcitation   = "%%CITATION = ARXIV:1410.3012;%%"
}

@Article{Lange:2001uf,
      author         = "Lange, D. J.",
      title          = "{The EvtGen particle decay simulation package}",
      booktitle      = "{Proceedings, 7th International Conference on B physics
                        at hadron machines (BEAUTY 2000)}",
      journal        = "Nucl. Instrum. Meth. A",
      volume         = "462",
      year           = "2001",
      pages          = "152",
      doi            = "10.1016/S0168-9002(01)00089-4",
      SLACcitation   = "%%CITATION = NUIMA,A462,152;%%"
}

@Article{Alwall:2014hca,
      author         = "Alwall, J. and Frederix, R. and Frixione, S. and Hirschi,
                        V. and Maltoni, F. and Mattelaer, O. and Shao, H. -S. and
                        Stelzer, T. and Torrielli, P. and Zaro, M.",
      title          = "{The automated computation of tree-level and
                        next-to-leading order differential cross sections, and
                        their matching to parton shower simulations}",
      journal        = "JHEP",
      volume         = "07",
      year           = "2014",
      pages          = "079",
      doi            = "10.1007/JHEP07(2014)079",
      eprint         = "1405.0301",
      archivePrefix  = "arXiv",
      primaryClass   = "hep-ph",
      reportNumber   = "CERN-PH-TH-2014-064, CP3-14-18, LPN14-066, MCNET-14-09,
                        ZU-TH-14-14",
      SLACcitation   = "%%CITATION = ARXIV:1405.0301;%%"
}

@Article{Alioli:2010xd,
      author         = "Alioli, Simone and Nason, Paolo and Oleari, Carlo and Re,
                        Emanuele",
      title          = "{A general framework for implementing NLO calculations in
                        shower Monte Carlo programs: the POWHEG BOX}",
      journal        = "JHEP",
      volume         = "06",
      pages          = "043",
      doi            = "10.1007/JHEP06(2010)043",
      year           = "2010",
      eprint         = "1002.2581",
      archivePrefix  = "arXiv",
      primaryClass   = "hep-ph",
}

@Article{Gleisberg:2008ta,
      author         = "Gleisberg, T. and H{\"o}che, S. and Krauss, F. and Sch{\"o}nherr, M. and Schumann, S. and Siegert, F. and Winter J.",
      title          = "{Event generation with SHERPA 1.1}",
      journal        = "JHEP",
      volume         = "02",
      pages          = "007",
      doi            = "10.1088/1126-6708/2009/02/007",
      year           = "2009",
      eprint         = "0811.4622",
      archivePrefix  = "arXiv",
      primaryClass   = "hep-ph",
}

@Article{Gleisberg:2008fv,
      author         = "Gleisberg, Tanju and H{\"o}che, Stefan",
      title          = "{Comix, a new matrix element generator}",
      journal        = "JHEP",
      volume         = "12",
      pages          = "039",
      doi            = "10.1088/1126-6708/2008/12/039",
      year           = "2008",
      eprint         = "0808.3674",
      archivePrefix  = "arXiv",
      primaryClass   = "hep-ph",
}

@Article{Schumann:2007mg,
      author         = "Schumann, Steffen and Krauss, Frank",
      title          = "{A parton shower algorithm based on Catani--Seymour dipole factorisation}",
      journal        = "JHEP",
      volume         = "03",
      pages          = "038",
      doi            = "10.1088/1126-6708/2008/03/038",
      year           = "2008",
      eprint         = "0709.1027",
      archivePrefix  = "arXiv",
      primaryClass   = "hep-ph",
}

@article{Cascioli:2011va,
      author         = "Cascioli, Fabio and Maierh{\"o}fer, Philipp and Pozzorini, Stefano",
      title          = "{Scattering Amplitudes with Open Loops}",
      journal        = "Phys. Rev. Lett.",
      volume         = "108",
      year           = "2012",
      pages          = "111601",
      doi            = "10.1103/PhysRevLett.108.111601",
      eprint         = "1111.5206",
      archivePrefix  = "arXiv",
      primaryClass   = "hep-ph",
      reportNumber   = "ZU-TH-23-11, LPN11-66",
      SLACcitation   = "%%CITATION = ARXIV:1111.5206;%%"
}

@article{Hoeche:2012yf,
      author         = "H{\"o}che, Stefan and Krauss, Frank and Sch{\"o}nherr, Marek and
                        Siegert, Frank",
      title          = "{QCD matrix elements + parton showers. The NLO case}",
      journal        = "JHEP",
      volume         = "04",
      year           = "2013",
      pages          = "027",
      doi            = "10.1007/JHEP04(2013)027",
      eprint         = "1207.5030",
      archivePrefix  = "arXiv",
      primaryClass   = "hep-ph",
      reportNumber   = "SLAC-PUB-15191, IPPP-12-52, DCPT-12-104, LPN12-081,
                        FR-PHENO-2012-017, MCNET-12-09, --FR-PHENO-2012-017",
      SLACcitation   = "%%CITATION = ARXIV:1207.5030;%%"
}

@article{Agostinelli:2002hh,
      author         = "Agostinelli, S. and others",
      title          = "{\textsc{Geant4} -- a simulation toolkit}",
      journal        = "Nucl. Instrum. Meth. A",
      volume         = "506",
      year           = "2003",
      pages          = "250",
      doi            = "10.1016/S0168-9002(03)01368-8",
      reportNumber   = "SLAC-PUB-9350, FERMILAB-PUB-03-339",
      SLACcitation   = "%%CITATION = NUIMA,A506,250;%%"
}

@Article{Read:2002hq,
    Author = {Read, Alexander L.},
    Journal = {J. Phys. G},
    Pages = {2693},
    doi = "10.1088/0954-3899/28/10/313",
    Title = {{Presentation of search results: the \(CL_S\) technique}},
    Volume = {28},
    Year = {2002}
}

@Article{PERF-2007-01,
    author         = "{ATLAS Collaboration}",
    title          = "{The ATLAS Experiment at the CERN Large Hadron Collider}",
    journal        = "JINST",
    volume         = "3",
    year           = "2008",
    pages          = "S08003",
    doi            = "10.1088/1748-0221/3/08/S08003",
    primaryClass   = "hep-ex",
}

@Article{SOFT-2010-01,
    author         = "{ATLAS Collaboration}",
    title          = "{The ATLAS Simulation Infrastructure}",
    journal        = "Eur. Phys. J. C",
    volume         = "70",
    year           = "2010",
    pages          = "823",
    doi            = "10.1140/epjc/s10052-010-1429-9",
    eprint         = "1005.4568",
    archivePrefix  = "arXiv",
    primaryClass   = "physics.ins-det",
}

@Article{PERF-2011-04,
    author         = "{ATLAS Collaboration}",
    title          = "{Jet energy resolution in proton--proton collisions at \(\sqrt{s} = 7\,\text{TeV}\) recorded in 2010 with the ATLAS detector}",
    journal        = "Eur. Phys. J. C",
    volume         = "73",
    year           = "2013",
    pages          = "2306",
    doi            = "10.1140/epjc/s10052-013-2306-0",
    reportNumber   = "CERN-PH-EP-2012-191",
    eprint         = "1210.6210",
    archivePrefix  = "arXiv",
    primaryClass   = "hep-ex",
}

@Article{PERF-2014-03,
    author         = "{ATLAS Collaboration}",
    title          = "{Performance of pile-up mitigation techniques for jets in \(pp\) collisions at \(\sqrt{s} = 8\,\text{TeV}\) using the ATLAS detector}",
    journal        = "Eur. Phys. J. C",
    volume         = "76",
    year           = "2016",
    pages          = "581",
    doi            = "10.1140/epjc/s10052-016-4395-z",
    reportNumber   = "CERN-PH-EP-2015-206",
    eprint         = "1510.03823",
    archivePrefix  = "arXiv",
    primaryClass   = "hep-ex",
}

@Article{PERF-2014-07,
    author         = "{ATLAS Collaboration}",
    title          = "{Topological cell clustering in the ATLAS calorimeters and its performance in LHC Run~1}",
    journal        = "Eur. Phys. J. C",
    volume         = "77",
    year           = "2017",
    pages          = "490",
    doi            = "10.1140/epjc/s10052-017-5004-5",
    reportNumber   = "CERN-PH-EP-2015-304",
    eprint         = "1603.02934",
    archivePrefix  = "arXiv",
    primaryClass   = "hep-ex",
}

@Article{PERF-2015-10,
    author         = "{ATLAS Collaboration}",
    title          = "{Muon reconstruction performance of the ATLAS detector in proton--proton collision data at \(\sqrt{s} = 13\,\text{TeV}\)}",
    journal        = "Eur. Phys. J. C",
    volume         = "76",
    year           = "2016",
    pages          = "292",
    doi            = "10.1140/epjc/s10052-016-4120-y",
    reportNumber   = "CERN-EP-2016-033",
    eprint         = "1603.05598",
    archivePrefix  = "arXiv",
    primaryClass   = "hep-ex",
}

@Article{STDM-2015-05,
    author         = "{ATLAS Collaboration}",
    title          = "{Measurement of the Inelastic Proton--Proton Cross Section at \(\sqrt{s} = 13\,\text{TeV}\) with the ATLAS Detector at the LHC}",
    journal        = "Phys. Rev. Lett.",
    volume         = "117",
    year           = "2016",
    pages          = "182002",
    doi            = "10.1103/PhysRevLett.117.182002",
    reportNumber   = "CERN-EP-2016-140",
    eprint         = "1606.02625",
    archivePrefix  = "arXiv",
    primaryClass   = "hep-ex",
}

@Article{EXOT-2016-08,
    author         = "{ATLAS Collaboration}",
    title          = "{Searches for scalar leptoquarks and differential cross-section measurements in dilepton--dijet events in proton--proton collisions at a centre-of-mass energy of \(\sqrt{s} = 13\,\text{TeV}\) with the ATLAS experiment}",
    journal        = "Eur. Phys. J. C",
    volume         = "79",
    year           = "2019",
    pages          = "733",
    doi            = "10.1140/epjc/s10052-019-7181-x",
    reportNumber   = "CERN-EP-2018-262",
    eprint         = "1902.00377",
    archivePrefix  = "arXiv",
    primaryClass   = "hep-ex",
}

@Article{HIGG-2016-29,
    author         = "{ATLAS Collaboration}",
    title          = "{Evidence for the \(H \rightarrow b\bar{b}\) decay with the ATLAS detector}",
    journal        = "JHEP",
    volume         = "12",
    year           = "2017",
    pages          = "024",
    doi            = "10.1007/JHEP12(2017)024",
    reportNumber   = "CERN-EP-2017-175",
    eprint         = "1708.03299",
    archivePrefix  = "arXiv",
    primaryClass   = "hep-ex",
}

@Article{PERF-2016-04,
    author         = "{ATLAS Collaboration}",
    title          = "{Jet energy scale measurements and their systematic uncertainties in proton--proton collisions at \(\sqrt{s} = 13\,\text{TeV}\) with the ATLAS detector}",
    journal        = "Phys. Rev. D",
    volume         = "96",
    year           = "2017",
    pages          = "072002",
    doi            = "10.1103/PhysRevD.96.072002",
    reportNumber   = "CERN-EP-2017-038",
    eprint         = "1703.09665",
    archivePrefix  = "arXiv",
    primaryClass   = "hep-ex",
}

@Article{PERF-2016-05,
    author         = "{ATLAS Collaboration}",
    title          = "{Measurements of \(b\)-jet tagging efficiency with the ATLAS detector using \(t\bar{t}\) events at \(\sqrt{s} = 13\,\text{TeV}\)}",
    journal        = "JHEP",
    volume         = "08",
    year           = "2018",
    pages          = "089",
    doi            = "10.1007/JHEP08(2018)089",
    reportNumber   = "CERN-EP-2018-047",
    eprint         = "1805.01845",
    archivePrefix  = "arXiv",
    primaryClass   = "hep-ex",
}

@Article{PERF-2016-07,
    author         = "{ATLAS Collaboration}",
    title          = "{Performance of missing transverse momentum reconstruction with the ATLAS detector using proton--proton collisions at \(\sqrt{s} = 13\,\text{TeV}\)}",
    journal        = "Eur. Phys. J. C",
    volume         = "78",
    year           = "2018",
    pages          = "903",
    doi            = "10.1140/epjc/s10052-018-6288-9",
    reportNumber   = "CERN-EP-2017-274",
    eprint         = "1802.08168",
    archivePrefix  = "arXiv",
    primaryClass   = "hep-ex",
}

@Article{TRIG-2016-01,
    author         = "{ATLAS Collaboration}",
    title          = "{Performance of the ATLAS trigger system in 2015}",
    journal        = "Eur. Phys. J. C",
    volume         = "77",
    year           = "2017",
    pages          = "317",
    doi            = "10.1140/epjc/s10052-017-4852-3",
    reportNumber   = "CERN-EP-2016-241",
    eprint         = "1611.09661",
    archivePrefix  = "arXiv",
    primaryClass   = "hep-ex",
}

@Article{PIX-2018-001,
    author         = "Abbott, B. and others",
    title          = "{Production and integration of the ATLAS Insertable B-Layer}",
    journal        = "JINST",
    volume         = "13",
    year           = "2018",
    pages          = "T05008",
    doi            = "10.1088/1748-0221/13/05/T05008",
    eprint         = "1803.00844",
    archivePrefix  = "arXiv",
    primaryClass   = "physics.ins-det",
}

@Booklet{ATLAS-TDR-2010-19,
    author         = "{ATLAS Collaboration}",
    title          = "{ATLAS Insertable B-Layer Technical Design Report}",
    howpublished   = "ATLAS-TDR-19",
    url            = "https://cds.cern.ch/record/1291633",
    year           = "2010",
    related        = "ATLAS-TDR-2010-19-add",
    relatedstring  = "Addendum:",
}

@Article{CMS-B2G-16-016,
    author         = "{CMS Collaboration}",
    title          = "{Searches for \(W'\) bosons decaying to a top quark and a bottom quark in proton--proton collisions at \(13\,\text{TeV}\)}",
    journal        = "JHEP",
    volume         = "08",
    year           = "2017",
    pages          = "029",
    doi            = "10.1007/JHEP08(2017)029",
    reportNumber   = "CERN-EP-2017-090",
    eprint         = "1706.04260",
    archivePrefix  = "arXiv",
    primaryClass   = "hep-ex",
}

@Article{CMS-B2G-16-027,
    author         = "{CMS Collaboration}",
    title          = "{Search for Leptoquarks Coupled to Third-Generation Quarks in Proton--Proton Collisions at \(\sqrt{s} = 13\,\text{TeV}\)}",
    journal        = "Phys. Rev. Lett.",
    volume         = "121",
    year           = "2018",
    pages          = "241802",
    doi            = "10.1103/PhysRevLett.121.241802",
    reportNumber   = "CERN-EP-2018-233",
    eprint         = "1809.05558",
    archivePrefix  = "arXiv",
    primaryClass   = "hep-ex",
}

@Article{CMS-B2G-16-028,
    author         = "{CMS Collaboration}",
    title          = "{Search for third-generation scalar leptoquarks decaying to a top quark and a \(\tau\) lepton at \(\sqrt{s} = 13\,\text{TeV}\)}",
    journal        = "Eur. Phys. J. C",
    volume         = "78",
    year           = "2018",
    pages          = "707",
    doi            = "10.1140/epjc/s10052-018-6143-z",
    reportNumber   = "CERN-EP-2018-019",
    eprint         = "1803.02864",
    archivePrefix  = "arXiv",
    primaryClass   = "hep-ex",
}

@Article{CMS-EXO-17-003,
    author         = "{CMS Collaboration}",
    title          = "{Search for pair production of second-generation leptoquarks at \(\sqrt{s} = 13\,\text{TeV}\)}",
    journal        = "Phys. Rev. D",
    volume         = "99",
    year           = "2019",
    pages          = "032014",
    doi            = "10.1103/PhysRevD.99.032014",
    reportNumber   = "CERN-EP-2018-218",
    eprint         = "1808.05082",
    archivePrefix  = "arXiv",
    primaryClass   = "hep-ex",
}

@Article{CMS-EXO-17-009,
    author         = "{CMS Collaboration}",
    title          = "{Search for pair production of first-generation scalar leptoquarks at \(\sqrt{s} = 13\,\text{TeV}\)}",
    journal        = "Phys. Rev. D",
    volume         = "99",
    year           = "2019",
    pages          = "052002",
    doi            = "10.1103/PhysRevD.99.052002",
    reportNumber   = "CERN-EP-2018-265",
    eprint         = "1811.01197",
    archivePrefix  = "arXiv",
    primaryClass   = "hep-ex",
}

@Booklet{ATLAS-CONF-2019-021,
    author         = "{ATLAS Collaboration}",
    title          = "{Luminosity determination in \(pp\) collisions at \(\sqrt{s} = 13\,\text{TeV}\) using the ATLAS detector at the LHC}",
    howpublished   = "{ATLAS-CONF-2019-021}",
    url            = "https://cds.cern.ch/record/2677054",
    year           = "2019",
}

@Booklet{ATL-PHYS-PUB-2014-021,
    author         = "{ATLAS Collaboration}",
    title          = "{ATLAS Pythia~8 tunes to \(7~\text{TeV}\) data}",
    howpublished   = "{ATL-PHYS-PUB-2014-021}",
    url            = "https://cds.cern.ch/record/1966419",
    year           = "2014",
}

@Booklet{ATL-PHYS-PUB-2016-017,
    author         = "{ATLAS Collaboration}",
    title          = "{The Pythia~8 A3 tune description of ATLAS minimum bias and inelastic measurements incorporating the Donnachie--Landshoff diffractive model}",
    howpublished   = "{ATL-PHYS-PUB-2016-017}",
    url            = "https://cds.cern.ch/record/2206965",
    year           = "2016",
}

\clearpage 
 
\begin{flushleft}
\hypersetup{urlcolor=black}
{\Large The ATLAS Collaboration}

\bigskip

\AtlasOrcid[0000-0002-6665-4934]{G.~Aad}$^\textrm{\scriptsize 102}$,    
\AtlasOrcid[0000-0002-5888-2734]{B.~Abbott}$^\textrm{\scriptsize 128}$,    
\AtlasOrcid{D.C.~Abbott}$^\textrm{\scriptsize 103}$,    
\AtlasOrcid[0000-0002-2788-3822]{A.~Abed~Abud}$^\textrm{\scriptsize 36}$,    
\AtlasOrcid[0000-0002-1002-1652]{K.~Abeling}$^\textrm{\scriptsize 53}$,    
\AtlasOrcid[0000-0002-2987-4006]{D.K.~Abhayasinghe}$^\textrm{\scriptsize 94}$,    
\AtlasOrcid[0000-0002-8496-9294]{S.H.~Abidi}$^\textrm{\scriptsize 166}$,    
\AtlasOrcid[0000-0002-8279-9324]{O.S.~AbouZeid}$^\textrm{\scriptsize 40}$,    
\AtlasOrcid{N.L.~Abraham}$^\textrm{\scriptsize 155}$,    
\AtlasOrcid[0000-0001-5329-6640]{H.~Abramowicz}$^\textrm{\scriptsize 160}$
% ...
\bigskip
\\

% ...
$^{36}$CERN, Geneva; Switzerland.\\
% ...
$^{40}$Niels Bohr Institute, University of Copenhagen, Copenhagen; Denmark.\\
% ...
$^{53}$II. Physikalisches Institut, Georg-August-Universit\"{a}t G\"ottingen, G\"ottingen; Germany.\\
% ...
$^{94}$Department of Physics, Royal Holloway University of London, Egham; United Kingdom.\\
% ...
$^{102}$CPPM, Aix-Marseille Universit\'e, CNRS/IN2P3, Marseille; France.\\
$^{103}$Department of Physics, University of Massachusetts, Amherst MA; United States of America.\\
% ...
$^{128}$Homer L. Dodge Department of Physics and Astronomy, University of Oklahoma, Norman OK; United States of America.\\
% ...
$^{155}$Department of Physics and Astronomy, University of Sussex, Brighton; United Kingdom.\\
% ...
$^{160}$Raymond and Beverly Sackler School of Physics and Astronomy, Tel Aviv University, Tel Aviv; Israel.\\
% ...
$^{166}$Department of Physics, University of Toronto, Toronto ON; Canada.\\
% ...

\end{flushleft}


\end{document}